\documentclass[prd,tightenlines,floatfix,showpacs,preprintnumbers,nofootinbib,eqsecnum,superscriptaddress,twocolumn]{revtex4-1}

\usepackage{color}      
\usepackage{amsbsy}     
\usepackage{amsfonts}       
\usepackage{amsmath}        
\usepackage{amssymb}        
\usepackage{color}
\usepackage{wasysym}
\usepackage{multirow}
\usepackage{mciteplus}
\usepackage{psfrag}
\usepackage{slashed}
\usepackage{cancel}
\usepackage{graphicx}

\usepackage{hyperref}

\usepackage{placeins}

\newcommand{\MO}{{\tt micrOMEGAs}}

\newcommand{\beq}{\begin{equation}}
\newcommand{\eeq}{\end{equation}}
\newcommand{\bea}{\begin{eqnarray}}
\newcommand{\eea}{\end{eqnarray}}

\newcommand{\RM}[1]{\MakeUppercase{\romannumeral #1}}

\catcode`\@=11
\font\manfnt=manfnt
\def\Watchout{\@ifnextchar [{\W@tchout}{\W@tchout[1]}}
\def\W@tchout[#1]{{\manfnt\@tempcnta#1\relax%
  \@whilenum\@tempcnta>\z@\do{%
    \char"7F\hskip 0.3em\advance\@tempcnta\m@ne}}}
\let\foo\W@tchout
\def\dubious{\@ifnextchar[{\@dubious}{\@dubious[1]}}

\def\@dubious[#1]{%
  \color{red}\setbox\@tempboxa\hbox{\@W@tchout#1}
  \@tempdima\wd\@tempboxa
  \list{}{\leftmargin\@tempdima}\item[\hbox to 0pt{\hss\@W@tchout#1}]}
\def\@W@tchout#1{\W@tchout[#1]}
\catcode`\@=12

\begin{document}
\preprint{LAPTH-120/14, LCTS/2014-39, MS-TP-14-31}

\title{SUSY-QCD corrections to stop annihilation into electroweak final states\\ including Coulomb enhancement effects}

\author{J.~Harz}
 \email{j.harz@ucl.ac.uk}
 \affiliation{
	Department of Physics and Astronomy, University College London, London WC1E 6BT, United Kingdom
  }

\author{B.~Herrmann}
 \email{herrmann@lapth.cnrs.fr}
 \affiliation{
	LAPTh, Universit\'e de Savoie, CNRS, 9 Chemin de Bellevue, B.P.~110, F-74941 Annecy-le-Vieux, France
  }

\author{M.~Klasen}
 \email{michael.klasen@uni-muenster.de}
 \affiliation{
	Institut f\"ur Theoretische Physik, Westf\"alische Wilhelms-Universit\"at M\"unster, Wilhelm-Klemm-Stra{\ss}e 9, D-48149 M\"unster, Germany
  }

\author{K.~Kova\v{r}\'ik}
 \email{karol.kovarik@uni-muenster.de}
 \affiliation{
	Institut f\"ur Theoretische Physik, Westf\"alische Wilhelms-Universit\"at M\"unster, Wilhelm-Klemm-Stra{\ss}e 9, D-48149 M\"unster, Germany
  }
  
\author{M.~Meinecke}
 \email{mmein\_03@uni-muenster.de}
 \affiliation{
	Institut f\"ur Theoretische Physik, Westf\"alische Wilhelms-Universit\"at M\"unster, Wilhelm-Klemm-Stra{\ss}e 9, D-48149 M\"unster, Germany
  }

\date{\today}

\begin{abstract}
We present the full $\mathcal{O}(\alpha_s)$ supersymmetric QCD corrections for stop-antistop annihilation into electroweak final states within the 
Minimal Supersymmetric Standard Model. We also incorporate Coulomb corrections due to gluon exchange between
the incoming stops. Numerical results for the annihilation cross sections and the predicted neutralino relic density are presented.
We show that the impact of the radiative corrections on the cosmologically preferred region of the parameter space 
can become larger than the current experimental uncertainty, shifting the relic bands within the considered regions of the parameter space 
by up to a few tens of GeV.
\end{abstract}


\maketitle


\section{Introduction}
\label{Intro}

There exists convincing evidence today for a sizable cold dark matter (CDM) component in the Universe, stemming from a large variety of astronomical observations, 
such as rotation curves of galaxies, 
the Bullet Cluster, structure formation simulations on cosmological scales and the cosmic microwave background (CMB). 
The most recent measurement of the CMB carried out by the Planck collaboration \cite{Planck} in combination with WMAP data \cite{WMAP9} has led to a precise determination of the dark matter 
relic density
\beq
	\label{Planck}
	\Omega_{\mathrm{CDM}}h^2 = 0.1199 \pm 0.0027,
\eeq
with $h$ denoting the present Hubble expansion rate in units of 100 km $\mathrm{s}^{-1}$ $\mathrm{Mpc}^{-1}$.

Since within the Standard Model (SM) there is no dark matter (DM) candidate which could solely account for the 
correct value of $\Omega_{\mathrm{CDM}}h^2$, extensions of the SM, which can provide an adequate DM candidate are necessary. Among the most prominent candidates are the so called WIMPs, 
Weakly Interacting Massive Particles. 
WIMPs naturally arise within certain theories beyond the standard model, e.g., the four neutralinos $\tilde{\chi}^0_i$ ($i=\{1,...,4\}$) within 
the Minimal Supersymmetric Standard Model (MSSM). By further assuming $R$-parity 
conservation, the lightest neutralino $\tilde{\chi}^0_1$, which is for many realizations of the MSSM also the lightest supersymmetric particle (LSP),
can become stable and is therefore a viable DM candidate.

In the following, we will sketch a general way of calculating the neutralino relic density $\Omega_{\tilde{\chi}_1^0}h^2$.
We consider the case of $N$ species of unstable particles $\chi_i$ which are heavier than the lightest particle
denoted here by $\chi_0$. We further assume that the time evolution of their number densities $n_{i}$ 
is well described by a system of coupled Boltzmann equations \cite{GondoloGelmini},
\beq
	\label{Boltzmann1}
	\frac{\mathrm{d}n_{i}}{\mathrm{d}t} = -3 H n_{i} 
		- \left\langle\sigma_{ij}v_{ij}\right\rangle \Big[ n_{i} n_{j} 
		- \left( n_{i}^{\mathrm{eq}} n_{j}^{\mathrm{eq}} \right) \Big],
\eeq
for $i,j=0,1,\dots,N$.
The first term on the right-hand side of Eq.\ (\ref{Boltzmann1}) containing the Hubble parameter $H$ stands for the dilution of the 
particle number density due to the expansion of the Universe, while the second and third terms describe the creation and 
(co)annihilation of the particle species $\chi_{i}$ and $\chi_{j}$. $n^{\mathrm{eq}}_{i,j}$ stands for the equilibrium number density of the particle species 
$\chi_{i}$ or $\chi_{j}$, respectively, and $\left\langle \sigma_{ij} v_{ij}\right\rangle$ 
is the thermally averaged (co)annihilation cross section of $\chi_{i}$ and $\chi_{j}$ multiplied by their 
relative velocity $v_{ij}$.

\begin{figure*}
	\includegraphics[keepaspectratio=true,trim = 35mm 115mm -15mm 100mm, clip=true, width=0.8\textwidth]{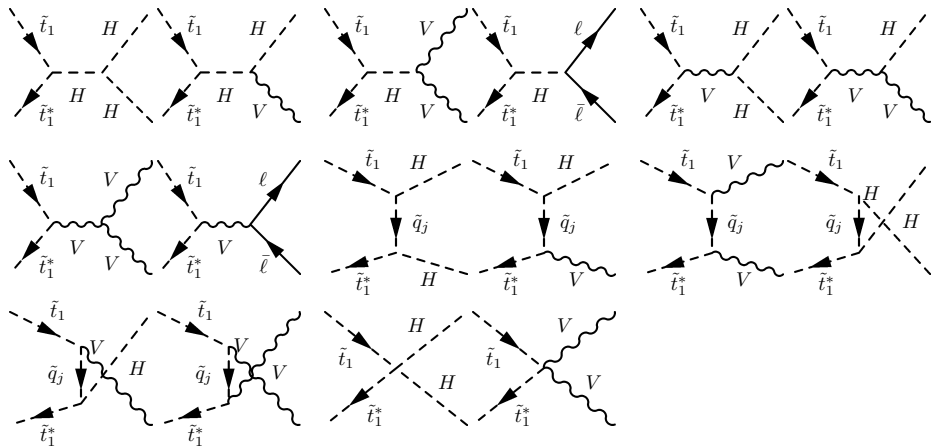}
	\caption{Tree-level diagrams contributing to the stop annihilation into electroweak SM final states.
	 Here, $V=\gamma, Z^0, W^{\pm}$, $H={h^0,H^0,A^0,H^{\pm}}$ and $\ell$ ($\bar{\ell}$) can be any (anti)lepton.}
	\label{fig1}
\end{figure*}

As all particles will at some point decay into the lightest particle $\chi_0$, the quantity relevant to estimate 
$\Omega_{\chi}h^2$ is the total number density $n_{\chi}= \sum_{i=0}^N n_{i}$. Using $n_{i}/n_{\chi}\approx n^{\mathrm{eq}}_{i}/n^{\mathrm{eq}}_{\chi}$ its time dependence can 
be expressed in the following form
\beq
	\label{Boltzmann2}
	\frac{\mathrm{d}n_{\chi}}{\mathrm{d}t} = -3 H n_{\chi} 
		- \left\langle\sigma_{\mathrm{ann}}v\right\rangle \Big[ n_{\chi}^2 
		- \left( n_{\chi}^{\mathrm{eq}}\right)^2 \Big].
\eeq
Here we have introduced the thermally averaged cross section \cite{EdsjoGondolo}
\bea
\label{Boltzmann3}
\langle\sigma_{\mathrm{ann}}v\rangle&=&\sum_{ij}\langle\sigma_{ij}v_{ij}\rangle 
\frac{n_{i}^{\mathrm{eq}}}{n^{\mathrm{eq}}_{\chi}}\frac{n_{j}^{\mathrm{eq}}}{n^{\mathrm{eq}}_{\chi}} \\
 &=&\frac{\int^{\infty}_0\mathrm{d}p_{\mathrm{eff}}\hspace{1mm}p_{\mathrm{eff}}^2 W_{\mathrm{eff}}K_1(\sqrt{s}/T)}
 {m_0^4 T \bigg{[}\sum_{i}\frac{g_{i}}{g_0}\frac{m_{i}^2}{m_0^2}K_2(m_{i}/T)\bigg{]}^2}\,, \nonumber
\eea
with $K_{i}$ being the modified Bessel of the second kind of order $i$ and
\beq
\label{Boltzmann3b}
W_{\mathrm{eff}}=\sum_{ij}\frac{p_{ij}}{p_{\mathrm{eff}}}\frac{g_i g_j}{g_0^2} W_{ij}.
\eeq
In Eq.\ (\ref{Boltzmann3b}), $p_{ij}$ stands for the absolute value of the three-momentum of $\chi_{i}$ (or $\chi_{j}$) in the center-of-mass 
frame of the $(\chi_{i}$--$\chi_{j})$ pair ($p_\mathrm{eff}=p_{00}$) and
\bea
	W_{ij} &=&\frac{1}{g_ig_jS_f}\sum_{\substack{\mathrm{internal} \\ 
		\mathrm{d.o.f.}}}\int |\mathcal{M}|^2(2\pi)^4\nonumber\\
		& & \delta^{4}(p_i+p_j-\sum_fp_f)\prod_f\frac{\mathrm{d}^3\mathbf{p}_f}{(2\pi)^32E_f} 
\eea
for a general $n$-body final state with momenta $p_f$.
Finally, $S_f$ is a symmetry factor, which accounts for identical particles in the final state and $g_{i}$ ($g_{j}$) stands for the number of internal degrees of freedom of the particular 
species. As it will be important in the following analysis, we recall that the ratios $n_{i}^{\mathrm{eq}}/n^{\mathrm{eq}}_{\chi}$ 
in Eq.\ (\ref{Boltzmann3}) at temperature $T$ are Boltzmann suppressed via
\beq
	\label{Boltzmann4}
	\frac{n_{i}^{\mathrm{eq}}}{n^{\mathrm{eq}}_{\chi}} \sim \mathrm{exp}\Big[-\frac{m_{i}-m_{0}}{T}\Big].
\eeq
Thus, only particles with a mass close to $m_{0}$ can give important contributions to
$\left\langle\sigma_{\mathrm{ann}}v\right\rangle$ and are able to sizably alter the time dependence of $n_{\chi}$. 
After solving the Boltzmann equation, today's relic density is given by
\beq
	\Omega_\chi  ~=~ \frac{m_\chi n_\chi}{\rho_{\mathrm{crit}}}\,,
\eeq
with $n_{\chi}$ and $\rho_{\mathrm{crit}}$ being today's particle number density and the critical density of the Universe, respectively.

For large parts of the MSSM parameter space, an enhancement of the neutralino annihilation cross section is necessary to drive the relic density $\Omega_{\tilde{\chi}_{1}^0}h^2$ to the experimentally favored region of Eq.\ (\ref{Planck}). 
One mechanism, which can yield such an enhancement, is the so-called coannihilation between the LSP and the 
next-to-lightest supersymmetric particle (NLSP), see Eq.\ (\ref{Boltzmann3}) \cite{Salati1983, GriestSeckel}.

Over wide ranges of the MSSM parameter space the lighter stop $\tilde{t}_1$ is the NLSP. If 
$m_{\tilde{\chi}_{1}^0} \approx m_{\tilde{t}_1}$, the coannihilations are no longer suppressed (see Eq.\ (\ref{Boltzmann4}))
and so the coannihilations of the lightest neutralino with the light stop are the leading mechanism which determines the relic density of neutralino dark matter. 
This is not the whole story, though. If the mass difference between the stop and the lightest neutralino
is even smaller, the dominating processes actually turn out to be the stop-antistop annihilation although they are normally doubly suppressed
by the same factor as the coannihilations given by Eq.\ (\ref{Boltzmann4}) \cite{stopstop-Ellis}.

Furthermore, it is well known that the (co)annihilation cross sections can become quite sensitive to higher-order 
corrections. 
Therefore, the impact of next-to-leading-order (NLO) corrections on the neutralino relic density has been explored in 
many previous analyses, e.g., SUSY-QCD corrections to 
neutralino-pair annihilation and coannihilation with heavier neutralinos and charginos into quarks \cite{DMNLO,ntnt2bb,ChiChi2qq} 
or SUSY-QCD corrections to neutralino-stop coannihilation \cite{Freitas2007,DMNLO_Stop1,DMNLO_Stop2}. Electroweak (EW) 
corrections to neutralino-pair annihilation and coannihilation with another gaugino have been investigated 
in Ref.\ \cite{Sloops}. Further studies rely on effective
coupling approaches to capture certain classes of corrections to neutralino-pair annihilation or coannihilation with a tau slepton
\cite{EffCouplings1,EffCouplings2}. All these analyses have shown the significance of higher-order corrections to (co)annihilation 
channels for a precise prediction of $\Omega_{\tilde{\chi}_{1}^0}h^2$, which can even by far exceed the current 
experimental uncertainty given in Eq.\ (\ref{Planck})\footnote{See also Ref.\ \cite{finite_temperature} for a recent 
investigation on the applicability of the formalism presented here in the context of NLO calculations.}.

Motivated by these results, we have calculated the full $\mathcal{O}(\alpha_s)$ SUSY-QCD corrections to stop annihilation 
into electroweak final states (i.e.\ leptons, vector and Higgs bosons) 
\bea
	\label{channels1}
 	\tilde{t}_1 \tilde{t}^*_1 &\to& VV, \\
	\label{channels2}
 	\tilde{t}_1 \tilde{t}^*_1 &\to& VH,\\
	\label{channels3}
 	\tilde{t}_1 \tilde{t}^*_1 &\to& HH,\\
	\label{channels4}
 	\tilde{t}_1 \tilde{t}^*_1 &\to& \ell \bar{\ell}, 
\eea
with $V=\gamma, Z^0, W^{\pm}$ and $H={h^0,H^0,A^0,H^{\pm}}$. 
The corresponding Feynman diagrams at the tree-level are shown in Fig.\ \ref{fig1}. We further have taken into account
the corresponding Coulomb corrections due to the exchange of soft gluons between the initial stop-antistop pair. 
Their importance to $\Omega_{\tilde{\chi}_{1}^0}h^2$ has been discussed in Refs.\ \cite{Freitas2007,stopstrip-Ellis}.
Our corrections to the given channels enter the total cross section $\sigma_{\mathrm{ann}}$ in the Boltzmann 
equation (\ref{Boltzmann2}). They are included in our computer package called {\tt DM@NLO}, 
which can be linked to public codes like {\tt micrOMEGAs} 
\cite{micrOMEGAs} or {\tt DarkSUSY} \cite{DarkSusy} to obtain the final corrected relic density. Up to now both of these 
codes evaluate the (co)annihilation cross sections just at an effective tree-level such that the results of this work present 
a natural extension.

This paper is organized as follows: In Sec.\ \ref{Pheno} we specify the model framework, introduce our reference scenarios and discuss the phenomenology of stop annihilation into 
the electroweak final states mentioned in Eqs.\ (\ref{channels1}) -- (\ref{channels4}). Sec.\ \ref{Technical_details} contains technical details about the  actual cross section calculation. 
There, we discuss in particular our handling of ultraviolet (UV) and infrared (IR) divergences as well as the treatment of the Coulomb corrections. In Sec.\ \ref{Numerics} we present our 
numerical results to illustrate the impact of our corrections on the cross section and the relic density. Finally, our conclusions are given in Sec.\ \ref{Conclusion}.
\section{Phenomenology of stop annihilation}
\label{Pheno}

As discussed in Sec.\ \ref{Intro}, we study the impact of higher-order SUSY-QCD corrections to stop annihilation on the neutralino relic density $\Omega_{\tilde{\chi}_{1}^0}h^2$. We have already mentioned above that in order for these processes to become phenomenologically relevant, we have to assume the lightest scalar top to be almost mass degenerate with the lightest neutralino. This assumption is motivated by the latest LHC data, where SUSY scenarios with a light third squark generation are able to reduce fine-tuning while still evading the LHC exclusion limits \cite{lightstop_LHC1,lightstop_LHC2}.

The existence of a light stop gets further support from the LHC discovery of a new boson with a mass of $m \approx 125$ GeV \cite{ATLAS2012, CMS2012, ATLAS2012update}. If we interpret it 
as the light ``SM-like'' Higgs boson $h^0$, its mass has to be enhanced, e.g., by a large stop loop contribution, 
which in the decoupling limit $m_{A^0}\gg m_{Z^0 }$ takes the form \cite{Higgs_mass1, Higgs_mass2}
\begin{widetext}
\beq
	\label{Higgs_mass}
	m_{h^0}^2 ~\approx~ m_{Z^0}^2\cos^2 2\beta + \frac{3g^2m_{t}^4}{8\pi^2m_{W^{\pm}}^2}
		  \Bigg[ \ln \bigg( \frac{M_{\mathrm{SUSY}}^2}{m_t^2} \bigg)
 	   	+ \frac{X_t^2}{M_{\mathrm{SUSY}}^2}{\bigg(}1-\frac{X_t^2}{12M_{\mathrm{SUSY}}^2} \bigg) \Bigg], 
\eeq
\end{widetext}
where $X_{t}=A_{t} - \mu \tan\beta$ and $M_{\mathrm{SUSY}} = \sqrt{m_{\tilde{t}_1} m_{\tilde{t}_2}}$.
For these contributions to become sufficiently large, $|X_t|\approx \sqrt{6}\;M_{\mathrm{SUSY}}$ should be fulfilled, 
which hints toward a sizable $A_{t}$ and therefore toward a large stop mass splitting $m_{\tilde{t}_1} \ll m_{\tilde{t}_2}$ driving $\tilde{t}_1$ to be rather light.

Throughout this analysis we will work within the phenomenological MSSM (pMSSM), where the  soft breaking parameters are fixed at the input scale $Q=1$ TeV according to the SPA 
convention \cite{SPA_convention}. Out of the nineteen parameters, which usually span the pMSSM parameter space, we restrict ourselves to the following set of eleven free parameters: 
The Higgs sector is fixed by the Higgsino mass parameter $\mu$, the ratio $\tan\beta$ of the vacuum expectation values of the two Higgs doublets, and the pole mass $m_{A^0}$ of the 
pseudoscalar Higgs boson. For the first- and second-generation squarks we introduce a common soft breaking mass parameter $M_{\tilde{q}_{1,2}}$, while the mass parameters for the third-generation 
squarks are given by $M_{\tilde{q}_{3}}$ for sbottoms and left-handed stops as well as $M_{\tilde{u}_{3}}$ for right-handed stops. We further set all trilinear couplings to zero 
except for $A_{t}$, the trilinear coupling of the stop sector. In contrast to the three independent mass parameters in the squark sector, we only use a single parameter 
$M_{\tilde{\ell}}$ as a soft breaking mass for all sleptons. Finally, since we do not assume gaugino mass unification, the gaugino sector is defined by three independent parameters 
$M_1$, $M_2$ and $M_3$, the bino, wino and gluino masses, respectively.

\begin{table*}
	\caption{Input parameters for three selected reference scenarios in the pMSSM. All values except $\tan\beta$ are given in GeV. }
	\begin{tabular}{|c|ccccccccccc|}
		\hline
			$\quad$ & $\quad\tan\beta\quad$ & $\quad\mu\quad$ & $\quad m_{A^0}\quad$ & $\quad M_1\quad$ & $\quad M_2\quad$ & $\quad M_3\quad$ & $\quad M_{\tilde{q}_{1,2}}\quad$ & $\quad M_{\tilde{q}_3}\quad$ & $\quad M_{\tilde{u}_3}\quad$ & $\quad M_{\tilde{\ell}}\quad$& $\quad A_{t}\quad$\\ 
			\hline 
			\RM{1}a & 16.3 & 2653.1 & 1917.9 & 750.0 & 1944.1 & 5832.4 & 3054.3 & 2143.7 & 1979.0 & 2248.3 & -3684.1 \\		
			\RM{1}b & 16.3 & 2653.1 & 1917.9 & 989.0 & 1944.1 & 5832.4 & 3054.3 & 2143.7 & 2159.0  & 2248.3 & -3684.1 \\			
			\RM{2} & 27.0 & 2650.8 & 1441.5 & 1300.0 & 1798.4 & 1744.8 & 2189.7 & 2095.3 & 1388.0  & 1815.5 & -4097.9 \\			
			\hline
	\end{tabular}
	\label{ScenarioList}
\end{table*}

\begin{table*}
	\caption{Physical squark, neutralino, chargino and Higgs masses, the bino ($\tilde{B}$) contribution to $\tilde{\chi}^0_1$, the decomposition of $\tilde{t}_1$ into left- 
                 and right-handed parts, and selected observables corresponding to the reference scenarios of Tab.\ \ref{ScenarioList}. All masses are given in GeV.}
	\begin{tabular}{|c|cc|cccc|ccc|ccc|cc|c|}
		\hline
			$\quad$  & $m_{\tilde{\chi}^0_1}$ & $m_{\tilde{t}_1}$ & $m_{\tilde{t}_2}$&$m_{\tilde{b}_1}$&$m_{\tilde{\chi}^0_2}$ & $m_{\tilde{\chi}^{\pm}_1}$ & $m_{h^0 }$ & $m_{H^0 }$& $m_{H^{\pm}}$ & $|Z_{\tilde{\chi}^0,1\tilde{B}}|^2$ & $|Z_{\tilde{t},1L}|^2$ & $|Z_{\tilde{t},1R}|^2$ &$\mathrm{BR}(b\rightarrow s\gamma)$ & $\delta a_{\mu}$ &$\Omega_{\tilde{\chi}^0_1} h^2$\\
			\hline 
			\RM{1}a & 758.0 & 826.1 & 1435.1 & 1260.5 & 1986.7 & 1986.8 & 128.8 & 1917.4  & 1919.6 & 0.9996 & 0.27  & 0.74  &  $3.1\cdot10^{-4}$ & $284\cdot10^{-11}$ & 0.1146\\		
			\RM{1}b & 999.6 & 1079.6 & 1543.4 & 1265.8   & 1986.8 & 1986.9 & 129.4  & 1917.9 & 1919.6& 0.9995  & 0.55  & 0.46  &$3.1\cdot10^{-4}$ & $284\cdot10^{-11}$ & 0.1193\\		
			\RM{2} & 1306.3 & 1363.0 & 2128.8 & 2055.2 & 1826.9 & 1827.1 & 124.6 & 1440.7 & 1443.6 & 0.9992 & 0.08  & 0.92  & $3.1\cdot10^{-4}$ & $279\cdot10^{-11}$ & 0.1209\\		
			\hline		
	\end{tabular}
	\label{ScenarioProps}
\end{table*}

Phenomenologically interesting scenarios have to fulfill a certain number of constraints. For our scenario search we have considered the following prominent observables:
\begin{align}
 	\label{bound1}
	0.1145 &\leq \Omega_{\tilde{\chi}^0_1}h^2 \leq 0.1253, \\
 	\label{bound2}
	120~{\rm GeV} &\leq m_{h^0} \leq 130~{\rm GeV}, \\
 	\label{bound3}
	2.56\cdot 10^{-4} &\leq \mathrm{BR}(b\rightarrow s\gamma) \leq 4.54\cdot 10^{-4}, \\
	\label{bound4}
	|\delta a_{\mu}| &< 288\cdot 10^{-11}.
\end{align}
They have been selected for the following reasons. To work with scenarios, which respect the recent Planck measurements, we require the neutralino relic density to lie within the limits given in Eq.\ (\ref{bound1}) at 2$\sigma$ confidence level. This means that we expect the neutralino to account for the whole amount of dark matter in our 
Universe today. Second, we require the mass of the lightest Higgs boson to agree with the observation at the LHC. However, we allow for a rather large uncertainty of about 5 GeV on the Higgs 
mass value due to large theoretical uncertainties arising from not yet included higher-order corrections in its calculation (see e.g. Ref.\ \cite{Buchmueller:2013psa}). 
The third bound, Eq.\ (\ref{bound3}), concerns the inclusive branching ratio of the flavor changing neutral current decay $b \to s\gamma$. The imposed interval corresponds to the latest 
HFAG value \cite{HFAG} at $3\sigma$ confidence level. The fourth bound limits the supersymmetric corrections $\delta a_{\mu}$ to the muon g-factor 
$g_{\mu}$, where $a_{\mu}=(g_{\mu}-2)/2$ and $\delta a_{\mu}=a^{\mathrm{exp}}_{\mu}-a^{\mathrm{theo}}_{\mu}$ is the discrepancy between experiment and the predicted theoretical value. 
We expect the SUSY corrections to improve on this discrepancy compared to the SM prediction (see Ref.\ \cite{PDG}). 

To illustrate the numerical impact of our derived corrections, we introduce the three reference scenarios given in Tab.\ \ref{ScenarioList}, which have been found by performing a random scan of one million points within the previously defined pMSSM. 
Their parameter values are summarized in Tab.\ \ref{ScenarioList}. The corresponding particle masses, mixings and further observables are summarized in Tab.\ \ref{ScenarioProps}.

\begin{table}
	\caption{Most relevant stop annihilation channels into EW final states of the reference scenarios in Tab.\ \ref{ScenarioList}.}
	\begin{tabular}{|rl|ccc|}
		\hline
		 & & ~Scenario \RM{1}a~ & ~Scenario \RM{1}b~ & ~Scenario \RM{2}~\\
		\hline
		$\tilde{t}_1 \tilde{t}^*_1 \to$ & $h^0 h^0 $ & 46.1\% & 15.9\% &11.3\% \\
		                                        & $h^0 H^0 $     & -- &  46.6\% & 11.1\% \\
		\hline                                        

		$\tilde{t}_1 \tilde{t}^*_1 \to$& $Z^0 A^0$    & -- & 4.0\% & 7.4\% \\
		                                        & $W^{\pm}H^{\mp}$ & -- & 4.2\% & 13.6\% \\
		\hline                                        
		$\tilde{t}_1 \tilde{t}^*_1 \to$& $Z^0 Z^0$     & 8.7\%  & 4.3\% & 7.4\% \\
		                                        & $W^+ W^-$ & 12.5\% &  2.7\% & 13.6\% \\		
		\hline
		\multicolumn{2}{|c|}{Total} & 67.3\% & 77.7\% & 64.4\% \\
		\hline
	\end{tabular}
	\label{ScenarioChannels}
\end{table}

Throughout our analysis we have used {\tt SPheno~3.2.3} \cite{SPheno} to obtain the physical mass spectrum and related mixings from the given input parameters. The neutralino relic density, 
the contributions of individual (co)annihilation channels and the numerical values of further observables such as the branching fraction $b \to s \gamma$ have been obtained by using {\tt micrOMEGAs 2.4.1} with the standard {\tt CalcHEP} \cite{CalcHEP} implementation of the MSSM. We only have introduced slight changes to stabilize the numerical evaluation of the occurring phase-space integrals (see Sec.\ \ref{Further subtleties}). We have checked within our typical scenarios that these changes do not have a relevant impact on the predicted relic density. As can be seen in Tab.\ \ref{ScenarioProps} the three selected scenarios fulfill the demanded constraints given in Eqs.\ 
(\ref{bound1}) -- (\ref{bound4}).

To better understand the origin of the radiative corrections in our scenarios, we dissect
all scenarios and show which processes are important in which parameter point. Moreover, we look into each process so that we can identify 
the dominating contributions. We start by listing the stop annihilation processes that we correct and that contribute more than 1$\%$ 
to $\langle \sigma_{\mathrm{ann}}v \rangle$ in Tab.\ \ref{ScenarioChannels}. Then, for each process in Tab.\ \ref{ScenarioChannels}, 
we list the underlying structure of subchannel contributions in Tab.\ \ref{ScenarioSubChannels}, i.e.\ the contributions of different 
diagram classes as shown in Fig.\ \ref{fig1}. We have grouped the contributions from quartic couplings 
(contribution denoted as $Q$), $s$-channel scalar exchange (denoted $s_S$) and the squark exchange in the $t$- and 
$u$-channels ($t/u$). The vector contributions $s_V$ to the $s$-channel do not appear in Tab.\ \ref{ScenarioSubChannels} as they turn out to be negligible within our reference scenarios (see below). 
The contributions from the corresponding squared matrix elements are denoted by 
$Q \times Q$, $s_S \times s_S$ and $t/u \times t/u$, while the interference terms are denoted by 
$Q \times s_S$, $Q \times t/u$ and $s_S \times t/u$. Note that negative values refer to destructive interferences. 
The percentages in Tab.\ \ref{ScenarioSubChannels} are obtained for the center-of-mass momentum of the incoming 
particles $p_{\mathrm{cm}} = 200$ GeV, which is roughly the region where the thermal distribution in the integrand 
of Eq.\ (\ref{Boltzmann3}) peaks for the scenarios presented here.
All calculations are performed in the 't Hooft-Feynman gauge. Following the treatment of external vector bosons presented in App. B of Ref.\ \cite{CompHEP}, we add the contributions 
of Goldstone bosons and Faddeev-Popov ghosts to the particular vector boson final states.

Note that, as the incoming scalar-antiscalar configuration is $CP$-even and as all the relevant interactions are $CP$-conserving, every 
intermediate and final state has to be $CP$-even, too. This limits all possible final states such that
pseudoscalar Higgs bosons can appear only in pairs or together with a suitable vector boson 
and are otherwise partial-wave suppressed (see Tab.\ \ref{ScenarioChannels}). Moreover, the same argument prohibits any
exchange of pseudoscalars in the $s$-channel. Finally any $s$-wave annihilation through the $s$-channel exchange of vector 
bosons is forbidden due to conservation of total angular momentum (see Tab.\ \ref{ScenarioSubChannels}).

In scenario \RM{1}a, we correct processes which contribute 67.3$\%$ to $\Omega_{\tilde{\chi}_1^0}h^2$.
The scenario is characterized by a dominant contribution of the $h^0 h^0$ final state (46.1\%), while final states, 
which include one or more of the heavier Higgs bosons $H^0$, $A^0$, $H^{\pm}$, are too heavy to be kinematically accessible.
One further encounters a relative dominance of the Higgs-Higgs final state over the vector-vector final states, where the latter 
contribute roughly $21\%$ to the relic density. This can be traced back to an enhancement of the Higgs coupling to scalar top quarks
as compared to all other relevant couplings, e.g., the gauge interactions of EW vector bosons to squarks. It is caused by 
the large top mass and the large trilinear coupling $A_t$ needed to achieve a sizable stop-loop contribution to $m_{h^0 }$. 
It is especially important in the case of $t$- and $u$-channels where the enhanced stop-Higgs/Goldstone-boson coupling enters twice. 
This results in large contributions and explains the overall dominance of the $t/u$ subchannels as can be seen in Tab.\ \ref{ScenarioSubChannels}.
But although the massive vector final states get contributions from Goldstone bosons, which give rise to couplings as large as the usual 
Higgs couplings, their corresponding $t/u$-channels contributions are further suppressed by large propagators. This is due to the fact that 
$G^0$ as a pseudoscalar only couples light and heavy squark mass eigenstates. 
Furthermore, the charged Goldstone boson $G^{\pm}$ connects up- and down-type squarks, which leads in scenario \RM{1}a to contributions of $t$- and $u$-channel diagrams where the exchanged 
particle is much heavier than the lighter stop $\tilde{t}_1$ and therefore to an overall propagator suppression of the Goldstone boson contributions to vector-vector final states relative to, 
e.g., the $h^0 h^0$ final state.

\begin{table*}
	\caption{Subprocesses for the channels of Tab.\ \ref{ScenarioChannels} contributing individually at least 0.1\% at $p_{\mathrm{cm}} = 200$ GeV.}
	\begin{tabular}{|cc|cccccc|}
		\hline
	       &  & $\quad Q \times Q\quad $ & $\quad Q \times s_S\quad $ & $\quad Q \times t/u \quad $ & $\quad s_S \times s_S\quad $ &   $\quad s_S \times t/u\quad $ &  $\quad t/u \times t/u\quad $ \\
                \hline
		 Scenario \RM{1}a &  &  &  &   &   &  & \\
		 $\quad \tilde{t}_1 \tilde{t}^*_1\to$&$h^0  h^0 \quad $      & 0.7\% & -0.2\% & -17.5\% & -- & 2.4\% & 114.6\% \\
		                                                   &$Z^0  Z^0 \quad $      & 2.7\% & -0.3\% & -37.7\% & -4.8\% & 4.2\% & 135.9\% \\
		                                                   &$ W^+ W^-\quad $  & 2.2\% & -0.4\% & -32.7\% & -6.1\% & 6.1\% & 131.0\% \\
		\hline
		 Scenario \RM{1}b &  &  &  &   &   &  & \\
		 $\quad \tilde{t}_1 \tilde{t}^*_1\to$&$h^0  h^0 \quad $      & 2.1\% & -0.2\% & -32.9\% & --& 1.5\% & 129.6\% \\
		                                                   &$h^0  H^0 \quad $      & -- & -- & 0.6\% & -- & -0.6\% & 100.0\%\\
		                                                   &$Z^0  A^0\quad $      & -- & -- & 2.3\% & -21.7\% & 10.3\% & 109.0\% \\
		                                                   &$W^{\pm} H^{\mp}\quad $  & --  & -- & 1.8\% & -35.4\% & 32.9\% & 100.8\% \\
		                                                   &$Z^0  Z^0 \quad $      & 5.1\% & -0.3\% & -54.5\% & -5.3\% & 4.3\% & 150.7\% \\
		                                                   &$W^+ W^-\quad $     & 6.6\% & -1.2\% & -52.4\% & -19.2\% & 18.7\% & 147.7\% \\
		\hline
		 Scenario \RM{2} &  &  &  &   &   &  &\\
		 $\quad \tilde{t}_1 \tilde{t}^*_1\to$&$h^0  h^0 \quad $      & 8.0\% & -0.4\% & -72.2\% & --& 1.8\% & 162.7\% \\
		                                                   &$h^0  H^0 \quad $      & -- & -- & 2.4\% & -- & -0.6\% & 98.2\% \\
		                                                   &$Z^0  A^0\quad $      & --  & -- & 3.0\% & -2.1\% & 1.4\% & 97.7\% \\
		                                                   &$W^{\pm} H^{\mp}\quad $  & --  & --  & 2.9\% & -1.8\% & 0.8\% & 98.1\% \\
		                                                   &$Z^0  Z^0 \quad $      & 11.9\% & -0.3\% & -92.6\% & -3.5\% & 3.1\% & 181.4\% \\
		                                                   &$W^+ W^-\quad $  & 11.4\% & -0.3\% & -90.1\% & -3.1\% & 3.0\% & 179.2\% \\
		\hline
	\end{tabular}
	\label{ScenarioSubChannels}
\end{table*}
In scenario \RM{1}b, we correct diagrams which contribute 77.7$\%$ to $\Omega_{\tilde{\chi}_1^0}h^2$. The situation is quite similar to scenario \RM{1}a except for the lightest stop 
being heavy enough so that also heavier Higgs bosons are kinematically accessible. As the final state has to be $CP$-even, the only additional sizable contributions stem from the 
$h^0 H^0$, $Z^0 A^0$ as well as from the $W^{\pm}H^{\mp}$ final states (see Tab.\ \ref{ScenarioChannels}).
Comparing the scenarios \RM{1}a and \RM{1}b, one can see a shift of the main contribution to the relic density away from the $h^0 h^0$ 
final state over to the $h^0 H^0$ final state, which is with 46.6$\%$ the most important channel of scenario \RM{1}b. This shift is 
mainly driven by the dominant $t/u$-channel contributions in Tab.\ \ref{ScenarioSubChannels}. The special feature of the 
$h^0 H^0$ final state is that it is just kinematically allowed ($m_{h^0}+m_{H^0} \approx 2m_{\tilde{t}_1}$), so that the final-state 
Higgs bosons do not have large momenta. Furthermore, the dominant contribution to any cross section contribution to 
$\Omega_{\tilde{\chi}_1^0}h^2$ comes from the region $\sqrt{s}\approx 2 m_{\tilde{t}_1}$, which further limits the momenta of the 
incoming and also outgoing particles. For the $h^0 H^0$ final states the $t$- and $u$-channel propagators are therefore close to their mass shells whereas for the $h^0 h^0$ final state these propagators 
are still far off their mass shells, which translates into the $h^0 H^0$ final state being the leading contribution.

In scenario \RM{2}, 64.4$\%$ of all contributions to $\Omega_{\tilde{\chi}_1^0}h^2$ are affected by our corrections.
The mass difference between the squarks and the heavier Higgs boson leads to the same structure of relevant processes as in scenario \RM{1}b but in contrast to the two previously 
encountered scenarios, scenario \RM{2} is chosen such that it gets roughly equal contributions from all possible vector and Higgs boson combinations in the final state.

It can further be seen in Tab.\ \ref{ScenarioChannels} that for all three scenarios there are no sizable contributions to $\Omega_{\tilde{\chi}_1^0}h^2$ from lepton-antilepton final states. However, our scans over the pMSSM parameter space 
will later show that leptonic final states are indeed important when their contribution is enhanced by a resonant Higgs exchange.
This happens if $2 m_{\tilde{t}_1}\approx m_{H^0}$.

The absence of final states in Tab.\ \ref{ScenarioChannels} containing one or more photons is due to the fact that the photon as the massless gauge boson of the Abelian $U(1)$ does not 
possess any $s$-channel contributions. Furthermore, there are no Goldstone boson contributions to photons in the final state, which turned out to be the dominant contributions to the $Z^0 Z^0$ and
$W^+ W^-$ final states as explained above. Finally, as the photon coupling to sfermions is diagonal in the squark mass eigenbasis, the $\tilde{t}_1$-annihilation lacks all contributions of 
photon-Higgs final states, which altogether leads to the absence of final states containing one or two photons as encountered in Tab.\ \ref{ScenarioChannels}.
All other (co)annihilation channels as, e.g., coannihilation with heavier neutralinos,
charginos, sbottoms, etc. are irrelevant in our scenarios \RM{1}a/b and \RM {2} as the mass gaps between all these particles and the lightest neutralino are already too large 
(see Tab.\ \ref{ScenarioProps}). This prevents these particles from significantly changing $\Omega_{\tilde{\chi}^0_1}h^2$ due to the Boltzmann suppression of Eq.\ (\ref{Boltzmann4}).


\section{Technical details}
\label{Technical_details}

\subsection{Calculation of $\mathcal{O}(\alpha_s)$ corrections}
\label{Technical}

\begin{figure*}[tbp]
	\includegraphics[keepaspectratio=true,trim = 45mm 130mm 25mm 105mm, clip=true, width=0.62\textwidth]{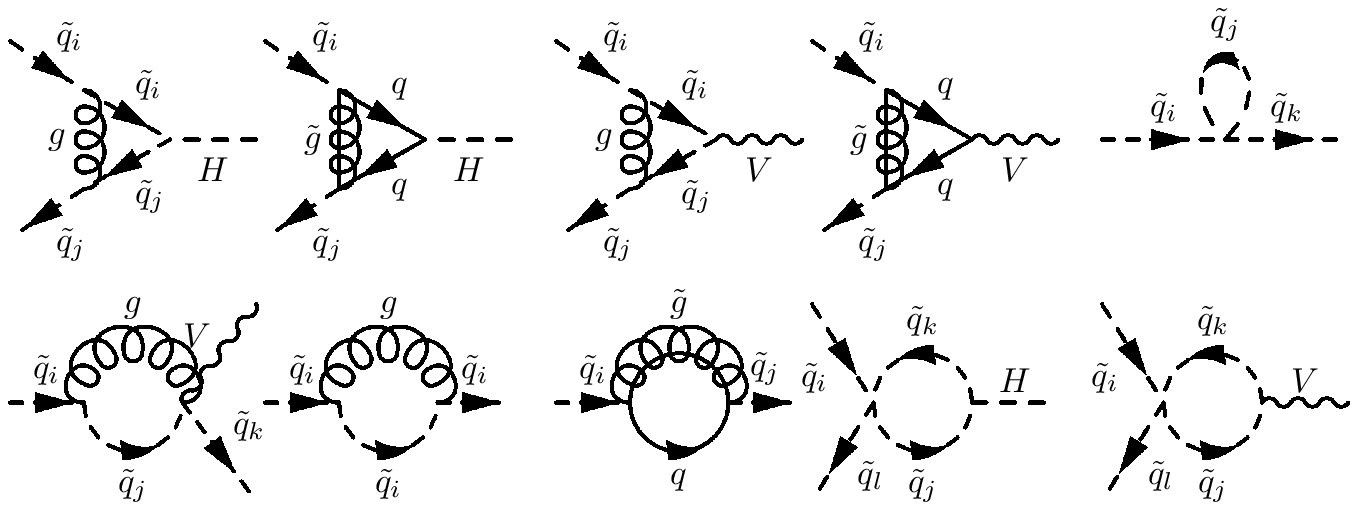}
	\caption{Vertex and propagator insertions depicting schematically the one-loop corrections of ${\cal O}(\alpha_s)$ to the stop-annihilation processes shown in Fig.\ \ref{fig1}. Here, $V=\gamma, Z^0, W^{\pm}$ and $H={h^0,H^0,A^0,H^{\pm}}$.}
	\label{fig2}
\end{figure*}

\begin{figure*}[tbp]
	\includegraphics[keepaspectratio=true,trim = 45mm 113mm 0mm 100mm, clip=true, width=0.7\textwidth]{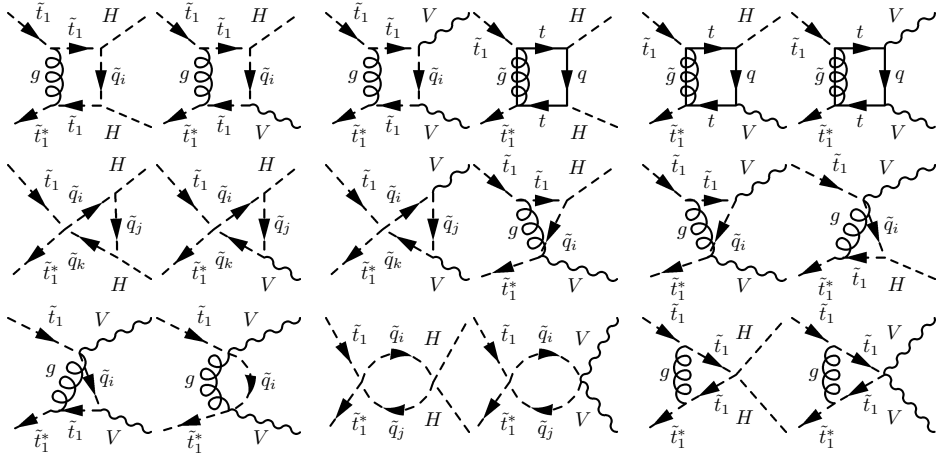}
	\caption{Diagrams depicting corrections of ${\cal O}(\alpha_s)$  to the stop-annihilation processes shown in Fig.\ \ref{fig1}. As before, $V=\gamma, Z^0, W^{\pm}$ and $H={h^0,H^0,A^0,H^{\pm}}$. The diagrams in the first row are in the following referred to as box contributions, whereas we subsume the diagrams of the second and third rows under vertex corrections. $u$-channel processes are not explicitly shown, as they can be obtained by crossing from the corresponding $t$-channel diagrams.}
	\label{fig3}
\end{figure*}

\begin{figure*}[tbp]
	\includegraphics[keepaspectratio=true,trim = 40mm 75mm 0mm 50mm, clip=true, width=0.7\textwidth]{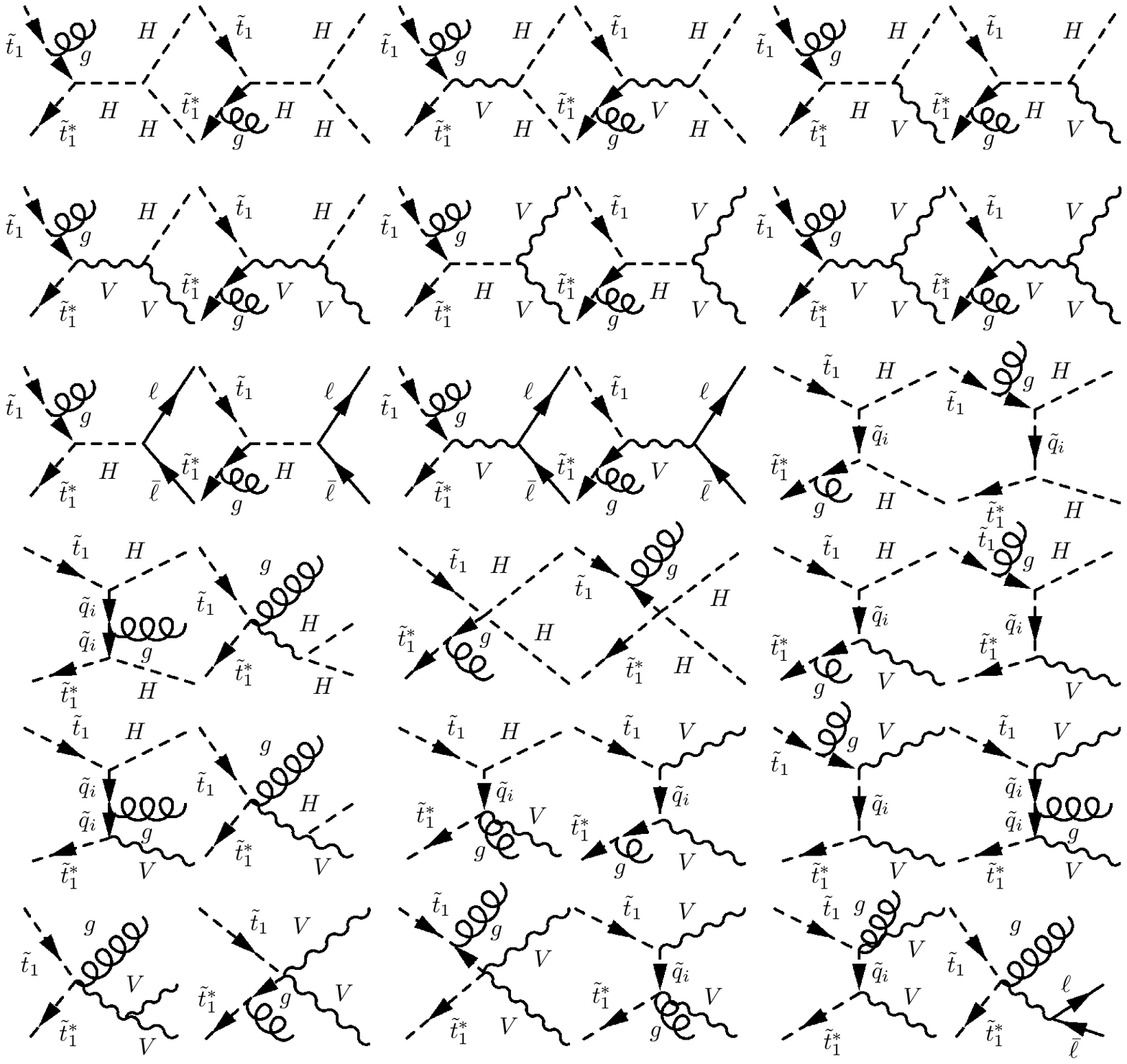}
	\caption{Diagrams depicting the real gluon emission corrections of ${\cal O}(\alpha_s)$ to the stop-annihilation processes shown in Fig.\ \ref{fig1}. As before, $V=\gamma, Z^0, W^{\pm}$ and $H={h^0,H^0,A^0,H^{\pm}}$. The corrections to the $u$-channel processes are not explicitly shown, as they can be obtained by crossing from the corresponding $t$-channel diagrams.}
 	\label{fig4}
\end{figure*}

The NLO cross section
\begin{equation}
	\label{sigNLO}
	\sigma_{\text{NLO}} ~=~ \int_{2} \mathrm{d}\sigma^{\text{V}} + \int_{3} \mathrm{d}\sigma^{\text{R}}
\end{equation}
consists of the virtual ($\mathrm{d}\sigma^{\text{V}}$) and the real emission contributions ($\mathrm{d}\sigma^{\text{R}}$), which are integrated over the two- and three-particle phase-space, 
respectively. Figs.\ \ref{fig2} and \ref{fig3} show the relevant one-loop diagrams for stop annihilation contributing to the virtual part $\mathrm{d}\sigma^{\text{V}}$. 
In Fig.\ \ref{fig4} the corresponding real gluon emission diagrams corresponding to $\mathrm{d}\sigma^{\text{R}}$ are depicted.

The virtual SUSY-QCD corrections to stop annihilation include contributions from the exchange of gluons and gluinos as well as from pure squark loops. 
These corrections, calculated using the SUSY-preserving dimensional reduction ($\overline{\tt DR}$) scheme, can be all reduced via the Passarino-Veltman reduction to the well-known 
scalar integrals $A_0$, $B_0$, $C_0$, and $D_0$ \cite{scalar_integrals}. The UV divergences, which appear in the resulting expressions, can then be cancelled by 
properly chosen counterterms.

In our calculation, the latter are defined in a hybrid on-shell\,/\,$\overline{\tt DR}$ renormalization scheme, where $A_t$, $A_b$, $m^2_{\tilde{t}_1}$, $m^2_{\tilde{b}_1}$ and $m^2_{\tilde{b}_2}$ are chosen as input parameters along with the heavy quark masses $m_b$ and $m_t$. 
The strong coupling $\alpha_s$, the trilinear couplings $A_t$, $A_b$ and the bottom quark mass $m_b$ are defined in the $\overline{\tt DR}$ scheme at the scale $\mu_R = 1$ TeV, 
which corresponds to the scale where the soft breaking parameters are defined.  All remaining input masses are defined 
on-shell. A more detailed discussion of this particular renormalization scheme as well as of our treatment of $\alpha_s$ can 
be found in Refs.\ \cite{DMNLO_Stop1,DMNLO_Stop2}.

Apart from the UV divergences, one-loop matrix elements also contain IR divergences which arise due to the exchange of soft gluons in the loop. 
These IR divergences are also dimensionally regularized using the $\overline{\tt DR}$-scheme. The associated poles cancel against IR poles of the same form, 
but opposite sign stemming from the real corrections shown in Fig.\ \ref{fig4} \cite{IR-finite}. Since a completely analytic integration of Eq.\ (\ref{sigNLO}) 
is in practice impossible for all but the simplest integrands, one usually makes use of numerical integration. However, to render Eq.\ (\ref{sigNLO}) numerically integrable, a matching of the 
IR singularities residing in the differential cross sections $\mathrm{d}\sigma^{\text{V}}$ and $\mathrm{d}\sigma^{\text{R}}$ is necessary. As these differential cross sections have to be 
integrated separately over different phase-spaces, one cannot take advantage of the direct cancellation of the IR divergences between the real and virtual parts. 
Especially, as the singularities of the real corrections actually arise during the integration over the $2 \to 3$ phase-space, whereas the IR singularities of the virtual corrections can 
already be separated as poles before performing any $2 \to 2$ phase-space integration, this matching is far from being trivial. Multiple possibilities exist to integrate Eq.\ (\ref{sigNLO}). 
One is the dipole subtraction method \cite{Dipole} and a second one is the so-called phase-space slicing method \cite{scalar_integrals}. In this work we made use of the latter.

The phase-space slicing method isolates the IR divergence in the real corrections by slicing the $2 \to 3$ phase-space 
into two parts using a cut $\Delta E$ on the energy $|\vec{k}|$ of the additional gluon. 
In the soft-gluon region, where $|\vec{k}| \leq \Delta E$, we can approximate the $2 \to 3$ amplitudes and factorize them according to
\beq
	\label{factorization}
	\left(\frac{\mathrm{d}\sigma}{\mathrm{d}\Omega}\right)_{\mathrm{soft}} = 
	F \times \left(\frac{\mathrm{d}\sigma}{\mathrm{d}\Omega}\right)_{\mathrm{tree-level}},
\eeq
where $F$ already contains the integration over the gluon phase-space with $|\overrightarrow{k}|\leq\Delta E$ and therefore all IR divergences. Furthermore, 
the integration in $F$ can be performed analytically in $D=4-2\epsilon$ dimensions such that a cancellation of the arising singularities against the IR singularities of the virtual 
corrections is already possible at the integrand level. The remaining part of the $2 \to 3$ phase-space integration in Eq.\ (\ref{sigNLO}), where $|\vec{k}| > \Delta E$, 
can then be performed numerically in $D = 4$ dimensions. Note that no collinear divergences occur in our case, since the additional gluon can be radiated only off a massive scalar.

The final sum of the soft-gluon approximation and the remaining $2 \to 3$ part should be independent of the unphysical cutoff $\Delta E$ on the gluon energy. 
In practice one has to choose a convenient value for $\Delta E$. On the one hand, it should not be too small, because the phase-space integration of the real corrections would 
be numerically unstable. On the other hand, the cut should also not be too large, not to invalidate the soft-gluon approximation of the cross section for $|\vec{k}| \leq \Delta E$. We verified that the full $2 \to 3$ cross sections are insensitive to a variation of $\Delta$E around our choice of this cut. In addition, 
there are logarithms of the dimensional regularization scale $\mu$, which we set equal to the renormalization scale $\mu=\mu_R=1$ TeV. These logarithms, which arise in the soft-gluon 
approximation of the $2 \to 3$ processes as well as in the corresponding virtual contributions can give rise to an enhancement of both contributions separately, but 
cancel in the final sum of Eq.\ (\ref{sigNLO}).

\subsection{Coulomb corrections}
\label{Coulomb corrections}

\begin{figure}
	\includegraphics[keepaspectratio=true,trim = 57mm 140mm 20mm 120mm, clip=true, width=0.51\textwidth]{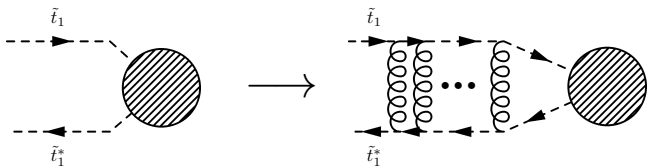}
	\caption{Ladder diagram for a leading-order (LO) Coulomb potential.}
	\label{fig5}
\end{figure}

In the previous subsection, we have discussed the fixed-order corrections due to the exchange of one gluon, squark or gluino for the  annihilation into electroweak final states. 
There are, however, additional potentially important corrections stemming from the exchange of multiple gluons between the stops in the initial state, which will be discussed in the 
following.\\ 
During the calculation of the $\mathcal{O}(\alpha_s)$ corrections of the previous subsection, we encounter terms which 
are proportional to $1/v$, where $v$ is the relative velocity of the incoming 
 pair. 
It is well known that the exchange of $n$ gluons generates a correction factor proportional to
$(\alpha_s/v)^{\mathrm{n}}$, within the perturbative expansion in $\alpha_s$.\footnote{The divergence at $v \to 0$ is the well-known Coulomb 
singularity signaling the production of a  quasibound state, called stoponium.} 

Since during freeze-out the stops are moving slowly ($E_{\mathrm{kin},\tilde{t}_1} \approx T_{\mathrm{freeze-out}} \ll m_{\tilde{t}_1}$), this fraction can become large,
\beq
	\label{Coulomb_1}
	\alpha_s/v \gtrsim \mathcal{O}(1) ,
\eeq
and spoil the convergence of the perturbative series \cite{Coulomb1,Coulomb1b}.
Hence these so-called Coulomb corrections need to be resummed to all orders to get a reliable 
result (see Fig.\ \ref{fig5}). This can be done in the framework of nonrelativistic QCD (NRQCD) 
\cite{NRQCD}. Following Ref.\ \cite{Coulomb2}, the Coulomb-corrected result can be cast into the form 
\bea
	\label{Coulomb_2}
	\sigma^{\mathrm{Coul.}} \big(\tilde{t}_1 \tilde{t}^*_1 \rightarrow \mathrm{EW} \big) & = & 
		\frac{4\pi}{v m_{\tilde{t}_1}^2}
		\Im \Big\{ G^{[1]} \big(\mathbf{r}=0;\sqrt{s}+i\Gamma_{\tilde{t}_1} \big) \Big\} \nonumber \\
	& & ~~~~\times \sigma^{\mathrm{LO}} \big(\tilde{t}_1\tilde{t}^*_1 \rightarrow \mathrm{EW} \big),
\eea
where $\sigma^{\mathrm{LO}}(\tilde{t}_1\tilde{t}^*_1\rightarrow \mathrm{EW})$ is the annihilation cross section of the  color singlet into EW final states. 
$G^{[1]}\big(\mathbf{r};\sqrt{s}+i\Gamma_{\tilde{t}_1}\big)=
G^{[1]}\big(\mathbf{r},\mathbf{r'}=0;\sqrt{s}+i\Gamma_{\tilde{t}_1}\big)$ stands for the color-singlet Green's 
function of the Schr\"odinger equation at $\mathbf{r'}=0$. It governs the dynamics of the would-be stoponium
evaluated at distance $\mathbf{r}$. 
More precisely, $G^{[1]}\big(\mathbf{r};\sqrt{s}+i\Gamma_{\tilde{t}_1}\big)$ is the solution to
\beq
	\label{Coulomb_3}
	\Big[ H^{[1]} - \big( \sqrt{s}+i\Gamma_{\tilde{t}_1} \big) \Big]
		G^{[1]} \big( \mathbf{r};\sqrt{s}+i\Gamma_{\tilde{t}_1}\big)
		= \delta^{(3)}(\mathbf{r}), 
\eeq
with $H^{[1]}$ being the Hamilton operator of the  system,
\beq
	\label{Coulomb_4}
	H^{[1]} = -\frac{1}{m_{\tilde{t}_1}}\Delta + 2 m_{\tilde{t}_1}+V^{[1]}(\mathbf{r}).
\eeq
The Fourier transform of the color-singlet Coulomb potential $V^{[1]}(\mathbf{r})$ can be written
at NLO  as \cite{Sommerfeld1,Sommerfeld2}
\bea
	\label{Coulomb-potential}
	\tilde{V}^{[1]}(\mathbf{q}) &=& -\frac{4\pi\alpha_s(\mu_G) C^{[1]}}{\mathbf{q}^2} ~~~ \\
	& & ~~~ \times \bigg[ 1 + \frac{\alpha_s(\mu_G)}{4\pi} \bigg( \beta_0 \ln\frac{\mu_G^2}
	{\mathbf{q}^2}+a_1 \bigg)  \bigg] \nonumber
\eea
with
\bea
	 C^{[1]} & = & C_{\mathrm{F}} = \frac{4}{3}, \hspace{2mm} C_{\mathrm{A}}  =  3, \nonumber\\
	 a_1 & = & \frac{31}{9} C_{\mathrm{A}}-\frac{20}{9}T_fn_f, \nonumber \\
	 \beta_0 & = & \frac{11}{3}C_A - \frac{4}{3} T_fn_f\,,
\eea
and $T_f=\frac{1}{2}$ for top squarks. The zero-distance NLO Green's function is known in a compact analytic form, 
\bea
	\label{Coulomb_5}
	G^{[1]}(0;\sqrt{s}+i\Gamma_{\tilde{t}_1})&=&\frac{C^{[1]}\alpha_s(\mu_G)m_{\tilde{t}_1}^2} {4\pi} \\
	& & \times \Big[ g_{\mathrm{LO}} + \frac{\alpha_s(\mu_G)}{4\pi}g_{\mathrm{NLO}} + \dots \Big], \nonumber
\eea
where its UV-divergence at $\mathbf{r}=0$ has been removed via $\overline{\tt MS}$-subtraction \cite{Coulomb1c}. 
We work with $n_f=5$ active quark flavors and with $\alpha_s$ including additional top quark effects. We further 
renormalized $\alpha_s$ in the $\overline{\tt MS}$-scheme.
In Eq.\ (\ref{Coulomb_5}) we made use of the definitions
\begin{widetext}
\bea
	\label{Coulomb_6}
	g_{\mathrm{LO}}\hspace{2mm}&=&-\frac{1}{2\kappa} + L - \psi^{(0)}, \nonumber \\
	g_{\mathrm{NLO}}&=& 	\beta_0 \Big[ L^2 - 2L(\psi^{(0)} - \kappa\psi^{(1)}) + \kappa\psi^{(2)} + 
		(\psi^{(0)})^2-3\psi^{(1)} - 2\kappa\psi^{(0)}\psi^{(1)} 
		+ 4\hspace{2mm} _4F_3(1,1,1,1;2,2,1-\kappa;1)\Big] \nonumber \\
	& & + a_1 \Big[L-\psi^{(0)}+\kappa\psi^{(1)} \Big],
\eea
\end{widetext}
and

\bea
	\label{Coulomb_7}
	\kappa&=&\frac{iC^{[1]}\alpha_s(\mu_G)}{2v}, \nonumber\\
	v & = & \sqrt{\frac{\sqrt{s}+i\Gamma_{\tilde{t}_1}-2m_{\tilde{t}_1}}{m_{\tilde{t}_1}}},\nonumber\\
	L & = & \ln\frac{i\mu_G}{2m_{\tilde{t}_1}v}. 
\eea
Here, $\psi^{(n)}=\psi^{(n)}(1-\kappa)$ is the $n$-th derivative of 
$\psi(z) = \gamma_{\mathrm{E}} + \mathrm{d}/\mathrm{d}z\ln\Gamma(z)$ and $_4F_3(1,1,1,1;2,2,1-\kappa;1)]$ is 
a hypergeometric function (for further details see App.\ \ref{Appendix:Hypergeometric function}).
For the NLO Green's function in  Eq.\ (\ref{Coulomb_2}) $\mu_G$ can be chosen independently of the 
renormalization scale $\mu_R$. 
Since the Coulomb corrections are related to the exchange of potential gluons with momentum 
$|\bold{p}|\approx m_{\tilde{t}_1} v $, taking $\mu_G$ of the order 
\beq
\mu_G~\sim~m_{\tilde{t}_1} v~\sim~m_{\tilde{t}_1} \alpha_s
\eeq
is expected to be a natural choice (see Eq.\ (\ref{Coulomb_1})).
Hence, we define $\mu_G$ to be \cite{Coulomb1d}
\beq
	\label{Coulomb_10}
	\mu_G ~=~ \mathrm{max}\{C^{[1]} m_{\tilde{t}_1}\alpha_s(\mu_G),2 m_{\tilde{t}_1} v\},
\eeq
where $\mu_G ~=~ C^{[1]} m_{\tilde{t}_1}\alpha_s(\mu_G)$
corresponds to twice the inverse Bohr radius. 
It has been shown in Ref.\ \cite{Coulomb1e} for the color singlet 
top-antitop pair production near threshold, that, with $\mu_G$ set to this characteristic (s)quarkonium-energy 
scale, the Green's function possesses a well-convergent perturbative series.

To avoid double counting of NLO corrections, which are included in the Green's function as well as in our full NLO calculation (see, 
e.g., the first diagram of Fig.\ \ref{fig3}), we have to subtract the one-loop contribution  
\begin{align}
	\label{Coulomb_11}
	&\Im \Big\{ G^{[1]} \big(0;\sqrt{s} + i \Gamma_{\tilde{t}_1} \big) \Big\} = \\ 
	& m_{\tilde{t}_1}^2 \Im \Big\{ \frac{v}{4\pi} \Big[ i + \frac{\alpha_s(\mu_G) C^{[1]}}{v}
	\Big( \frac{i \pi}{2} + \ln \frac{\mu_G}{2m_{\tilde{t}_1}v} \Big)
		+ \mathcal{O}(\alpha_\mathrm{s}^2) \Big] \Big\}  \nonumber
\end{align}
from Eq.\ (\ref{Coulomb_5}). Eq.\ (\ref{Coulomb_11}) has been obtained by expanding Eq.\ (\ref{Coulomb_5}) 
up to $\mathcal{O}(\alpha_s^2)$.

Setting $\mu_G$ in Eq.\ (\ref{Coulomb_11}) to the hard scale $\mu_G=1$ TeV and renormalizing $\alpha_s$ 
according to Sec.\ \ref{Technical}, we find a matching between the Coulomb 
enhanced diagrams of the full NLO calculation and the Coulomb corrections expanded up to $\mathcal{O}(\alpha_s^2)$ in the threshold region with a precision better than 1\%.

\begin{figure*}
	\includegraphics[width=0.49\textwidth]{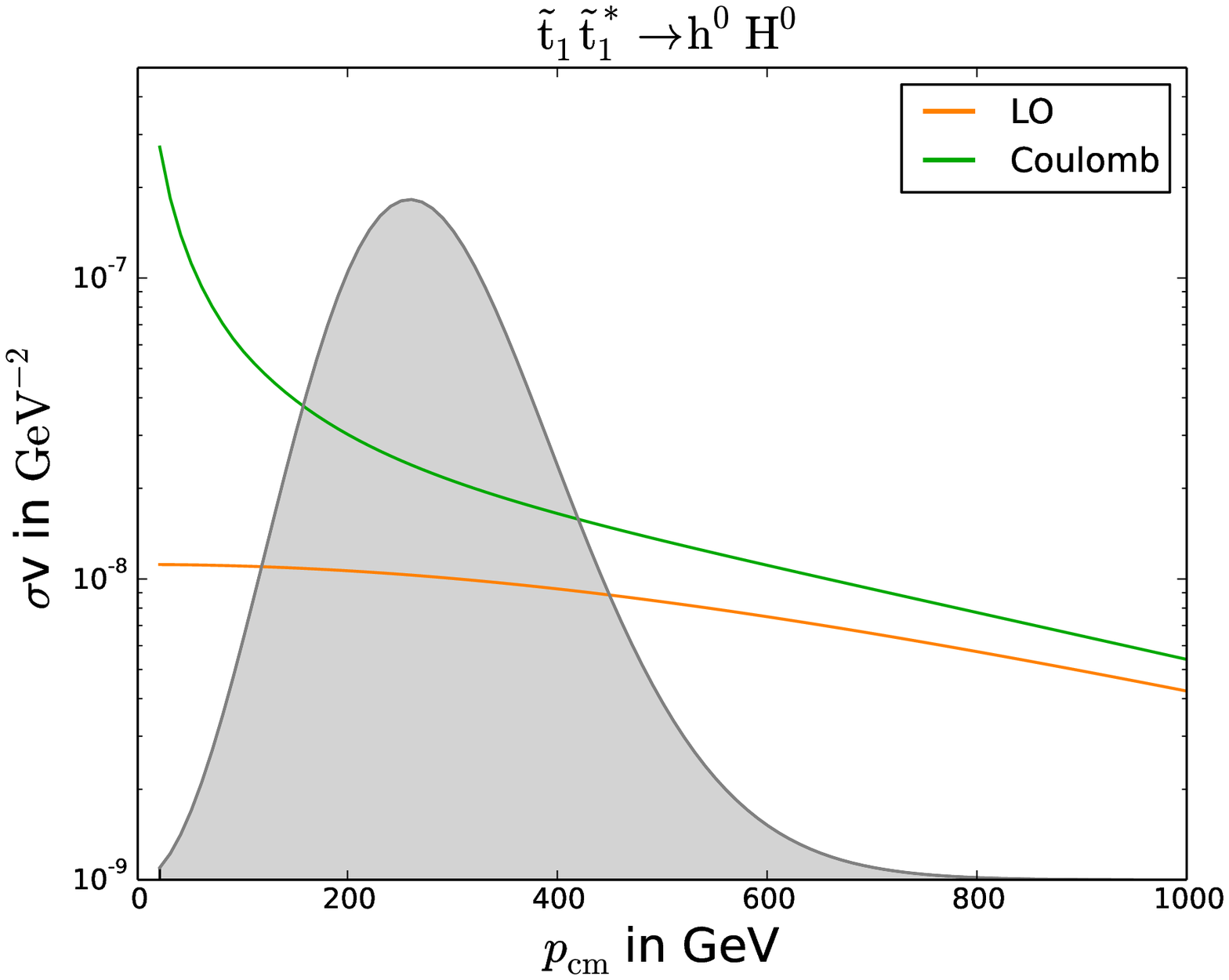}
	\includegraphics[width=0.49\textwidth]{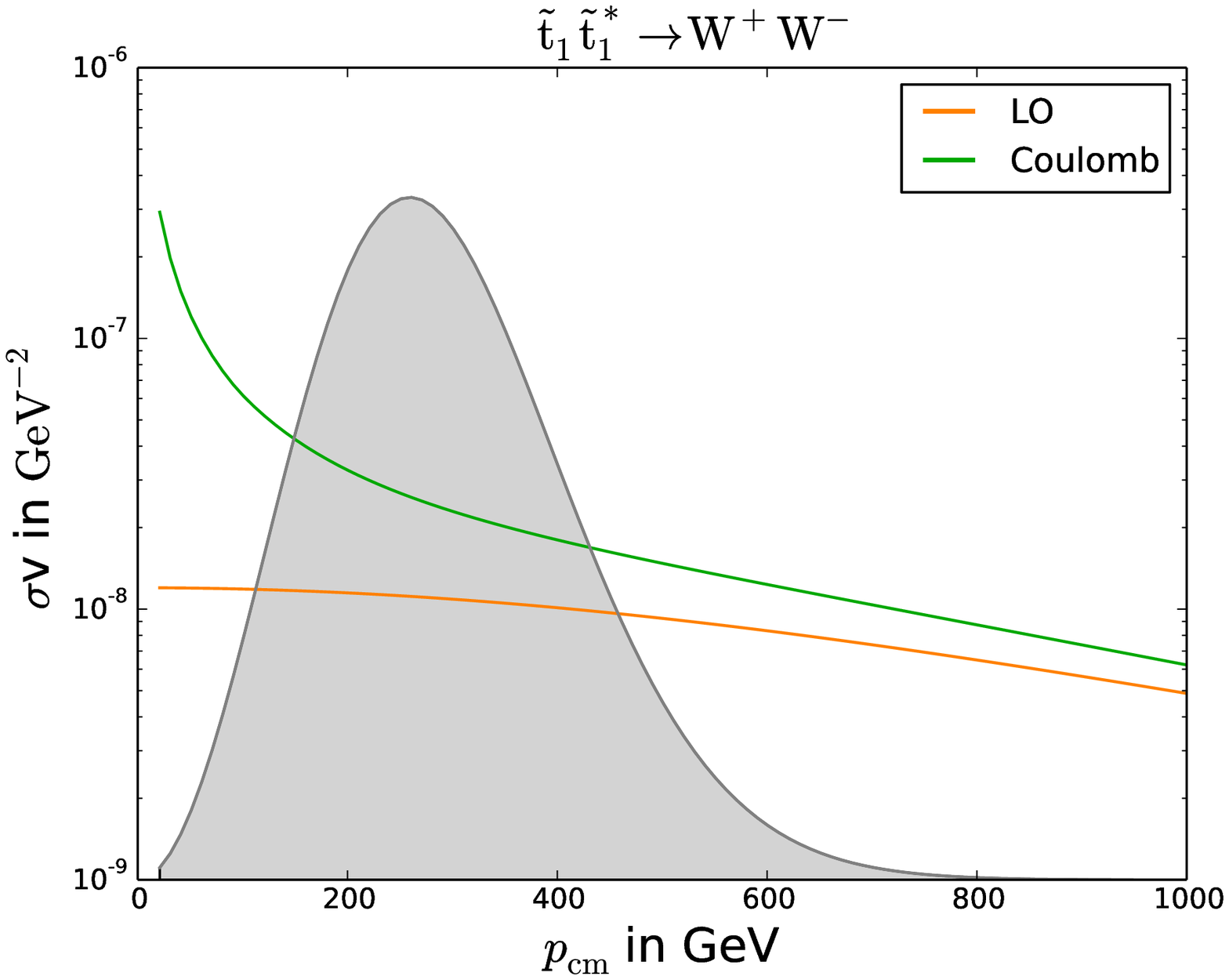}
	\vspace{0mm}
	\caption{The leading-order (orange line) and the Coulomb-corrected cross section (green line) multiplied with the relative velocity $v$ in dependence of the center-of-mass momentum $p_{\mathrm{cm}}$ for two selected channels of scenario \RM{2}. The grey areas indicate the thermal distribution (in arbitrary units).}
	\label{fig6}
\end{figure*}

Another subtlety arises as Eq.\ (\ref{Coulomb_6}) is only an expansion around the leading-order bound-state poles. It therefore induces 
poles in the Green's function of the general form $\big[ \alpha_s E_n^{\mathrm{LO}}/(E^{\mathrm{LO}}_n-\sqrt{s}-i\Gamma_{\tilde{t}_1}) \big]^k$ ($k$=1,2 at NLO), 
which differ by an $\mathcal{O}(\alpha_s^2)$ correction from an exact treatment 
\cite{Sommerfeld1, Coulomb4}. Hence, this difference only becomes relevant in the vicinity of the associated bound-state 
poles. But as their production is suppressed by the nonzero temperature during freeze-out\footnote{See also the vanishing weighting factor of the thermal distribution 
for $v \approx 0$ ($m_{\tilde{t}_1}v \ll T_{\mathrm{freeze-out}}$), e.g., in Fig.\ \ref{fig6}.}, 
there is no need for a more elaborated treatment in terms of a precise calculation of $\Omega_{\tilde{\chi}_1^0}h^2$.

Finally note that the approach presented here implicitly assumes that the amplitudes, which enter 
$\sigma^{\mathrm{LO}}$ in Eq.\ (\ref{Coulomb_2}), do not depend on the momenta of the 
annihila\-ting particles. In the case of dominant $s$-wave annihilation in the nonrelativistic limit 
this is a well justified approximation but turns out to be misleading for cross sections dominated by, e.g., the $p$-wave 
contribution. For these cases the Coulomb corrections for a leading-order Coulomb potential can be found in 
Refs.\ \cite{partial1,partial2}. Since we provide a complete NLO calculation, the error turns out to be 
of the order $\mathcal{O}(\alpha_{\mathrm{s}}^2)$ for $\alpha_s \ll v$ and remains of this order relative
to the leading $\mathcal{O}((\alpha_s/v)^n)$ Coulomb corrections even 
in the limit $\alpha_s \gtrsim v$. Hence, we choose to rely on this simplified treatment.

In Fig.\ \ref{fig6}, we compare cross sections which include the Coulomb corrections to the corresponding tree-level
cross sections for two processes of scenario \RM{2}. We chose scenario \RM{2} for presenting our results, but it should be noted that the basic qualitative behavior 
is scenario independent. The grey shaded areas represent the thermal averaging function in Eq.\ (\ref{Boltzmann3}) in arbitrary units 
and indicate the thermal weighting of the $\sigma v$ contribution to $\Omega_{\tilde{\chi}^0_1}h^2$. 

We show the stop-annihilation into the $h^0 H^0$ and $W^+ W^-$ final state. In both cases a steep rise of the 
Coulomb-corrected $\sigma v$ (green line) is observed for low $p_{\mathrm{cm}}$ due 
to the attractive force felt by the stop-antistop pair (see Eq.\ (\ref{Coulomb-potential})), 
whereas the tree-level (orange line), which is dominated by $s$-wave annihilation of the 
$\tilde{t}_1\tilde{t}^*_1$ pair, is roughly constant. 
For higher $p_{\mathrm{cm}}$ values, the $1/v$-enhancement becomes more and more subdominant, 
and the Coulomb corrections turn into a usual 
perturbative series in $\alpha_s$.  Although the Coulomb corrections become very large only in the region where 
the thermal distribution is small, Fig.\ \ref{fig6} can still elucidate the 
relevance of these corrections for a precision calculation of $\Omega_{\tilde{\chi}^0_1}h^2$.\\
\subsection{Further subtleties}
\label{Further subtleties}

\begin{figure}
	\includegraphics[keepaspectratio=true,trim = 60mm 143mm 50mm 120mm, clip=true, width=0.55\textwidth]{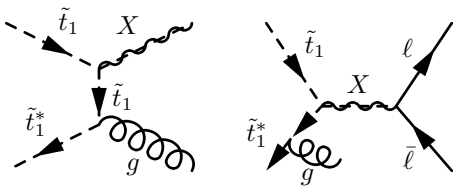}
	\caption{Example for a potential double-counting between $2\to2$ and $2\to3$ amplitudes for a $\ell\bar{\ell}$ final state.}
	\label{fig7}
\end{figure}

Some of the $2 \to 2$ amplitudes, which contribute to the final neutralino relic density $\Omega_{\tilde{\chi}_1^0}h^2$, contain a gluon and an unstable electroweak particle $X$, 
such as a Higgs or a $Z$-boson, in their final state. By further adding the $2 \to 3$ processes as, e.g., the diagrams of the first line of Fig.\ \ref{fig4}, we partly double-count some of these contributions. The reason is that in the case of an on-shell Higgs or vector boson propagator the $2 \to 3$ amplitude corresponds to the on-shell production of a gluon and a heavy boson $X$ followed by its decay, which is already included within the $2 \to 2$ processes (exemplified in Fig.\ \ref{fig7}). 

To avoid this double counting, we subtract from the usual $2 \to 2$ matrix element the $2 \to 2$ matrix element weighted by the fraction of the EW decay width $\Gamma_{\mathrm{X}\rightarrow\mathrm{EW}}$ divided by the total decay width $\Gamma_{\mathrm{X}\rightarrow\mathrm{tot}}$, both for a two-particle final state. More precisely, we have introduced the replacement
\beq
	\label{on_shell2}
	\big| M_{\tilde{t}_1\tilde{t}_1^*\rightarrow Xg} \big|^2 \rightarrow 	
	\left( 1 - \frac{\Gamma_{X \rightarrow \mathrm{EW} }}{\Gamma_{X \rightarrow \mathrm{tot}}} \right)  
	\times \big| M_{\tilde{t}_1\tilde{t}^*_1\rightarrow Xg} \big|^2 .
\eeq
Within our implementation it is in principal possible that in some rare cases a gluon-$X$ final state is corrected as in Eq.\ (\ref{on_shell2}) without the corresponding $2 \to 3$ amplitude having been 
taken into account. But as we correct all processes which contribute more than 1$\%$ to $\Omega_{\tilde{\chi}_1^0} h^2$, we expect this to be a minor error with respect to the aimed level of precision.

One more comment seems to be in order concerning the radiation of potentially soft photons. In the case of photons in the final state,
the $2 \to 3$ real radiation process is IR divergent as the photon can become soft. As for the gluon this soft behavior would cancel if one would take the corresponding virtual corrections into account. This is, however, beyond the scope of this work as it would require the inclusion of EW corrections. To regulate the divergence we have introduced a lower bound on the photon energy similar to $\Delta$E in Sec.\ \ref{Technical}, which did not much alter the final relic density but prevented the integration over the $2 \to 3$ phase-space from becoming numerically 
unstable\footnote{The $2\to 3$ corrections turn out to be only a tiny contribution to $\Omega_{\tilde{\chi}_1^0}h^2$ for most of the relevant channels (see Sec.\ \ref{CrossSectionTeil}), 
and channels with photon final states are in general less important (Sec.\ \ref{Pheno}).}.

Further, we have introduced electron and muon masses, $m_e = 5.1\cdot10^{-4}$ GeV and $m_{\mu}=0.106$ GeV, to keep the photon propagator in the 
last diagram of Fig.\ \ref{fig4} away from its mass shell. 

For consistency all changes including the associated lepton-Higgs couplings have been implemented in {\tt CalcHEP} and are used by {\tt micrOMEGAs} in
our analysis. Finally our {\tt DM@NLO} package includes a lower bound on the squark widths in order to stabilize the 
phase-space integration in the vicinity of squark-propagator poles. We set this bound to 0.01 GeV.
If the value of a particular squark width, by default taken from {\tt micrOMEGAs}, drops below this bound, 
we set its value to the 0.01 GeV, and keep the {\tt micrOMEGAs} value otherwise.

\section{Numerical results}
\label{Numerics}

\subsection{Impact on the cross section}
\label{CrossSectionTeil}

\begin{figure*}
	\includegraphics[width=0.49\textwidth]{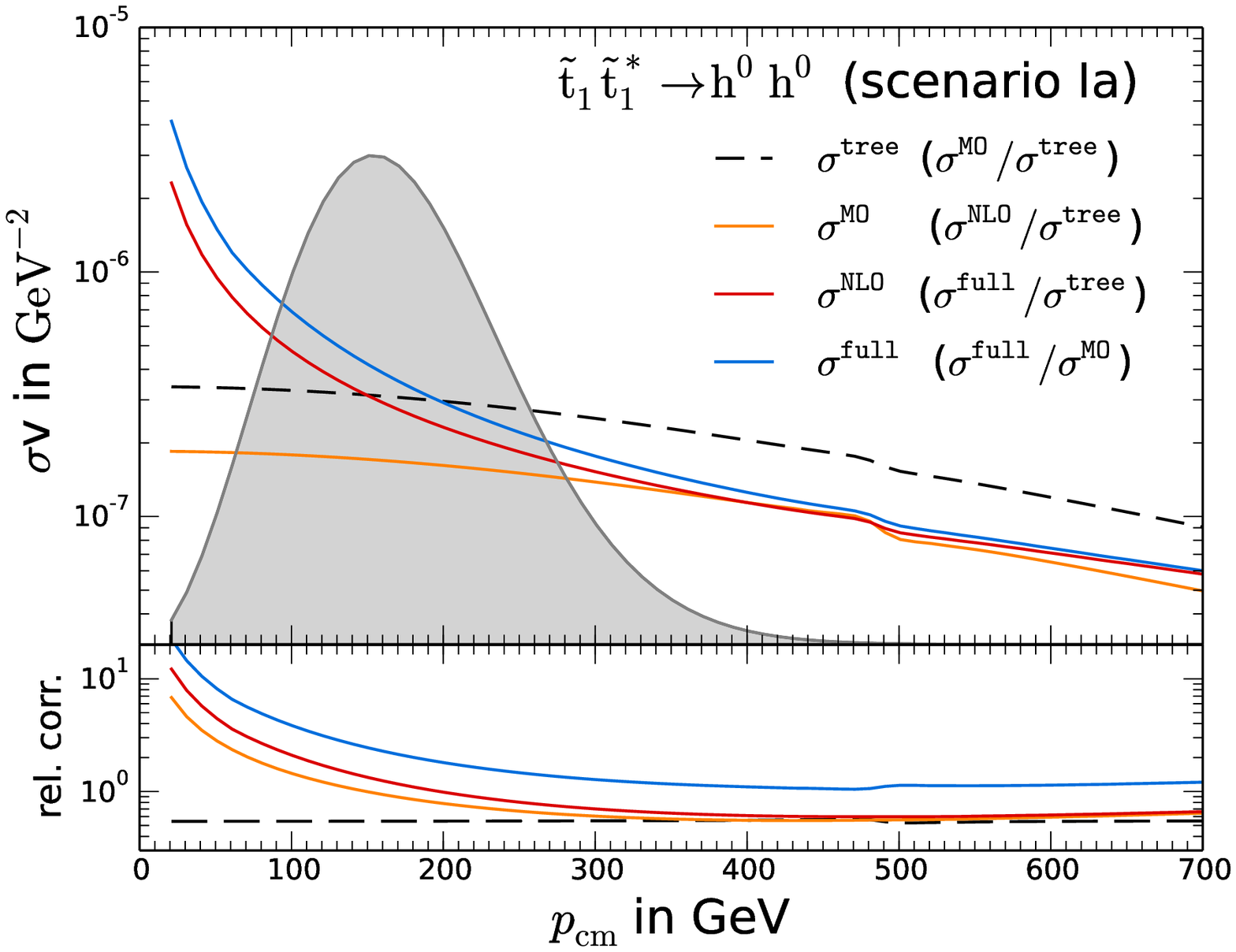}
	\includegraphics[width=0.49\textwidth]{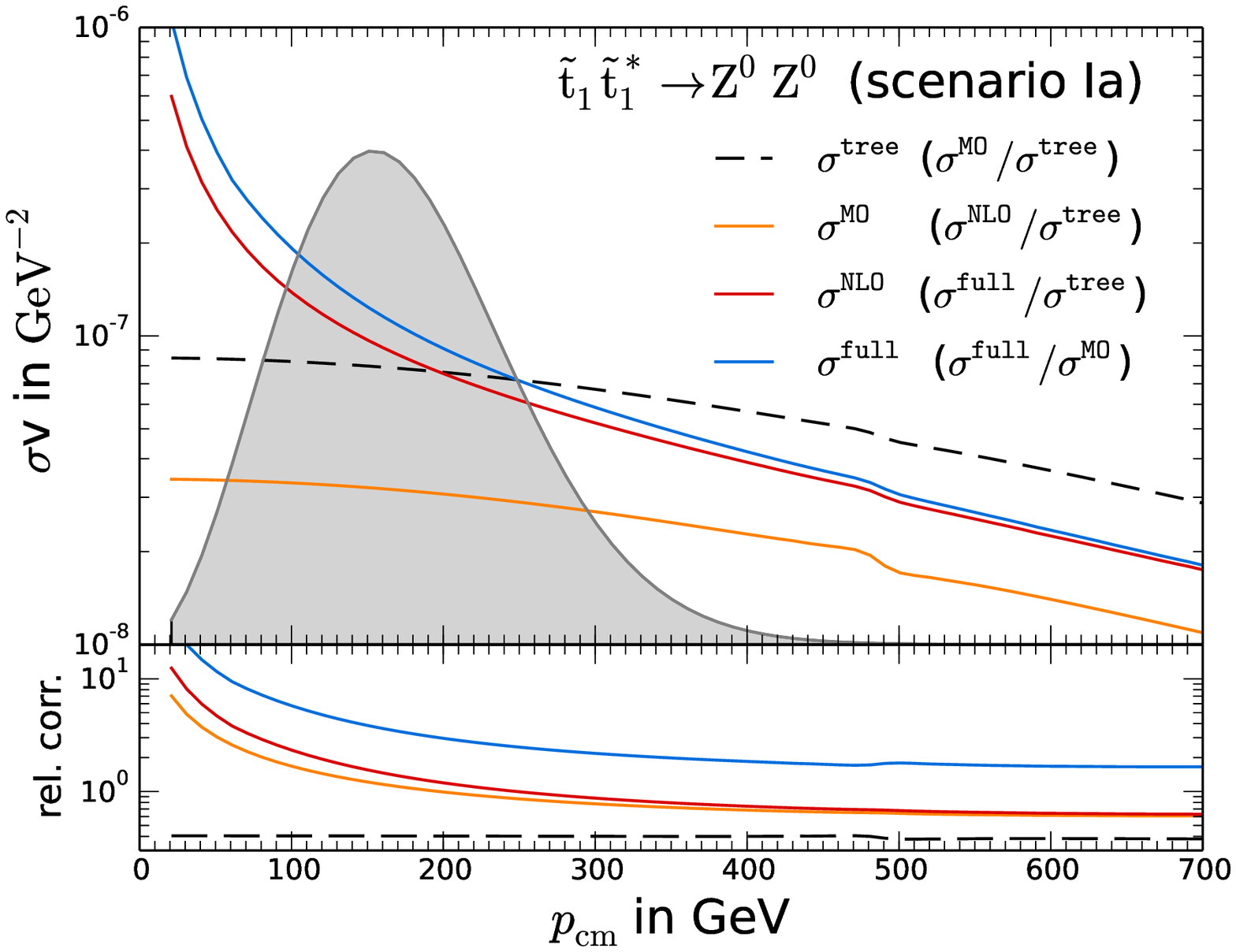}
	\includegraphics[width=0.49\textwidth]{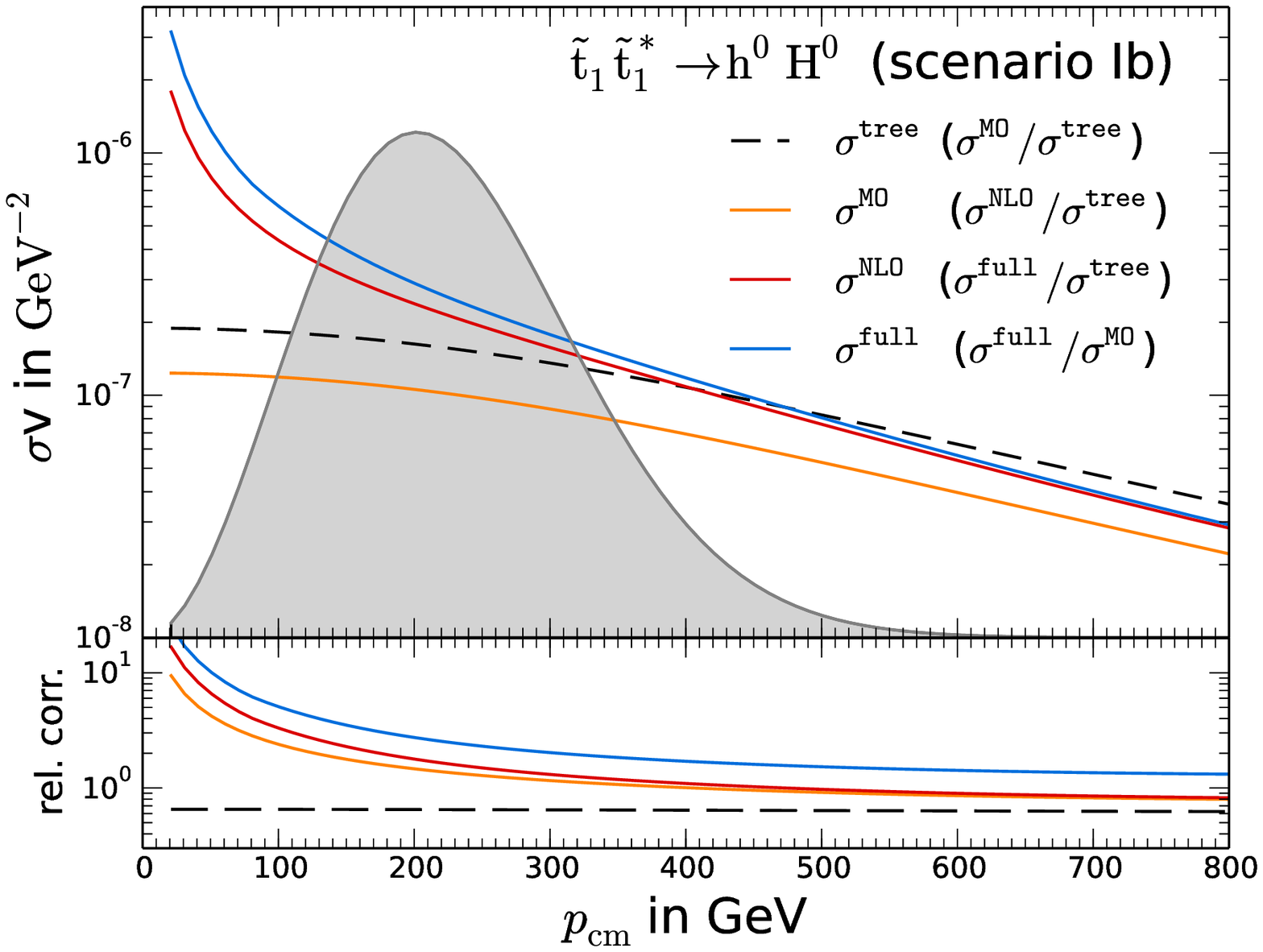}
	\includegraphics[width=0.49\textwidth]{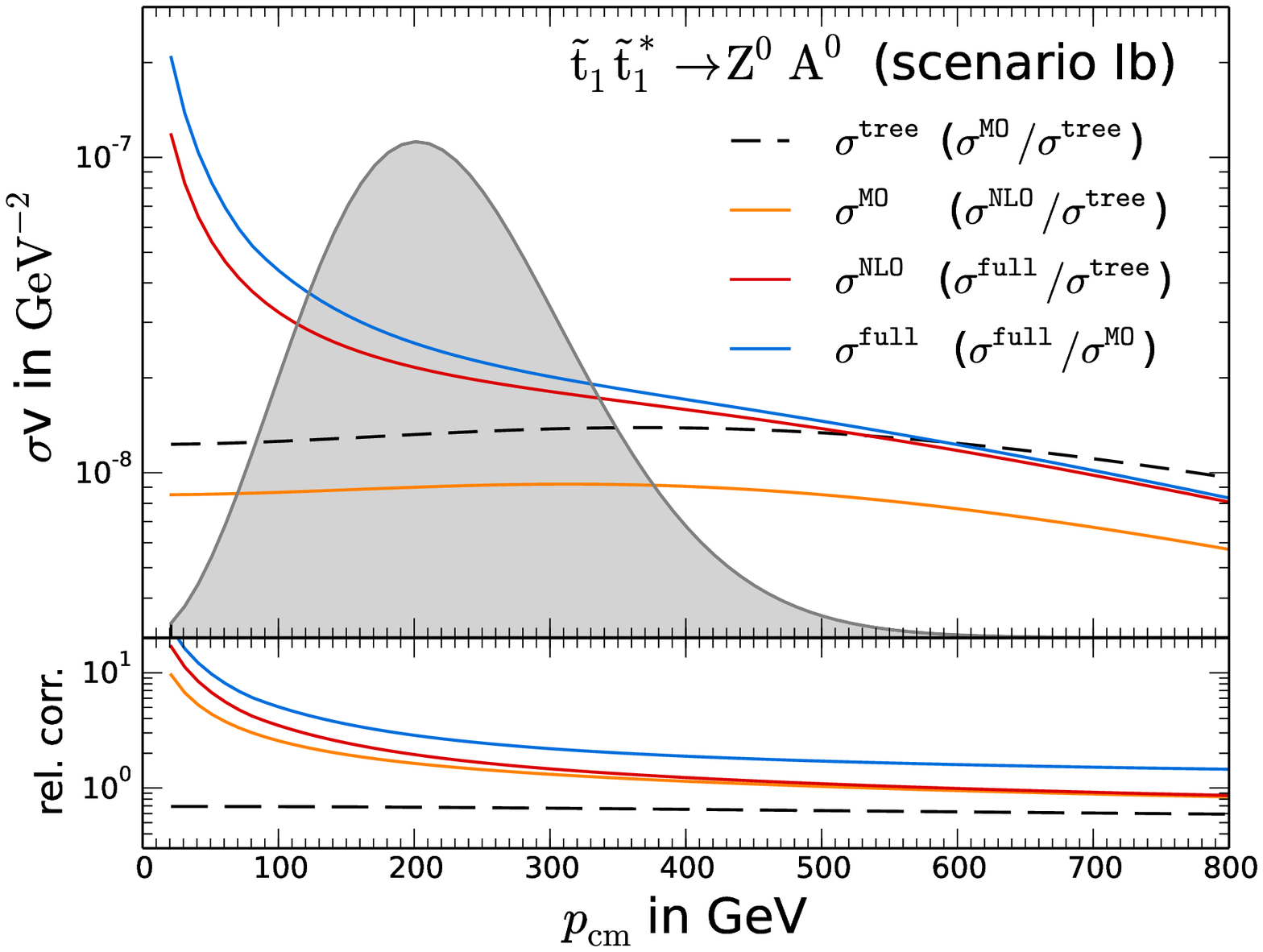}
	\includegraphics[width=0.49\textwidth]{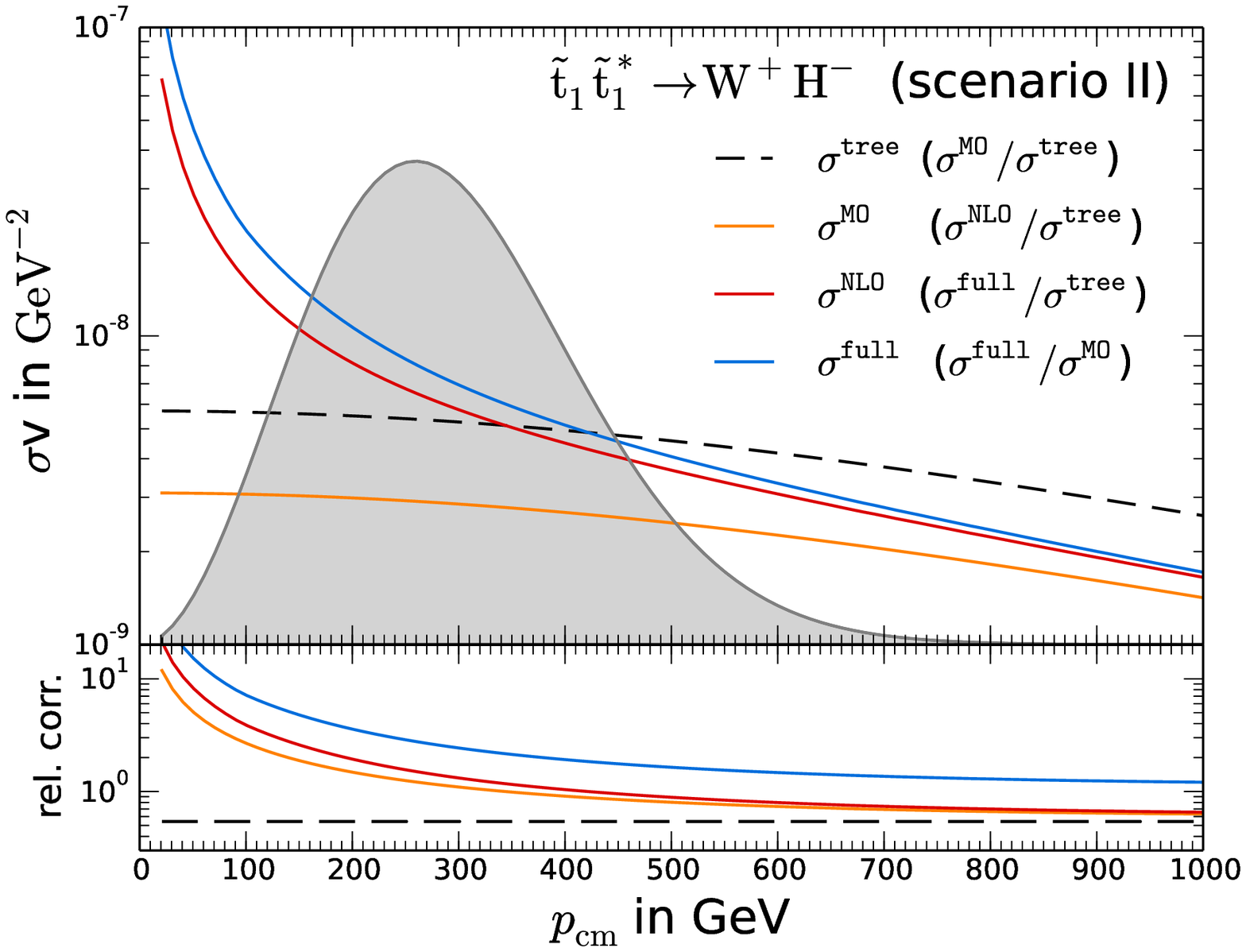}
	\includegraphics[width=0.49\textwidth]{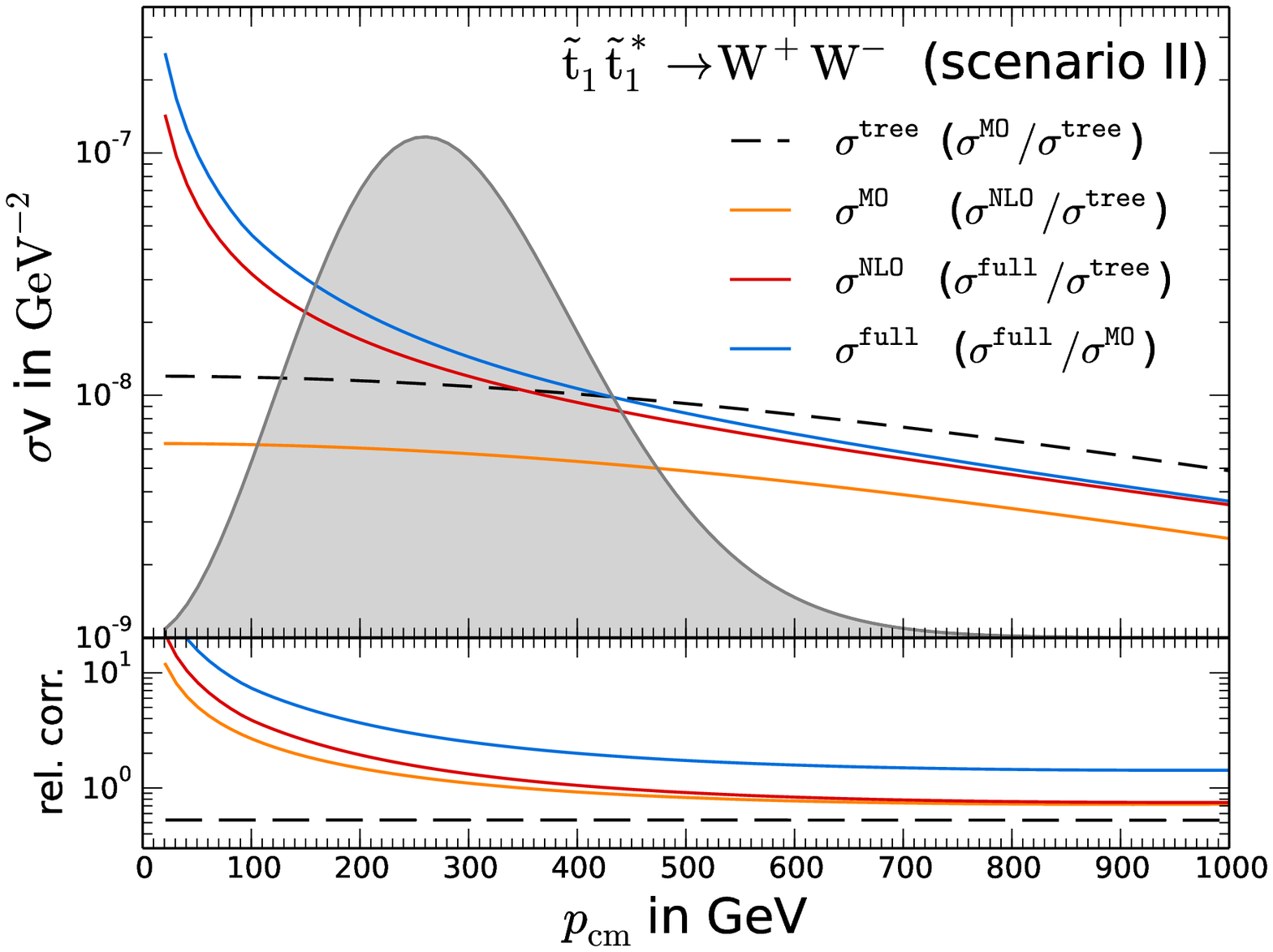}
	\caption{Tree level (black dashed line), \MO\ (orange solid line), NLO ($\mathcal{O}(\alpha_s)$) corrections 
	(red solid line) and full corrections of Sec.\ \ref{Technical_details} (blue solid line) for  selected 
	channels in the scenarios of Tab. \ref{ScenarioList}. The upper part of each plot shows $\sigma v$ 
	in GeV$^{-2}$ in dependence of the momentum in the center-of-mass frame $p_{\mathrm{cm}}$. The grey areas 
	indicate the thermal distribution (in arbitrary units). The lower parts of the plots show the corresponding 
	ratios of the cross sections (second item in the legends).}
	\label{fig8}
\end{figure*}

We now turn to the discussion of the impact of our full corrections presented in Sec.\ \ref{Technical_details} on the 
processes listed in Eqs.\ (\ref{channels1}) -- (\ref{channels4}). In Fig.\ \ref{fig8}, we show the cross 
sections multiplied by the relative velocity $v$ as a function of the center-of-mass momentum $p_{\mathrm{cm}}$ for selected 
annihilation channels of the three reference scenarios presented in Tab.\ \ref{ScenarioList}. More precisely, 
we show the cross section at tree-level (black dashed line), including the full $\mathcal{O}(\alpha_s)$ corrections as 
discussed in Sec.\ \ref{Technical} (red solid line), with the full corrections including the Coulomb corrections of 
Sec.\ \ref{Coulomb corrections} (blue solid line), and the corresponding value obtained by \MO/{\tt CalcHEP} 
(orange solid line). 
The lower part of each plot contains different ratios between the four cross sections (second item in the legend). 
As before, the grey shaded regions represent the thermal weighting of the $\sigma v$ contributions to 
$\langle \sigma_{\rm ann}v \rangle$ in Eq.\ (\ref{Boltzmann3}). 

The upper left plot of Fig.\ \ref{fig8} shows $\sigma v$ for the process 
$\tilde{t}_1\tilde{t}^*_1 \to h^0 h^0$, which is the dominant subchannel in scenario \RM{1}a. 
We observe that our prediction for the cross section at 
tree-level deviates by roughly $45\%$ from the \MO\ result. This deviation can be traced back to a different 
treatment of couplings as well as different input parameters used within \MO. In particular, \MO\ uses the 
$\overline{\mathrm{DR}}$-top mass $m^{\overline{\mathrm{DR}}}_{\mathrm{t}}=161.6$ GeV whereas we take the on-shell 
top mass $m^{\tt OS}_{\mathrm{t}}=172.3$ GeV. These enter the Yukawa couplings and in turn alter the 
important $t$- and $u$-channels (see Tab.\ \ref{ScenarioSubChannels}), which is the main reason for the 
observed shift between our tree-level and the \MO\ result. Due to the Coulomb corrections discussed in 
Sec.\ \ref{Coulomb corrections}, the higher-order corrections (red and blue curves) rise steeply for small velocities 
(i.e.,\ small $p_{\mathrm{cm}}$). For larger values of $p_{\mathrm{cm}} > 400$ GeV, the Coulomb 
corrections become less relevant, and the full correction converges against the $\mathcal{O}(\alpha_s)$ correction 
with growing $p_{\mathrm{cm}}$, whereas the $2\to 3$ processes become more and more important and already start to 
significantly alter the $p_{\mathrm{cm}}$ dependence of the NLO and full result. 
Here, the full correction leads to a change of around 35$\%$ compared to our tree-level calculation.

Comparing the ratios $\sigma_{\mathrm{full}}/\sigma_{\mathrm{tree}}$ (red line) and 
$\sigma_{\mathrm{NLO}}/\sigma_{\mathrm{tree}}$ (orange line) in the lower part of the plot within the most relevant 
region for the calculation of $\Omega_{\tilde{\chi}_1^0}h^2$ between $p_{\mathrm{cm}}=50$ and $p_{\mathrm{cm}}=350$ 
GeV, we observe that the Coulomb correction significantly contributes even 
beyond the NLO. Its contribution at next-to-nexto-to-leading order (NNLO) and higher amounts up to about half of the 
$\mathcal{O}(\alpha_s)$ contribution. Furthermore, our full result deviates from the tree-level by up to $300\%$ and 
from the \MO\ result even by up to a factor $7$ to $8$ within the interval between $p_{\mathrm{cm}}=50$ GeV and $350$ 
GeV.

\begin{figure*}
	\includegraphics[width=0.49\textwidth]{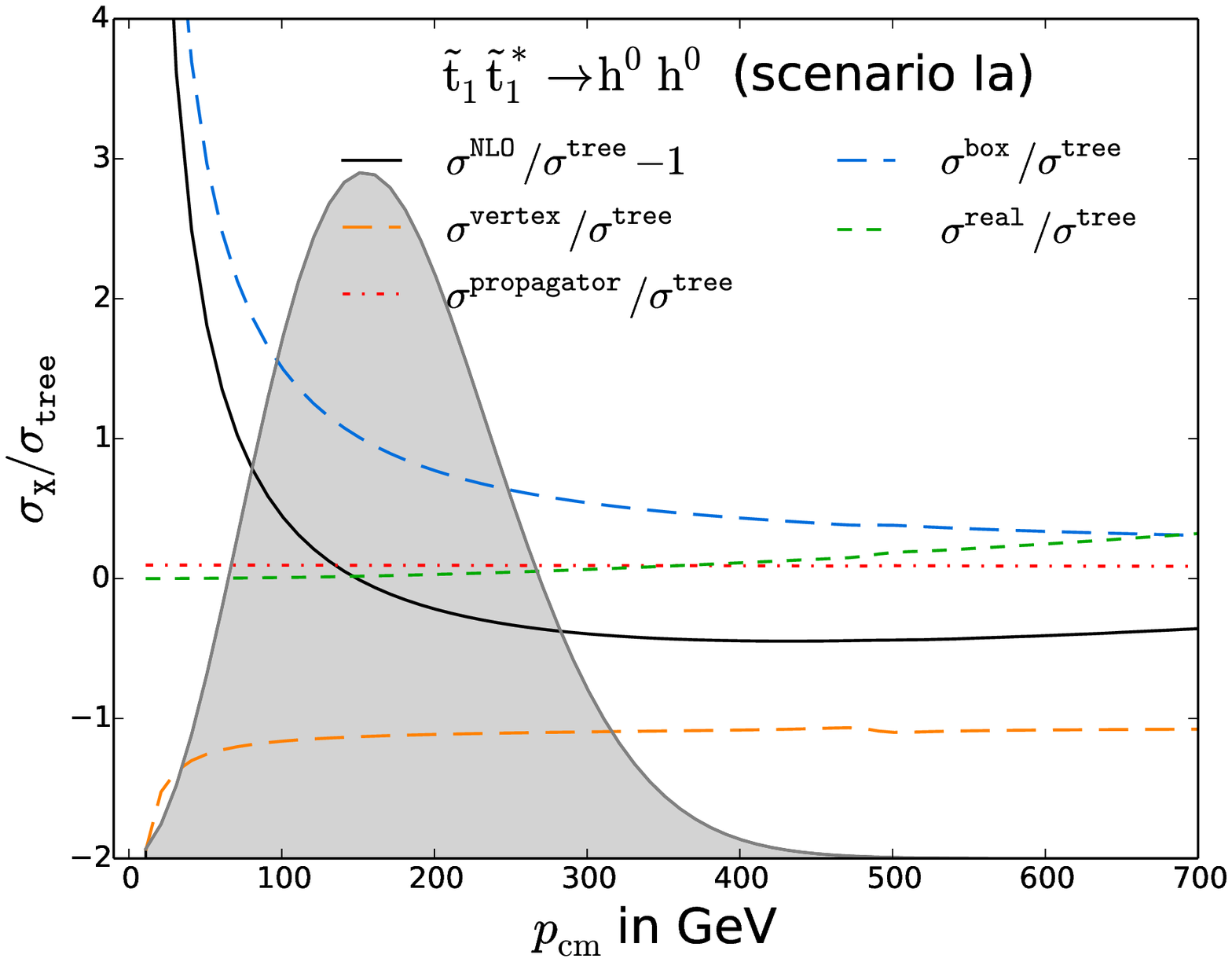}
	\includegraphics[width=0.49\textwidth]{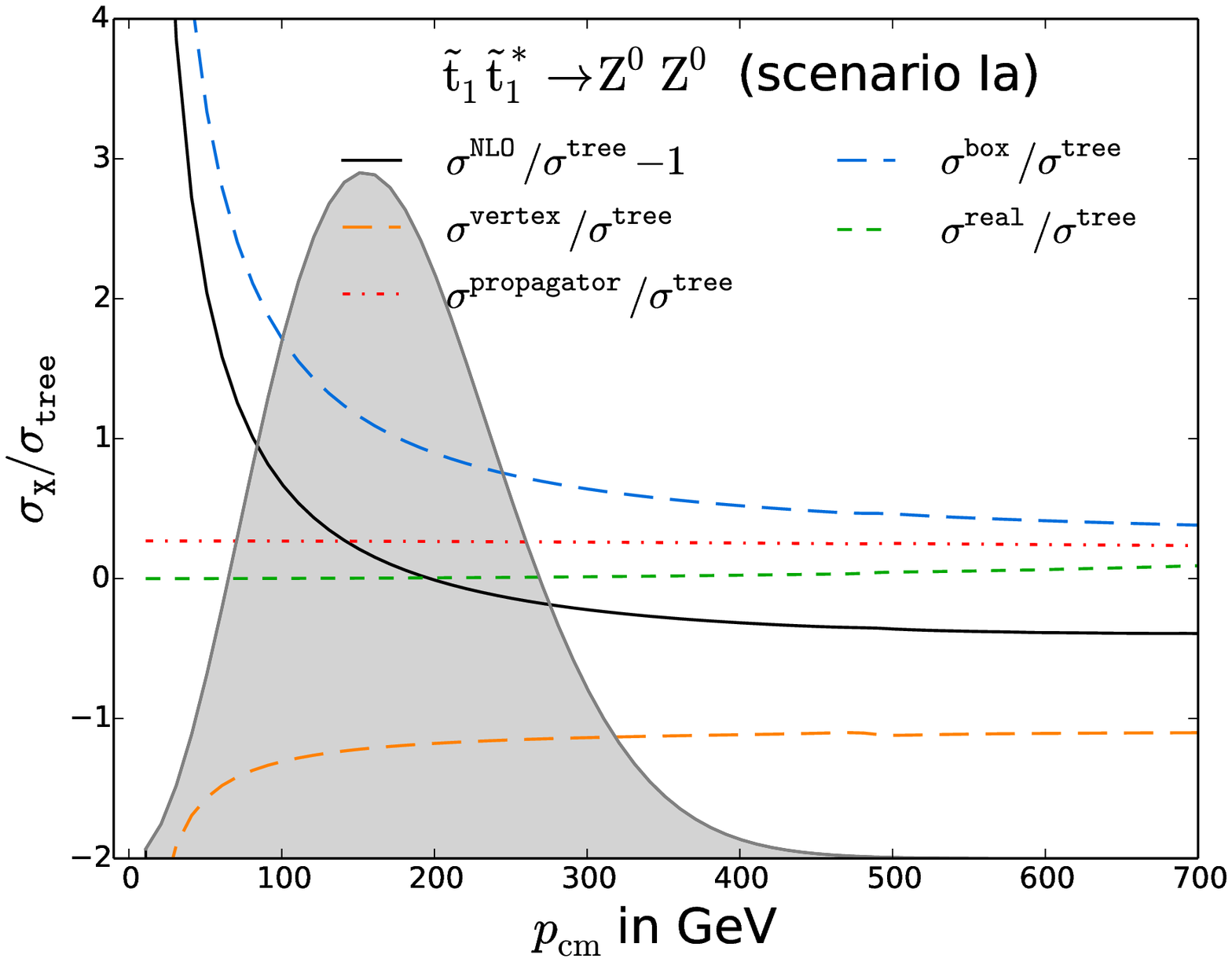}
	\includegraphics[width=0.49\textwidth]{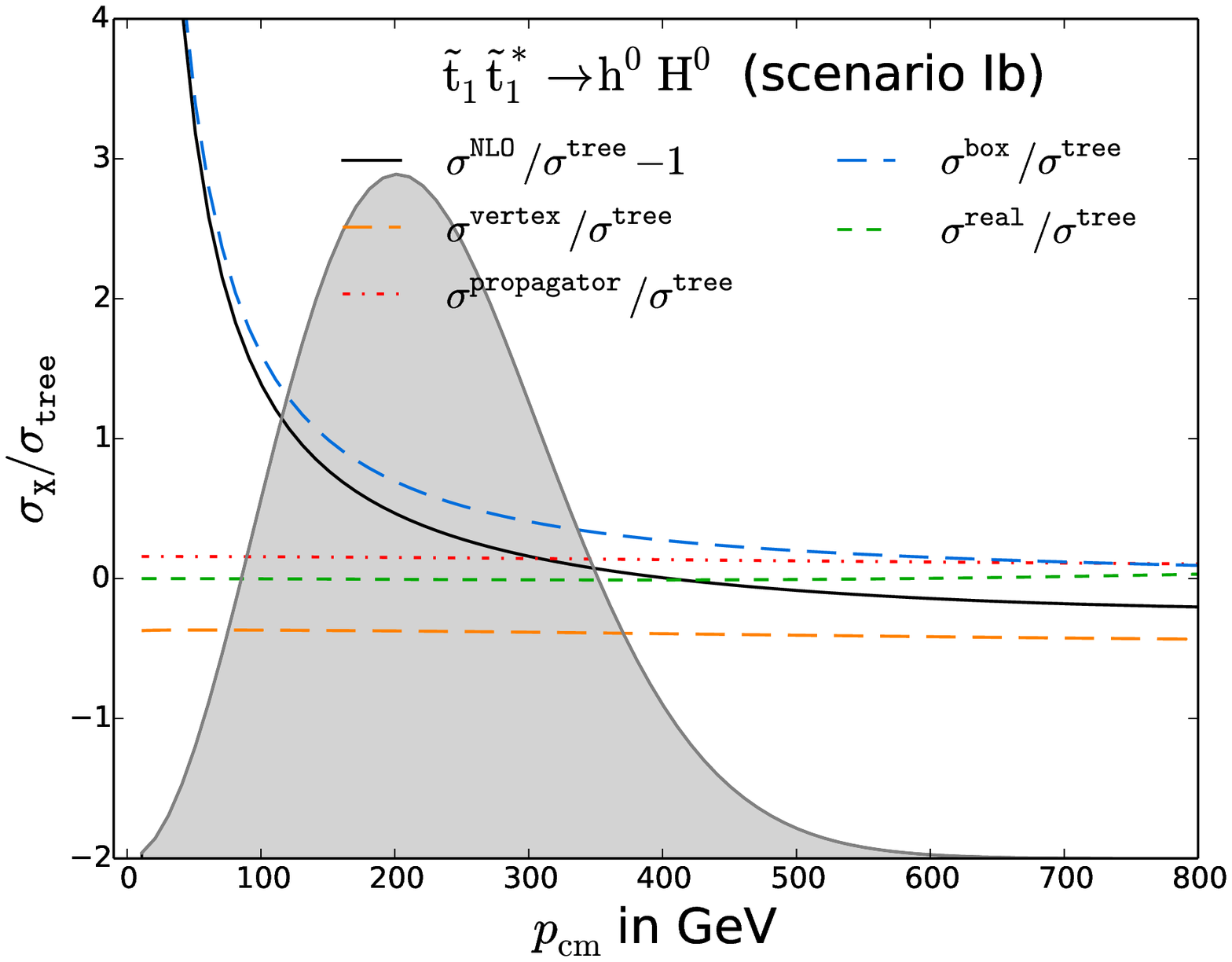}
	\includegraphics[width=0.49\textwidth]{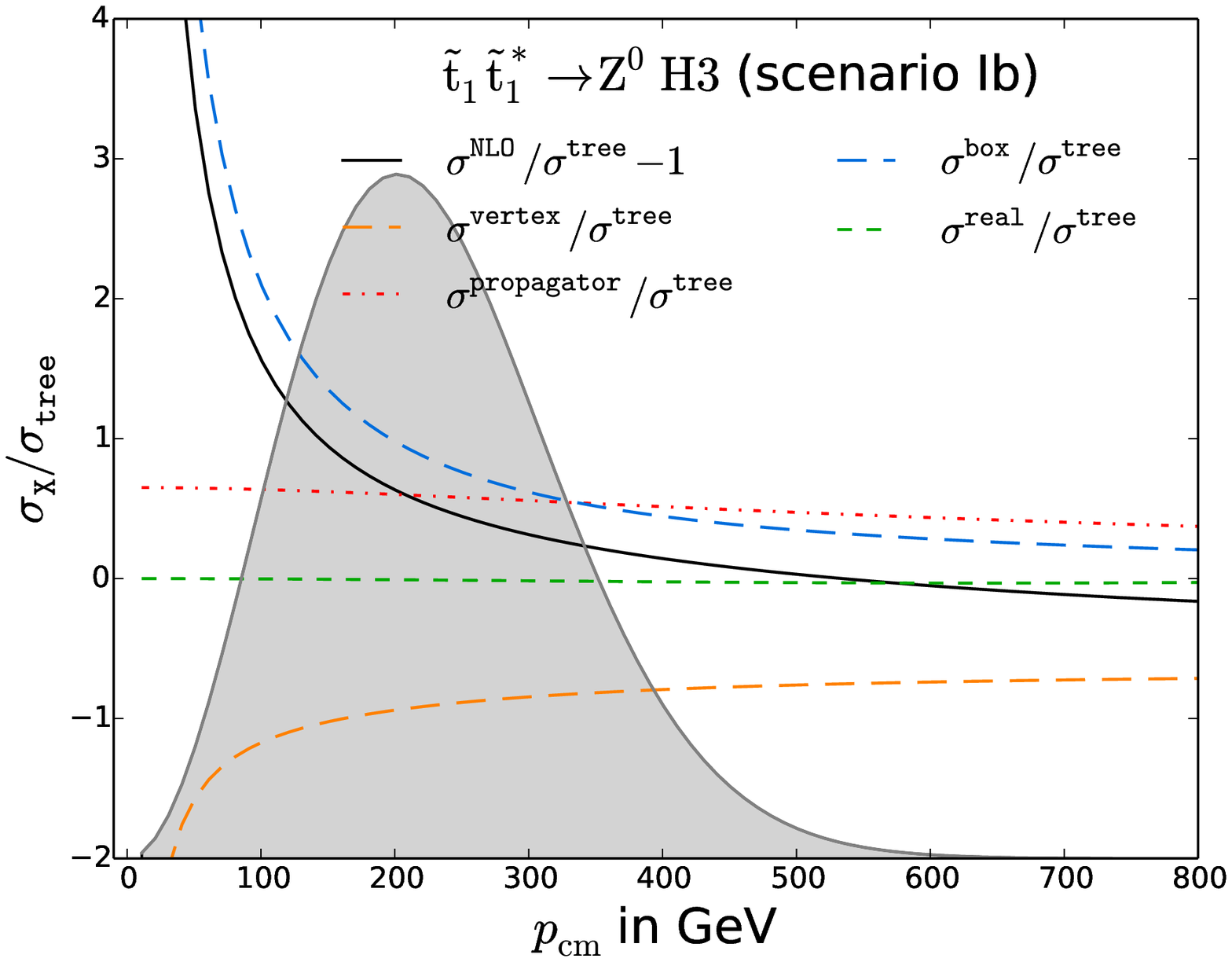}
	\includegraphics[width=0.49\textwidth]{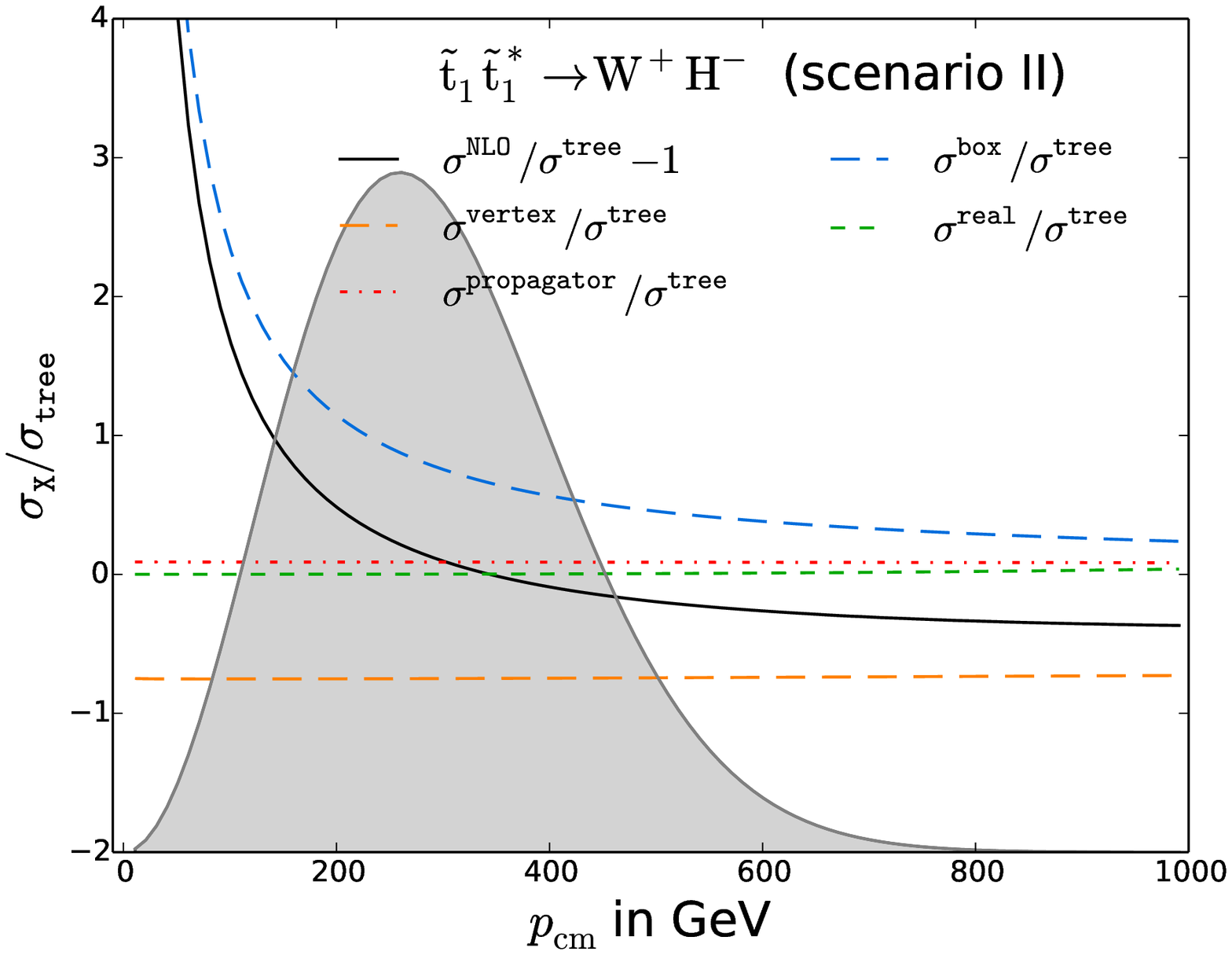}
	\includegraphics[width=0.49\textwidth]{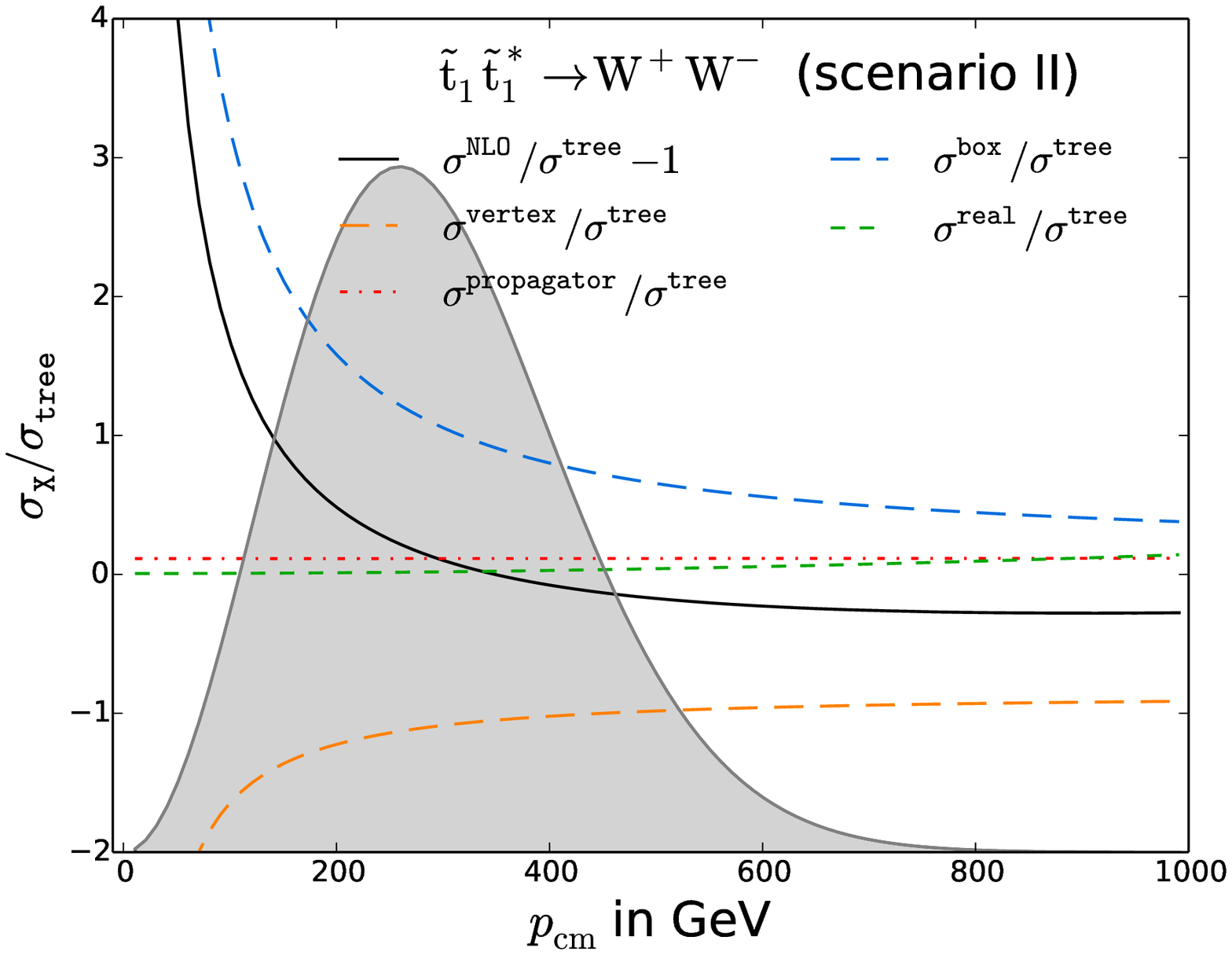}
	\caption{Results for NLO- (without tree-level, black), vertex- (orange), propagator- (red), box-  (blue) and  
	real plus soft photon corrections (green) of Sec.\ \ref{Technical} for selected channels in the scenarios of Tab. 
	\ref{ScenarioList}. The plots show ratios of the different corrections over tree-level cross sections dependent on 
	$p_{\mathrm{cm}}$. The grey areas indicate the thermal distribution (in arbitrary units).}
	\label{fig9}
\end{figure*}
In the upper right corner of Fig.\ \ref{fig8}, we show the analogous plot for the process 
$\tilde{t}_1\tilde{t}^*_1 \to Z^0 Z^0$ of scenario \RM{1}a. Here, our tree-level differs again quite strongly from the 
\MO\ result by about $60\%$. As before this deviation can be 
traced back to the different treatment of couplings and input parameters due to our choice of the renormalization 
scheme. For small $p_{\mathrm{cm}}$, however, the Coulomb enhancement takes over again and results in large 
corrections of a factor of 10 and more relative to our tree-level. In the important region between 
$p_{\mathrm{cm}} = 50$ and $350$ GeV, the deviation between our full correction and our tree-level amounts up to a factor 3 or 4,
whereas the ratio between the full result and \MO\ gets even larger by a factor 3 and more. 

With these two final states, $h^0 h^0$ and $Z^0 Z^0$, constituting around 55\% of the total annihilation cross 
section $\langle \sigma_{\rm ann}v \rangle$ (see Tab.\ \ref{ScenarioChannels}), the importance of our corrections to 
the neutralino relic density is already indicated at this point.

The small kinks in the upper two plots of Fig.\ \ref{fig8} around $p_{\mathrm{cm}} = 485$ GeV are due to 
a very broad $s$-channel resonance caused by the heavier $CP$-even Higgs $H^0$. Even though the pseudoscalar Higgs 
boson $A^0$ is similar in mass ($m_{A^0}\approx m_{H^0}=1917.4$ GeV), it does not contribute to the $s$-channel in 
the case of $\tilde{t}_1\tilde{t}^*_1$ annihilation (see Sec.\ \ref{Pheno}) as it is $CP$-odd.

The remaining four plots show $\tilde{t}_1\tilde{t}^*_1\to h^0 H^0$ and 
$\tilde{t}_1\tilde{t}^*_1 \to Z^0 A^0$ for scenario \RM{1}b and $\tilde{t}_1\tilde{t}^*_1 \to W^+ H^-$ and 
$\tilde{t}_1\tilde{t}^*_1 \to W^+ W^-$ for scenario \RM{2}. In all four cases our tree-level differs quite strongly from 
the \MO\ result by up to roughly $50\%$. But although the $Z^0 A^0$ final state is quite similar to the $Z^0 Z^0$ final 
state the deviation between our tree-level and \MO\ is in the former case only half as large as in the latter case. 
The large difference seen in the case of the $Z^0 Z^0$ final state comes, besides the different
treatment of the top mass, from the longitudinal polarized vector bosons which are in the Feynman gauge represented by 
the Goldstone bosons $G^0$. More accurately, it is the coupling $\tilde{t}_1 \tilde{t}_2 G^0$ that causes the large 
difference in Fig.\ \ref{fig8}. 
It is treated differently in \MO\ and enters the $t$- and $u$-channel contributions twice in the case of $Z^0 Z^0$ but 
only once, e.g., if the final state is $Z^0 A^0$.

In the last four plots, the Coulomb corrections dominate our higher-order corrections in the region of small 
$p_{\mathrm{cm}}$. For large values of $p_{\mathrm{cm}}$, 
however, the full $\mathcal{O}(\alpha_s)$ corrections become relevant and give rise to corrections between roughly 
15$\%$ and 35$\%$. 
In the region relevant for $\Omega_{\tilde{\chi}_1^0}h^2$, i.e. in the vicinity of the peak of the thermal distribution, 
the deviation between our full result and our tree-level accounts for roughly 50\% to 100\% and between our full result 
and \MO\ for around 200 \%.

In Fig.\ \ref{fig9}, we present the decomposition of the absolute value of the 
NLO cross section without tree-level contributions $\sigma^{\tt NLO}/\sigma^{\tt tree}-1$ (black) into the various 
types of 
UV finite $\mathcal{O}(\alpha_s)$ corrections for each of the processes of Fig.\ \ref{fig8}. 
More precisely, we show the vertex (orange), propagator (red), box (blue) and real corrections (green), where the 
latter also contain the soft 
gluon contribution as discussed in Sec.\ \ref{Technical}. All contributions are normalized to the tree-level cross 
section. 
Although all these contributions are UV finite, the vertex, box and real corrections 
are separately IR divergent as well as dependent on large logarithms of the regularization scale $\mu$. These 
logarithms  cancel between the individual contributions of Fig.\ \ref{fig9}.

Comparing the different contributions for each process, one can clearly identify the subclasses of  
$\mathcal{O}(\alpha_s)$ corrections, 
which are enhanced by the Coulomb corrections of Sec.\ \ref{Coulomb corrections}, namely the vertex and box corrections.
Only the vertex corrections of the processes $\tilde{t}_1\tilde{t}^*_1\to h^0 H^0$ and 
$\tilde{t}_1\tilde{t}^*_1 \to W^+ H^-$ show no significant rise at small $p_{\mathrm{cm}}$. This is due to
the dominant $t$- and $u-$ channels contributions for these cases, which turn out to be much larger than the 
Coulomb enhanced diagrams subsumed under the vertex corrections (see Tab.\ \ref{ScenarioSubChannels}). Hence, one has 
to go to much smaller $p_{\mathrm{cm}} \sim \mathcal{O}(10^{-3} \hspace{0.5mm}\mathrm{GeV})$ to see a significant rise in 
the vertex corrections, which is, however, not shown here. 

The sum of box and vertex corrections results in a positive correction at low $p_{\mathrm{cm}}$.
For large $p_{\mathrm{cm}}$, however, the situation is reversed, and the overall corrections are negative. 
The point where the overall correction changes its sign is clearly visible in each plot and is given by the point 
where the box and vertex contributions are roughly the same. 
The real emission corrections are subdominant in all cases and rise only for larger $p_{\mathrm{cm}}$, 
where the larger kinematically accessible phase-space of the $2\to 3$ processes enhances the associated total cross 
sections. 

\subsection{Impact on the relic density}

\begin{figure*}
	\includegraphics[width=0.49\textwidth]{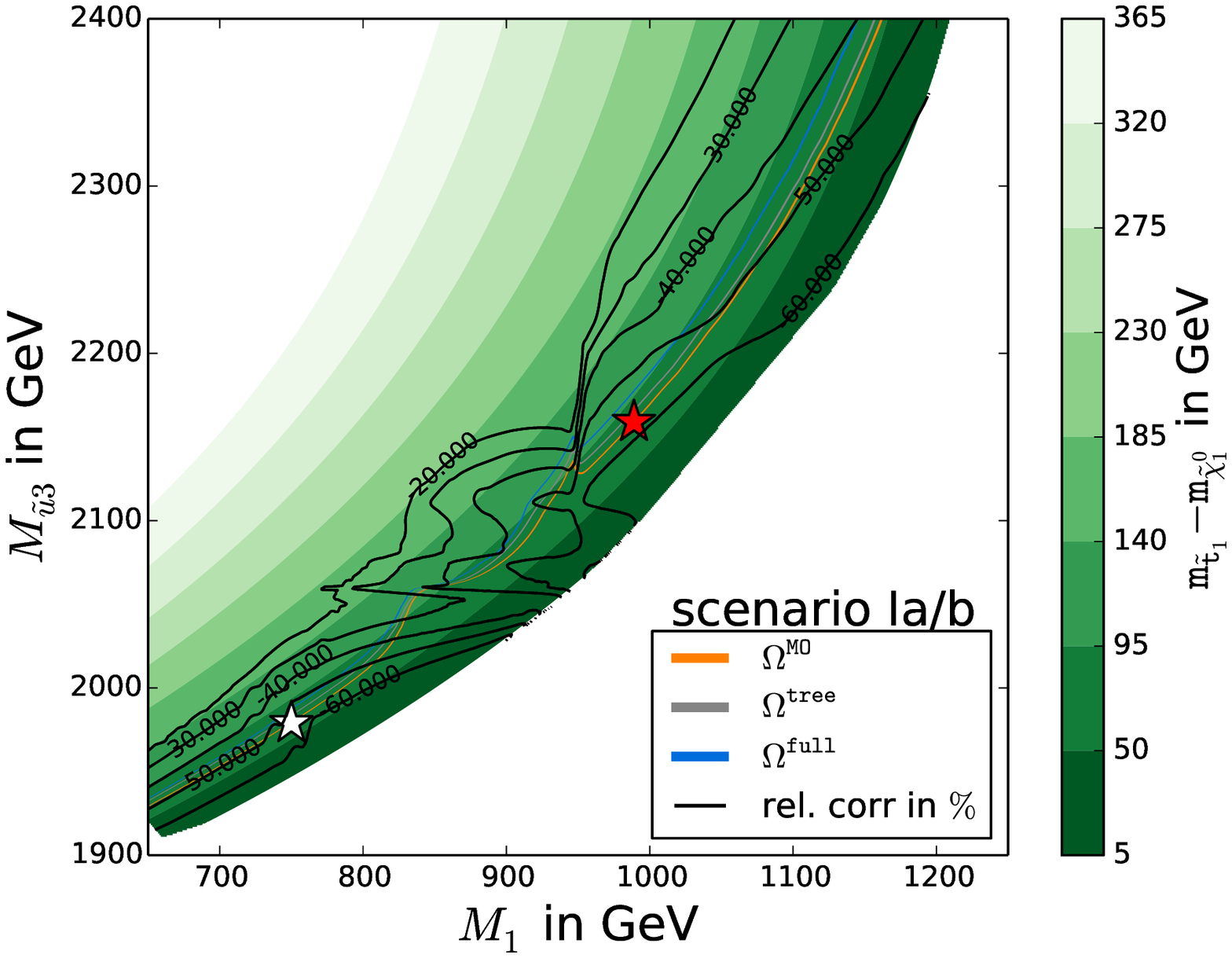}
	\includegraphics[width=0.49\textwidth]{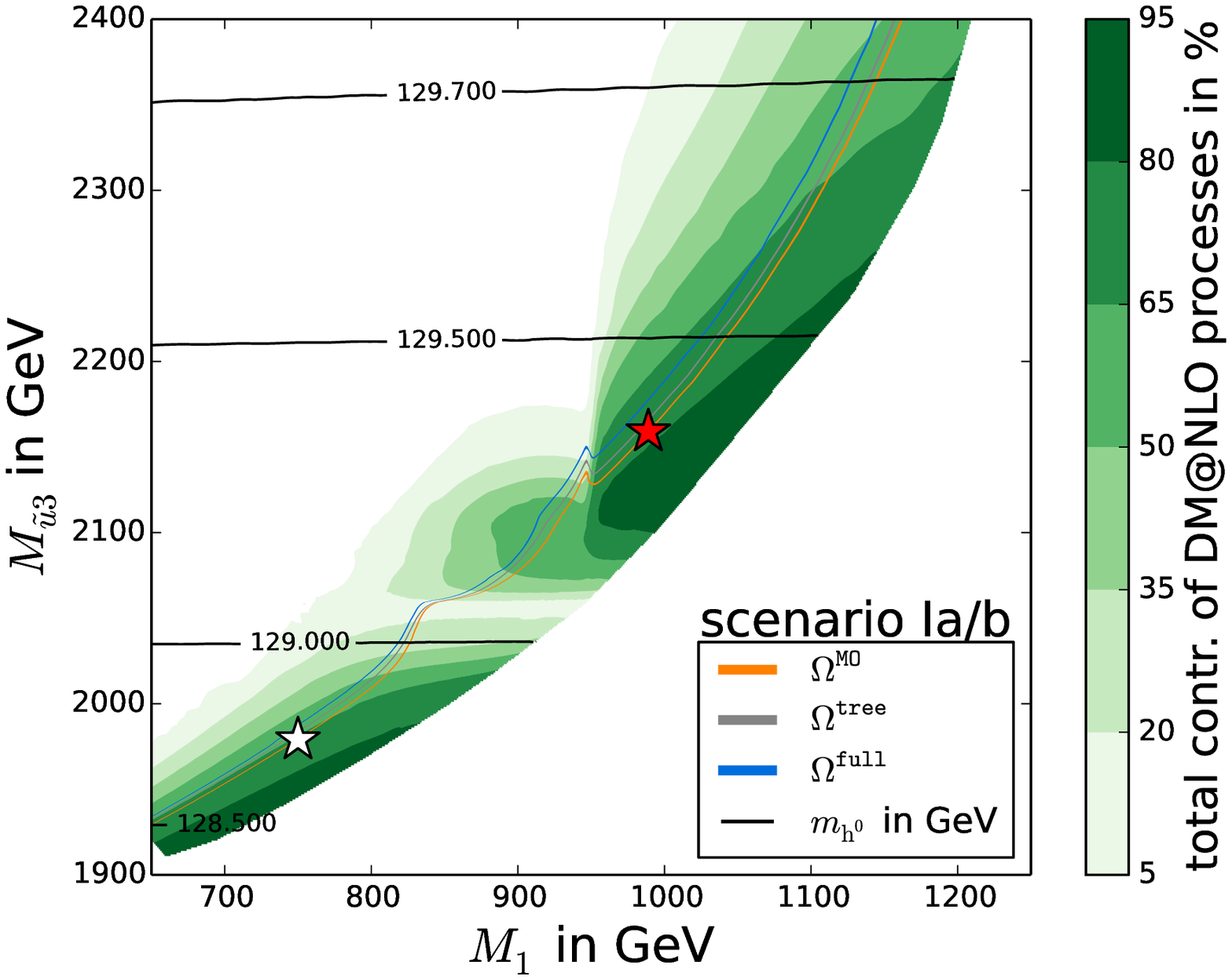}
 	\includegraphics[width=0.49\textwidth]{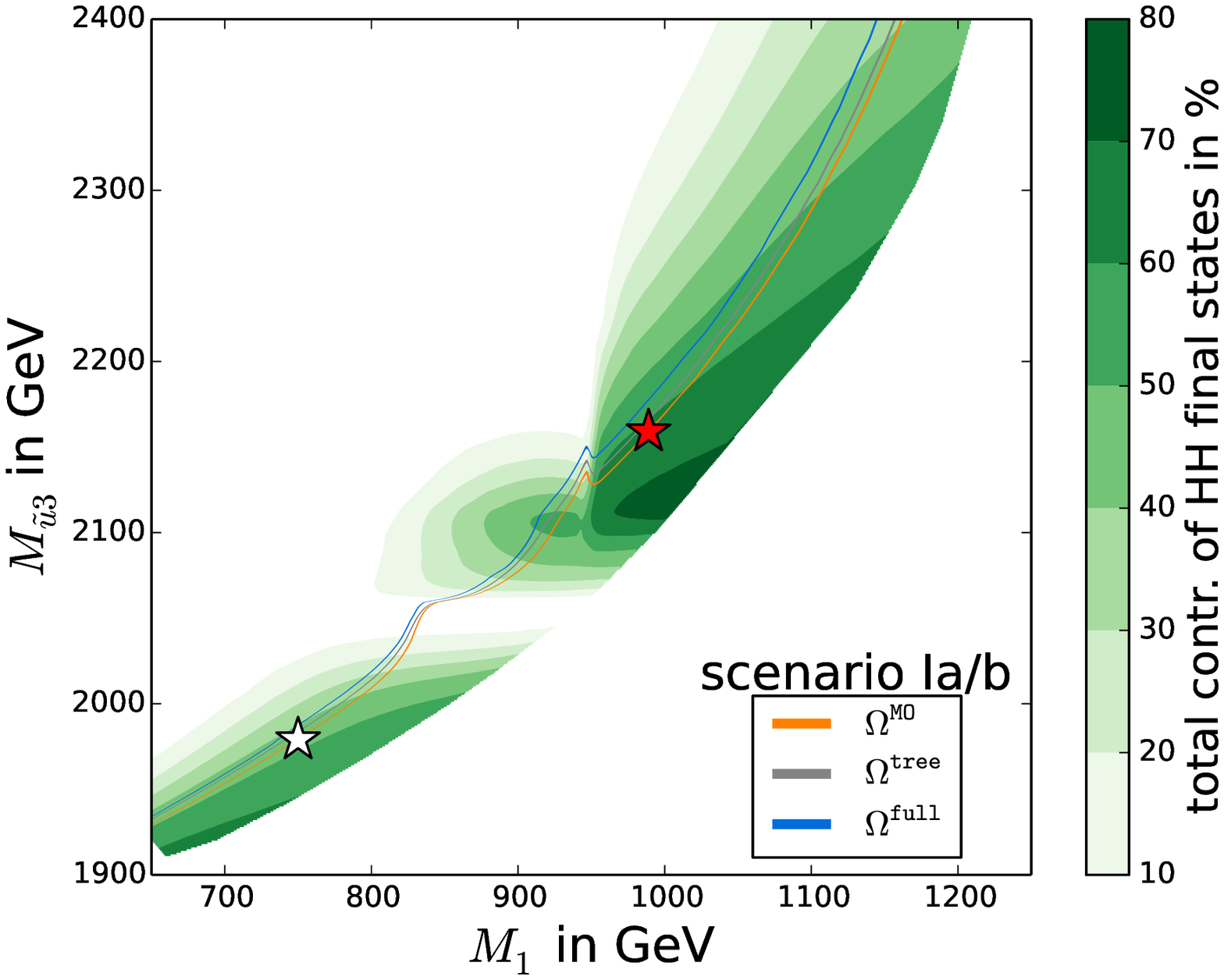}
 	\includegraphics[width=0.49\textwidth]{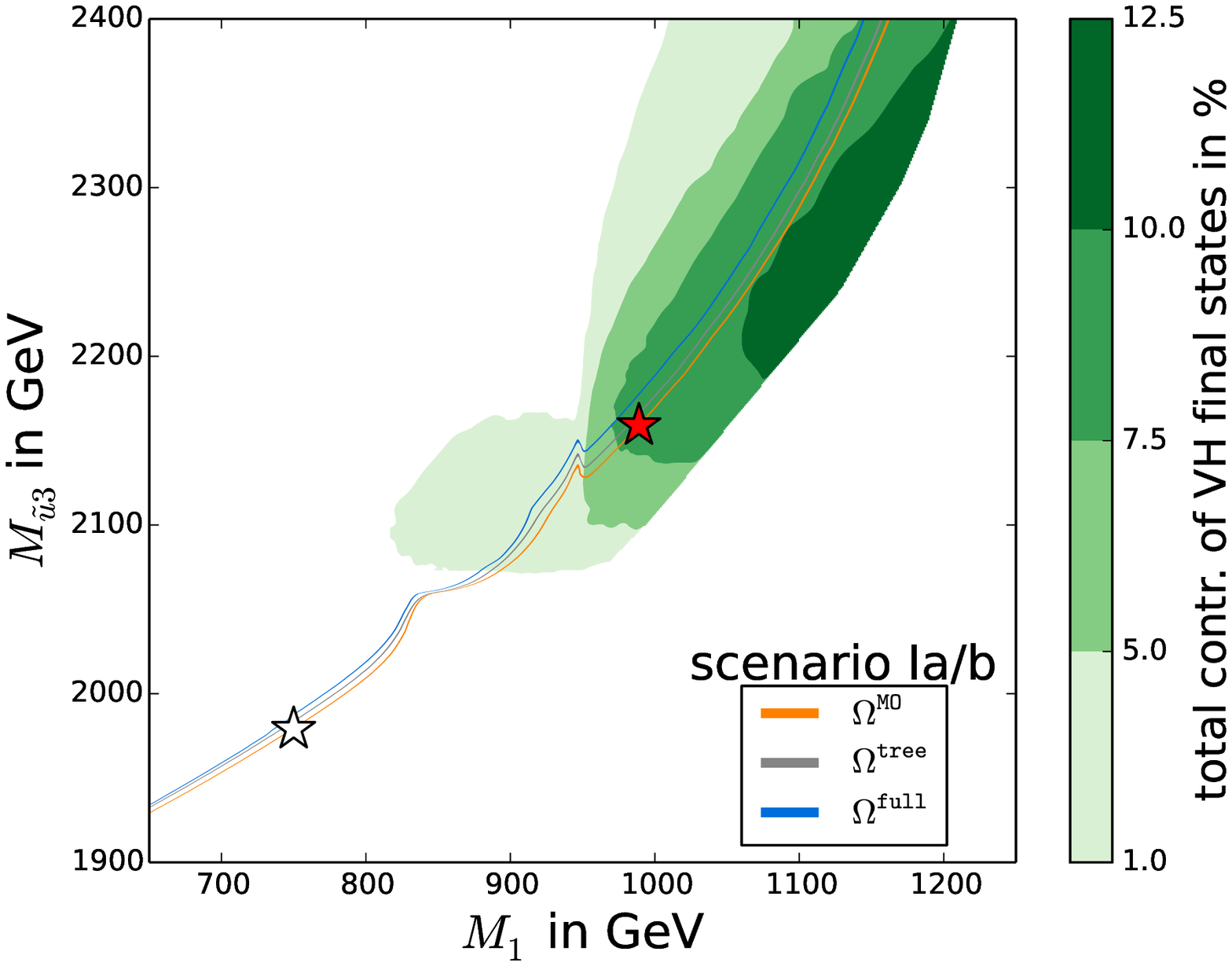}
 	\includegraphics[width=0.49\textwidth]{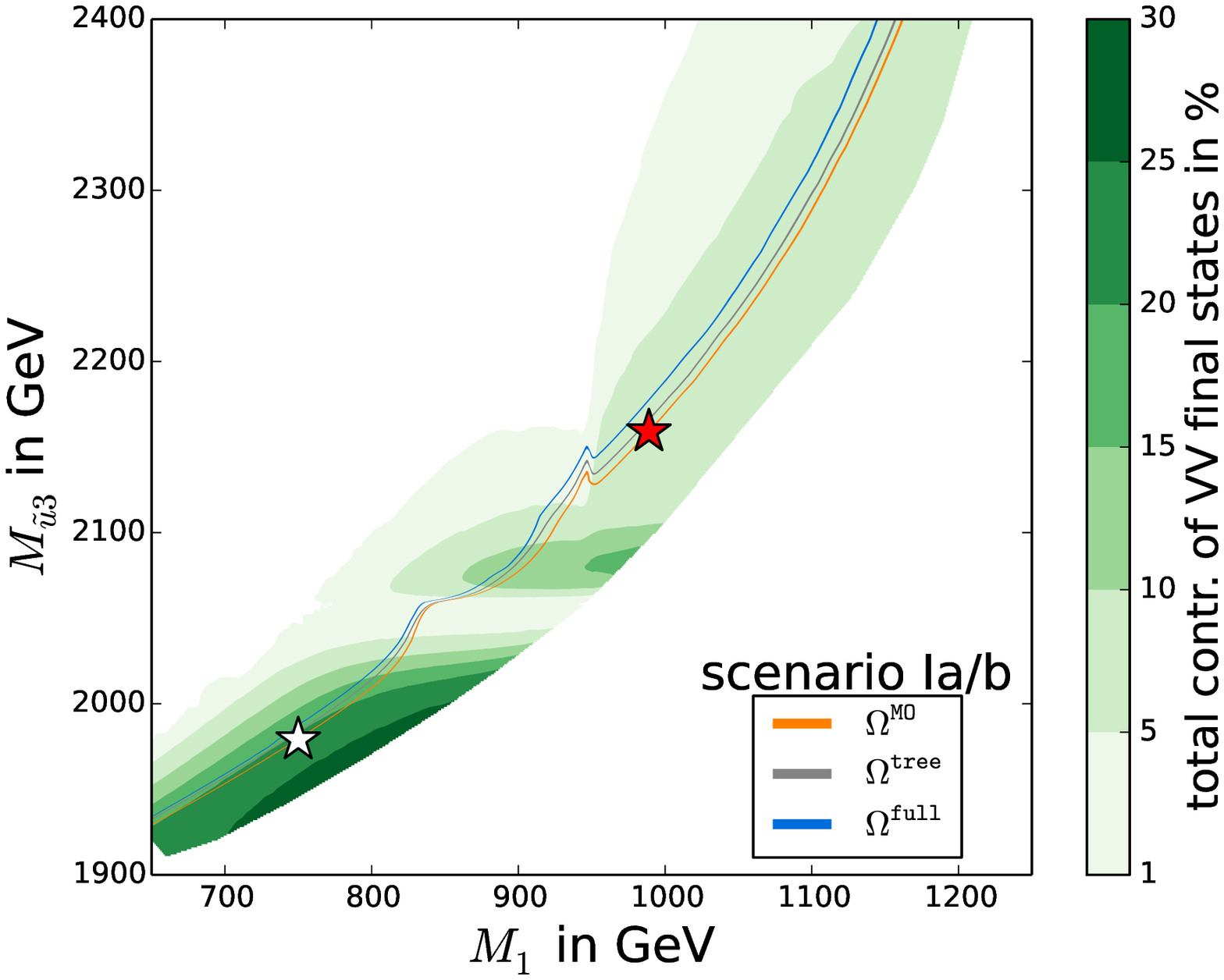}
 	\includegraphics[width=0.49\textwidth]{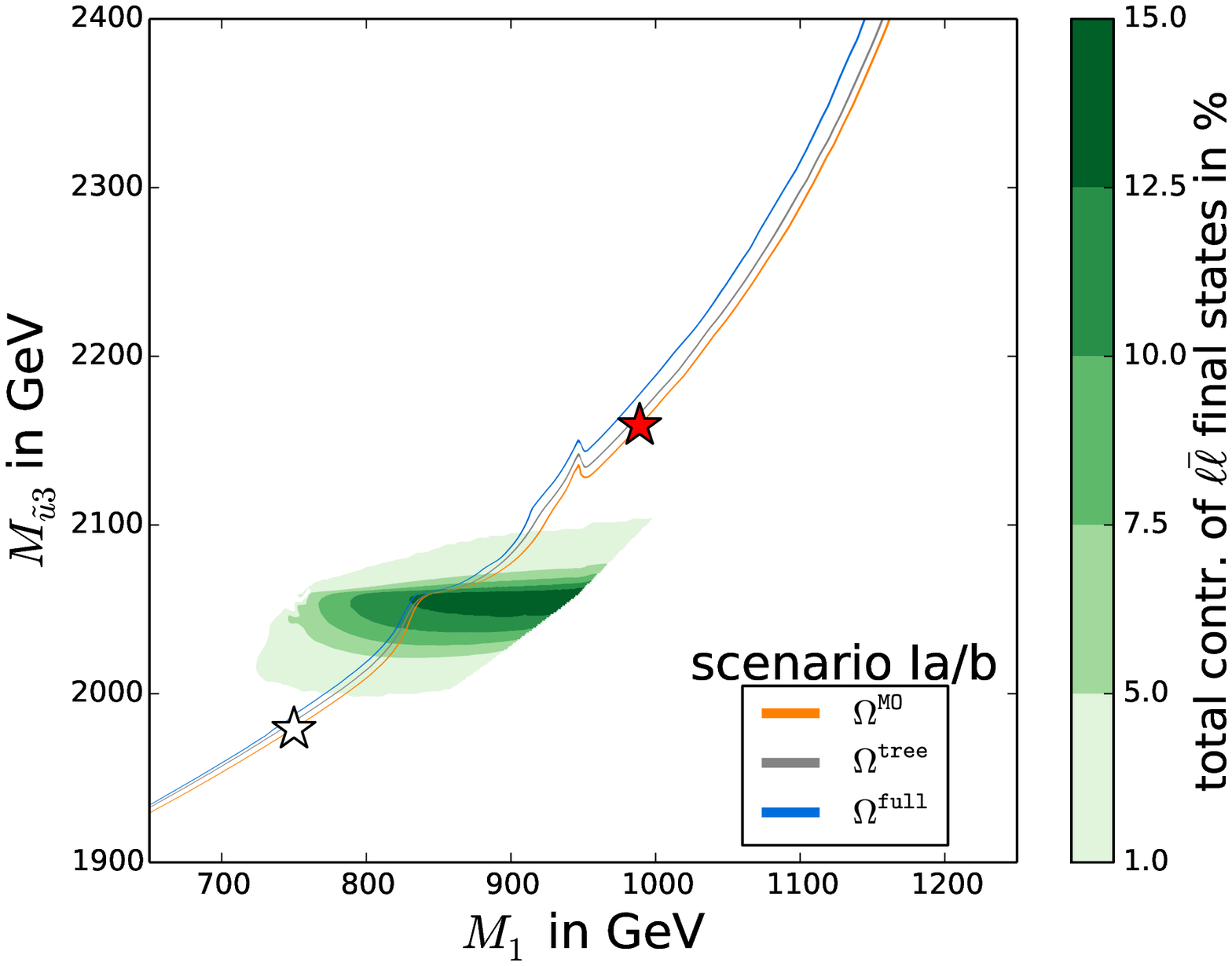}
	\caption{Planck-compatible relic density bands (see Eq.\ \ref{Planck}) in the $M_1$--$M_{\tilde{u}_3}$ plane 
	surrounding scenarios \RM{1}a and \RM{1}b. The calculation includes \MO\ (orange), our tree-level (grey) and 
	our full corrections (blue). The white and red stars mark the positions of our reference scenarios \RM{1}a and 
	\RM{1}b. The black lines in the upper left plot show the deviation between \MO\ and our full result in percent. 
	In the upper right plot the black lines stand for the mass of the lightest Higgs boson $m_{h^0}$ in GeV. For further explanations see the text.}
	\label{fig10}
\end{figure*}

\begin{figure*}
	\includegraphics[width=0.49\textwidth]{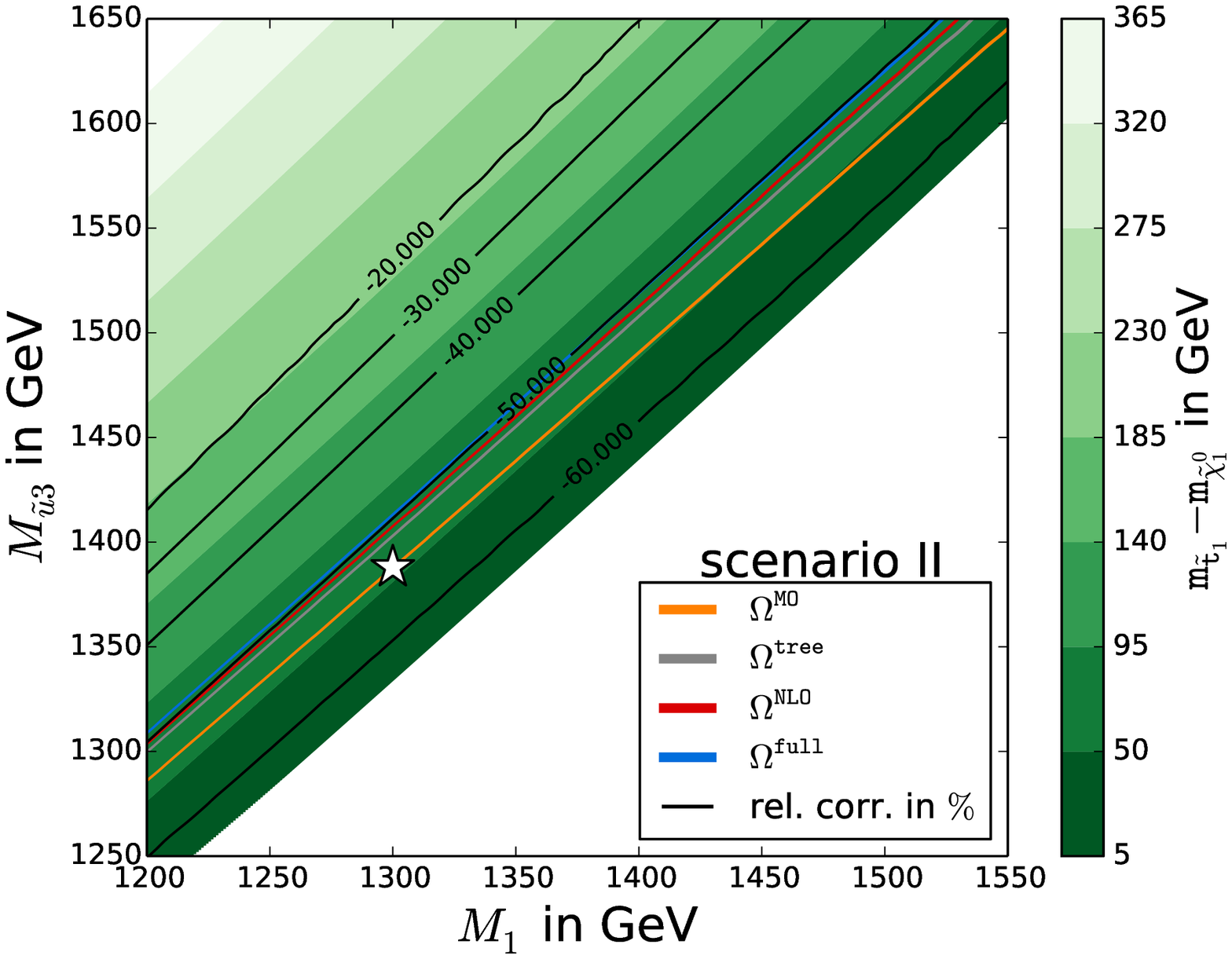}
	\includegraphics[width=0.49\textwidth]{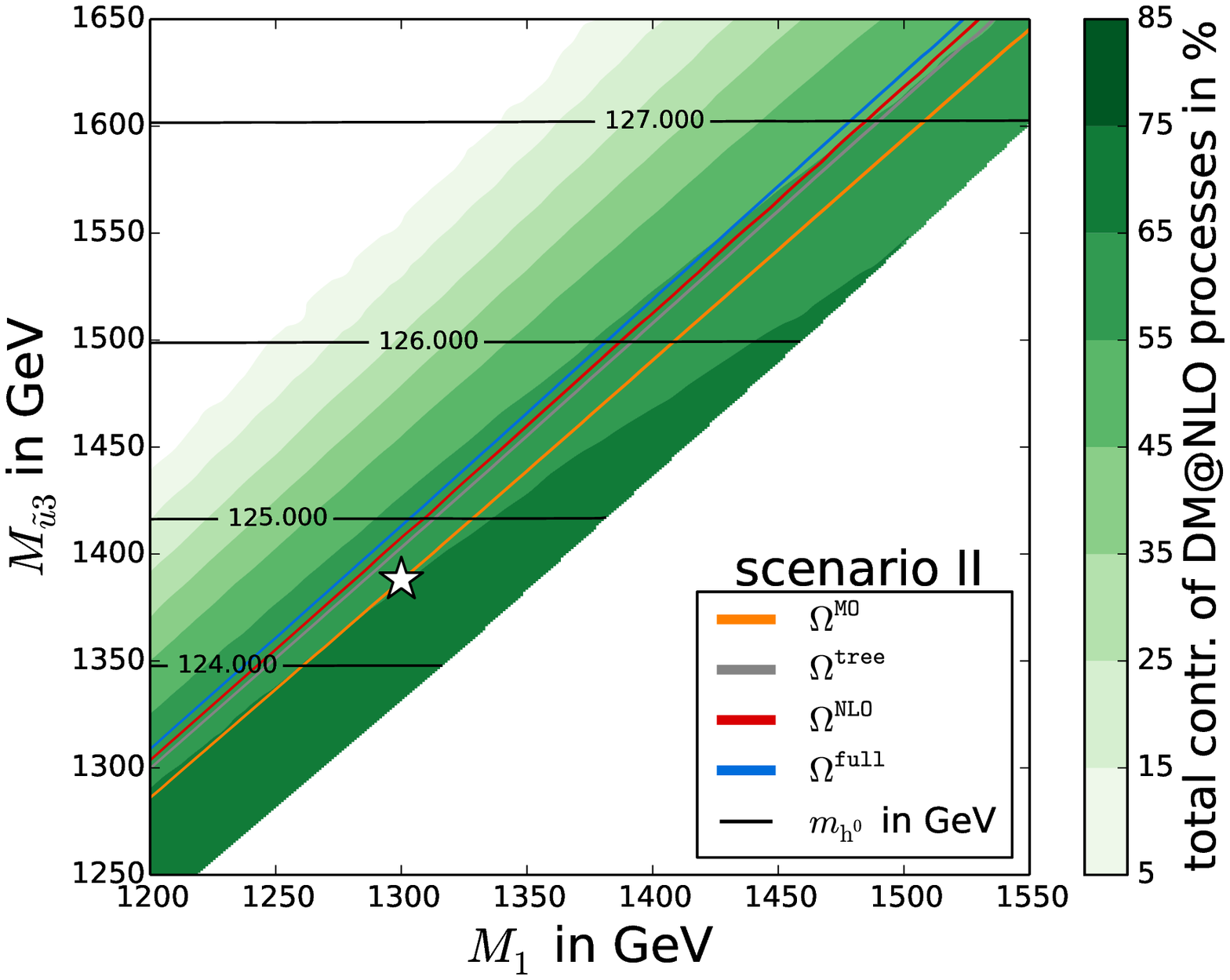}
	\includegraphics[width=0.49\textwidth]{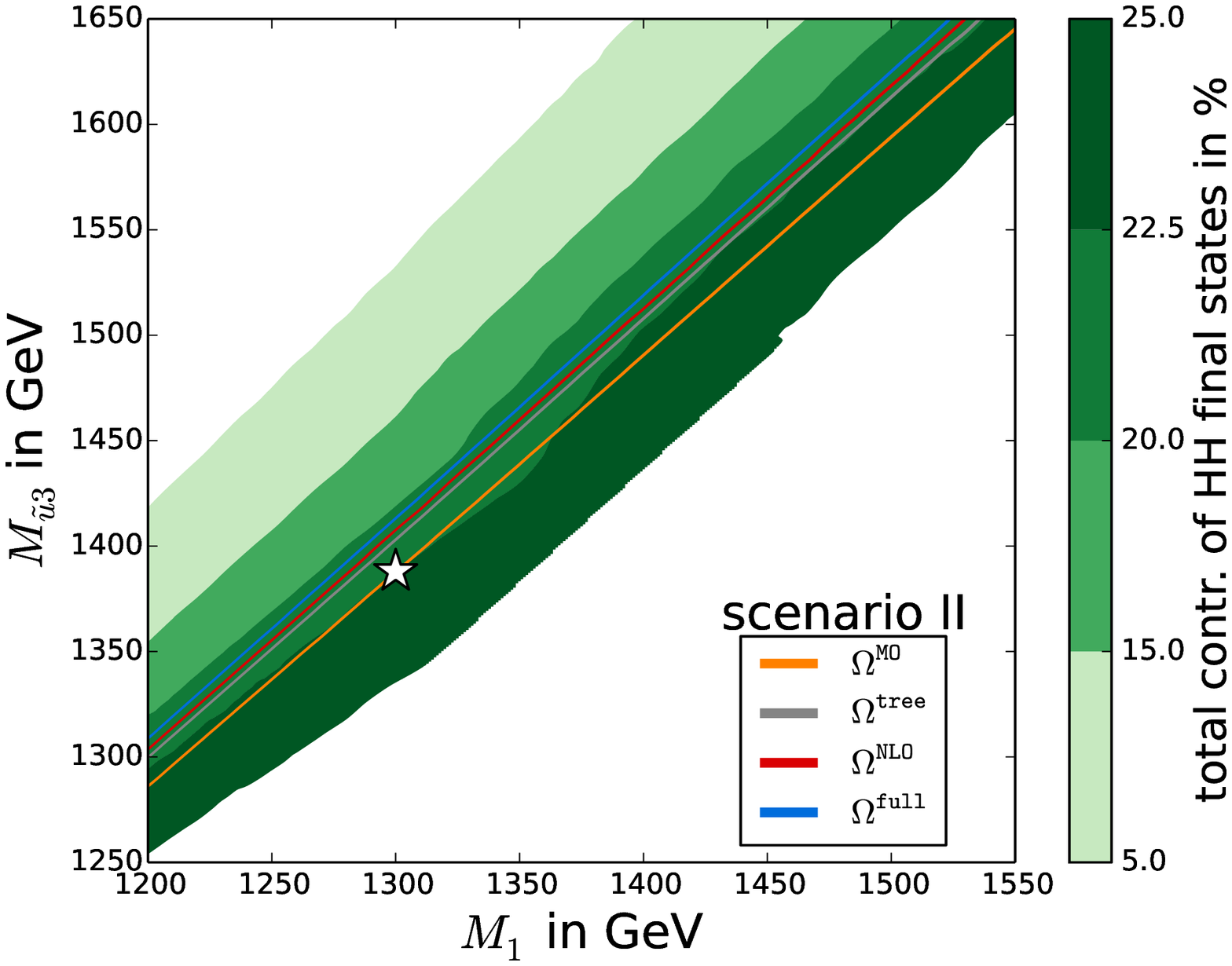}
	\includegraphics[width=0.49\textwidth]{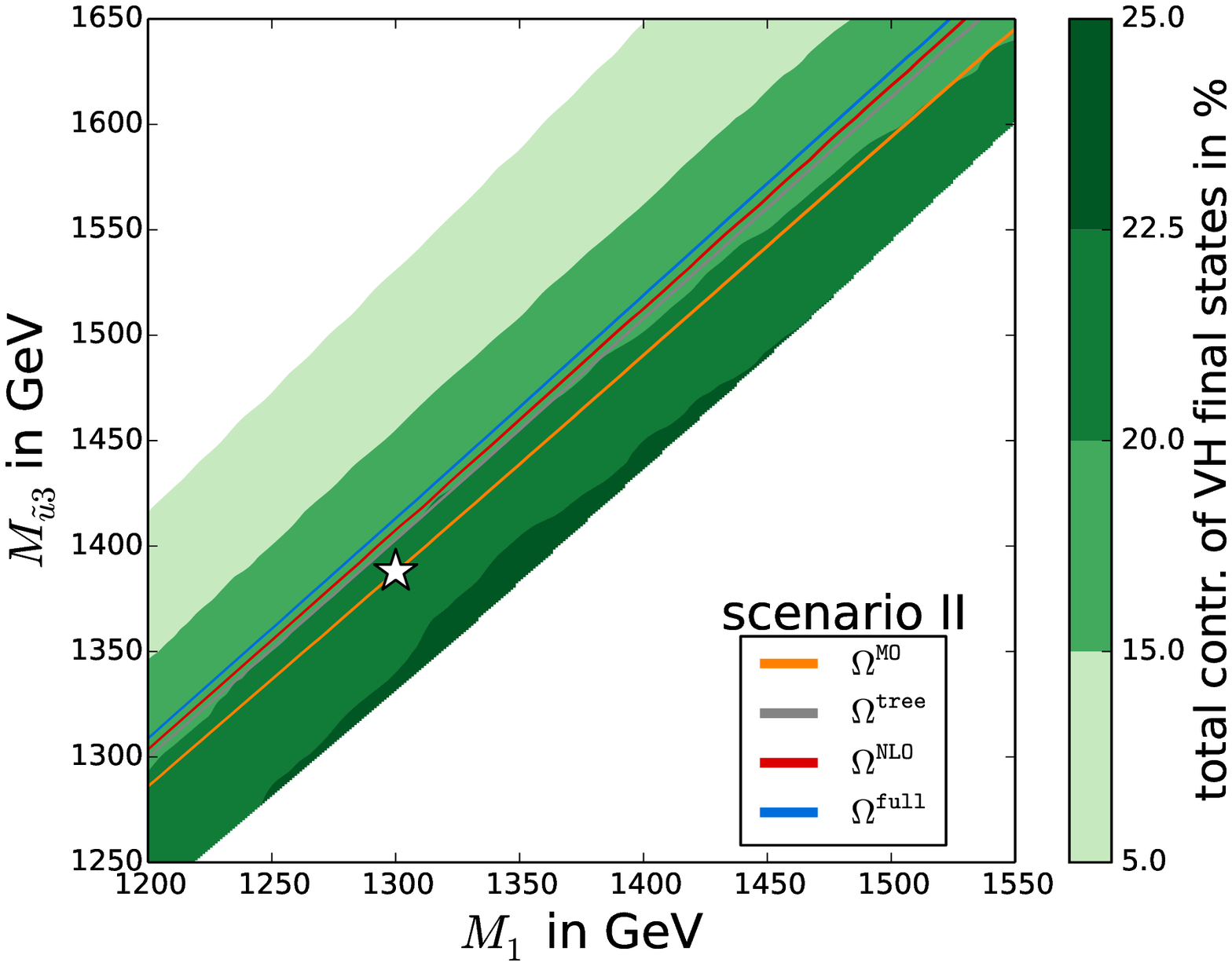}
	\includegraphics[width=0.49\textwidth]{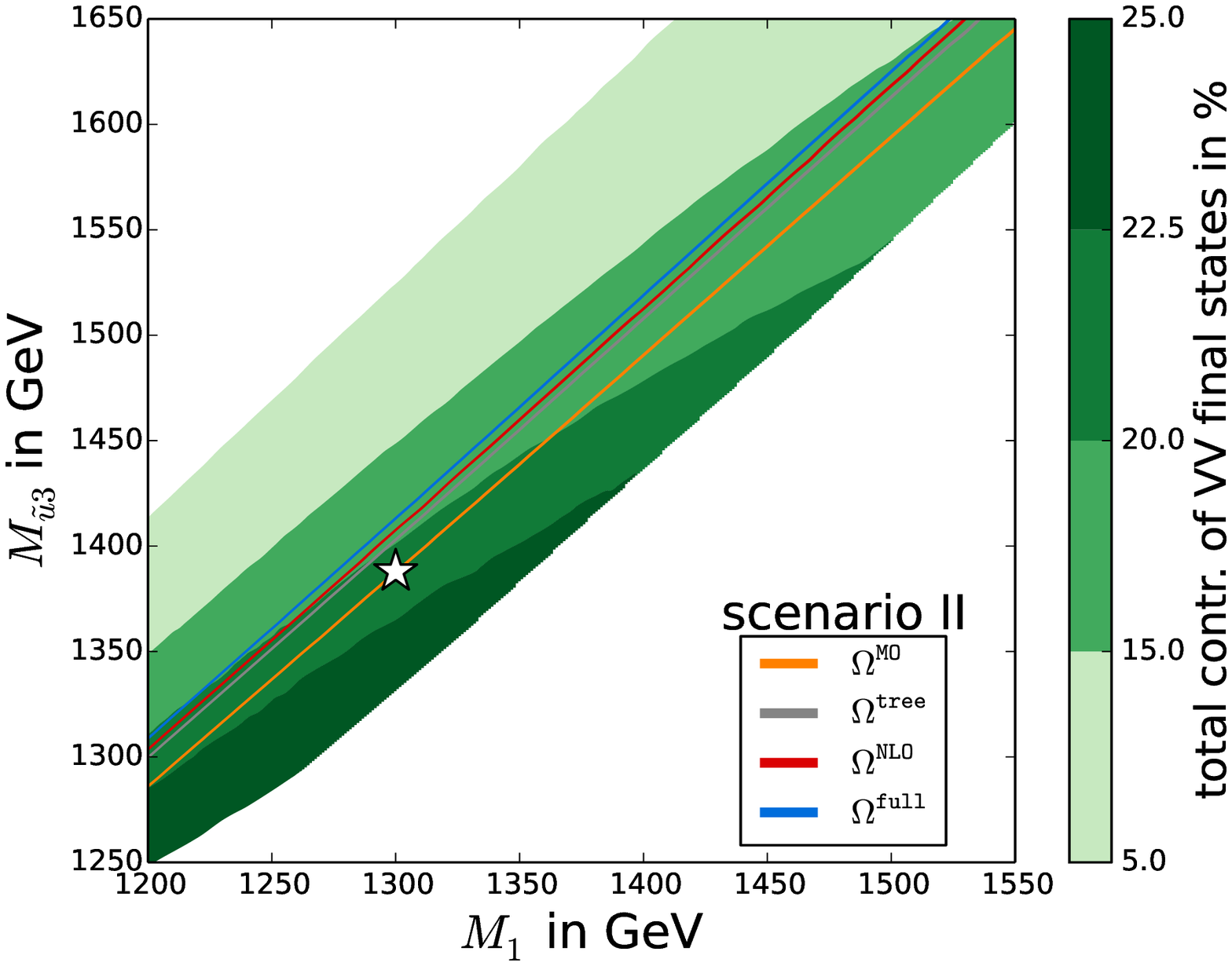}
	\caption{Same as Fig.\ \ref{fig10} for scenario \RM{2}, but here the plot for the $\ell \bar{\ell}$-final states is left out (see the text). We further added the NLO result in red.}
	\label{fig11}
\end{figure*}

In this subsection, we investigate the impact of our corrections on the neutralino relic density $\Omega_{\tilde{\chi}^0_1}h^2$. 
For the following analysis, we have implemented our results into a computer code called {\tt DM@NLO} that can be linked to \MO. In total we correct 24 different final states of $\tilde{t}_1\tilde{t}^*_1$ pair annihilation. 
All other processes, which do not subsume under the processes listed in Eqs.\ (\ref{channels1}) -- (\ref{channels4}) or the processes named in Sec.\ \ref{Further subtleties} (see Fig.\ \ref{fig7}) are provided by CalcHEP at effective tree-level.
Although most of them contribute only marginally to the final relic density, the relevance of each of the different processes is \textit{a priori} unknown as it depends strongly on the specific scenario. This makes a comprehensive study of each point of the parameter space necessary.

As the NLO corrections are more time consuming than the regular tree-level calculation,
we optimize our numerical evaluation by calculating the NLO corrections only for processes which contribute more than 1$\%$ to the total annihilation cross section.
This is in accordance with the current experimental precision of $\Omega_{\tilde{\chi}^0_1}h^2$, which is around 2$\%$ at 1$\sigma$ confidence level. 
The remaining channels are either replaced for consistency by our tree-level or are left unchanged.

We present our results in the $M_1$-$M_{\tilde{u}_3}$ plane of the pMSSM parameter space defined in Sec.\ \ref{Pheno}. These two parameters 
influence directly the masses of the lightest neutralino and the lightest scalar top quark, respectively, and thus the 
mass splitting $m_{\tilde{\chi}_1^0}-m_{\tilde{t}_1}$ to which $\tilde{t}_1\tilde{t}^*_1$ pair annihilation is extremely 
sensitive with respect to the relic density. 
In our scenarios the lightest neutralino is always binolike and hence is its mass predominantly determined 
by the $M_1$ parameter. The lightest scalar top quark possesses a large admixture of $\tilde{t}_R$, the superpartner of the right-handed part of the top quark, 
and so the mass is also sensitive to the right-handed supersymmetry breaking parameter $M_{\tilde{u}_3}$ (see Tab.\ \ref{ScenarioProps}).

In Figs.\ \ref{fig10} and \ref{fig11}, we present scans around our reference scenarios of 
Tab.\ \ref{ScenarioList}. The orange band ($\Omega^{\tt MO}$) refers to the relic density 
$\Omega_{\tilde{\chi}^0_1}h^2$ obtained by \MO/{\tt CalcHEP}, the grey band ($\Omega^{\tt tree}$) indicates the 
prediction of the relic density $\Omega_{\tilde{\chi}^0_1}h^2$ where our tree-level calculation replaces the 
{\tt CalcHEP} result for the processes specified in Eqs.\ (\ref{channels1}) -- (\ref{channels4}), and the blue band 
($\Omega^{\tt full}$) shows the neutralino relic density $\Omega_{\tilde{\chi}^0_1}h^2$ as a result of our full 
calculation discussed in Sec.\ \ref{Technical_details}. We further added to Fig.\ (\ref{fig11}) in red the relic 
density obtained by our NLO calculation.

The experimental 1$\sigma$-uncertainty is reflected by the width of the three bands in Figs.\ \ref{fig10} and 
\ref{fig11}.
The narrow band demonstrates how constraining the assumption that the lightest neutralino $\tilde{\chi}^0_1$ 
accounts for the whole cold dark matter in the Universe actually is. 
We encounter a distinct separation between the bands corresponding to our 
tree-level result (grey) and the default result of \MO\ (orange) in all plots nearly 
everywhere over the whole $M_1-M_{\tilde{u}_3}$ plane. This separation gets even enhanced 
if one takes the NLO (red) or full (blue) corrections into account.
The black contour lines in the top left plots 
of Figs.\ \ref{fig10} and \ref{fig11} quantify more 
precisely the magnitude of the corrections between \MO\ and our full result. 
They amount up to roughly 50\% in Fig.\ \ref{fig11} 
and reach even more than 50\% in the cosmologically favored 
region of the corresponding plot of Fig.\ \ref{fig10}.
Within the same regions, our fully corrected result deviates from our tree-level by up to $25\%$
in Fig.\ \ref{fig11} and by nearly $40\%$ in Fig.\ \ref{fig10}.
One can further see in Fig.\ \ref{fig11} the importance of the NNLO Coulomb corrections for a precise 
estimation of the relic density. The full result deviates by far more than one standard deviation from 
our NLO result, which is visible in the splitting of the associated blue and red bands.
The deviation due to Coulomb corrections of NNLO and beyond even exceeds the size of our full NLO corrections.
Besides the fact, that for $v \approx \alpha_s$ the higher-order Coulomb corrections are roughly of the same size as the
leading-order Coulomb corrections, this result can be further traced back to a cancellation among the NLO 
contributions to the relic density.
Fig.\ \ref{fig8} shows that the NLO corrections at large $v$ tend to lower the tree-level cross section, 
whereas at lower $v$ the Coulomb corrections start to alter the cross section turning the NLO corrections to positive 
values. Since this transition happens to be for certain processes relatively close to the peak of the 
thermal distribution, the associated cancellation significantly lowers the 
total contribution of the NLO corrections to the relic density and in turn raises the importance of the 
throughout positive higher-order Coulomb corrections\footnote{Note that this also increases the dependence of the 
final relic density on the choice of $\mu_G$. We postpone a more detailed analysis to later investigations.}.

Apart from the corrections discussed above, Figs.\ \ref{fig10} and \ref{fig11} highlight several regions of parameter space where different processes dominate the total 
annihilation cross section. The cosmologically preferred region of parameter space lies along a line of almost constant mass difference between the LSP and the NLSP. In both scenarios,  the 
regions where the processes investigated in this analysis are important stretch along the favored region of parameter space. For scenarios \RM{1}a/b, 
one observes that for higher values of $M_{\tilde{u}_3}$ (that means for heavier scalar top quarks) along the favored 
region the processes with Higgs bosons in the final state dominate. On the other end of the favored 
region where $M_{\tilde{u}_3}$ and $M_{1}$ are smaller, the processes with a vector boson in the final state take over to be most important. Here, the stops are 
lighter, and two Higgs bosons in the 
final state are no longer kinematically allowed or are at least largely suppressed. The same observation but less 
pronounced holds for scenario \RM{2}, where in the last plot 
of Fig.\ \ref{fig11} one encounters an increasing relevance of vector-vector final states toward lower values of 
$M_{\tilde{u}_3}$ and $M_{1}$.

Although both scenarios fulfill the experimental bounds on the Higgs boson mass, only scenario \RM{2} falls into the vicinity of the experimentally favored mass $m_{h^0}$ while
the scenarios \RM{1}a and \RM{1}b already lie at the edge of the experimental constraint as given in Eq.\ (\ref{bound2}). The mass of the lightest Higgs boson is mainly driven by the 
$M_{\tilde{u}_3}$ parameter as it determines the mass $m_{\tilde{t}_1}$ in our scenarios. The parameter $M_{\tilde{u}_3}$ therefore influences the mass splitting between the top quark and 
its superpartner $\tilde{t}_1$, which in turn enters the mass corrections of the mass of the lightest Higgs boson (see Eq.\ (\ref{Higgs_mass})).
\begin{figure*}
	\includegraphics[width=0.49\textwidth]{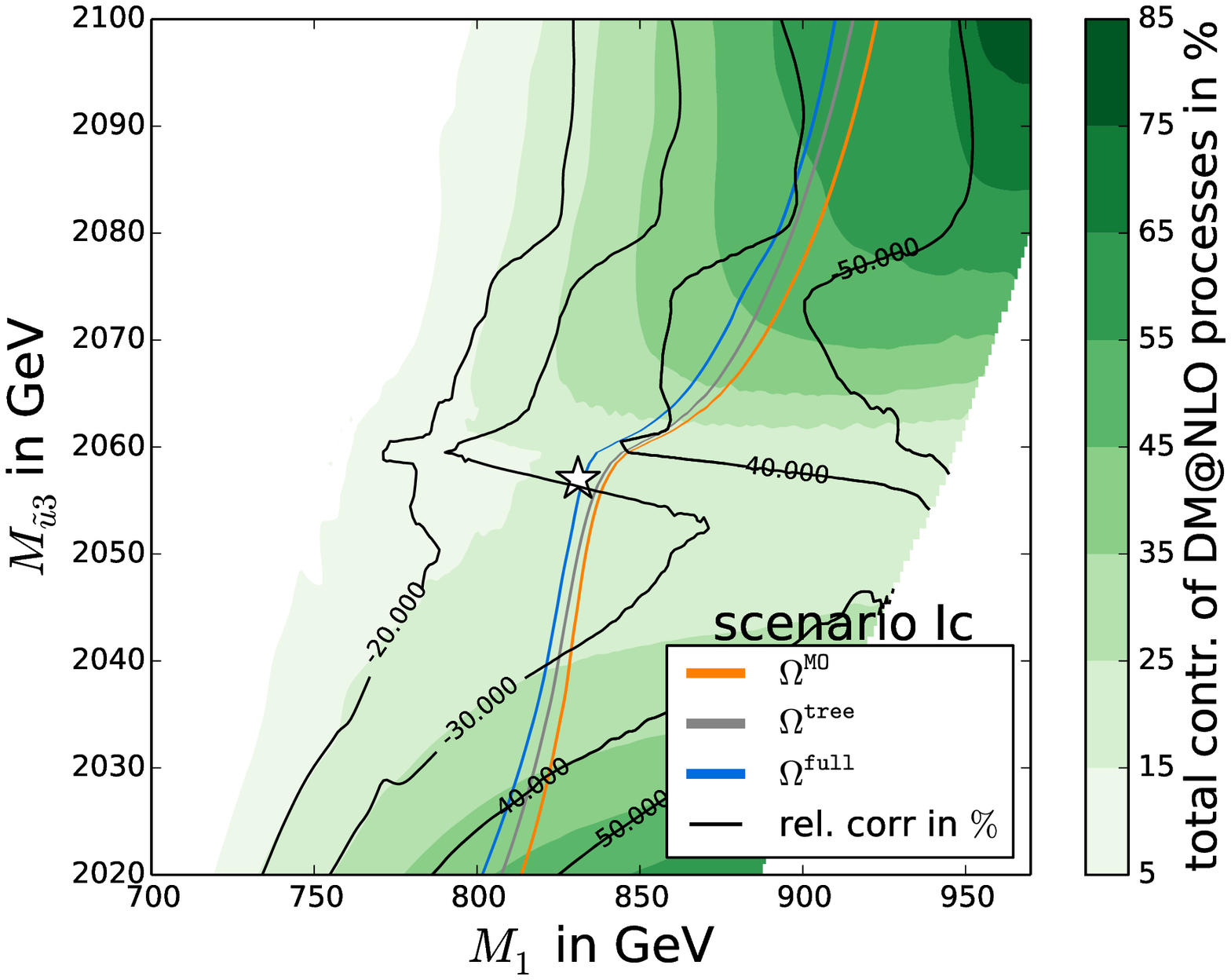}
	\includegraphics[width=0.49\textwidth]{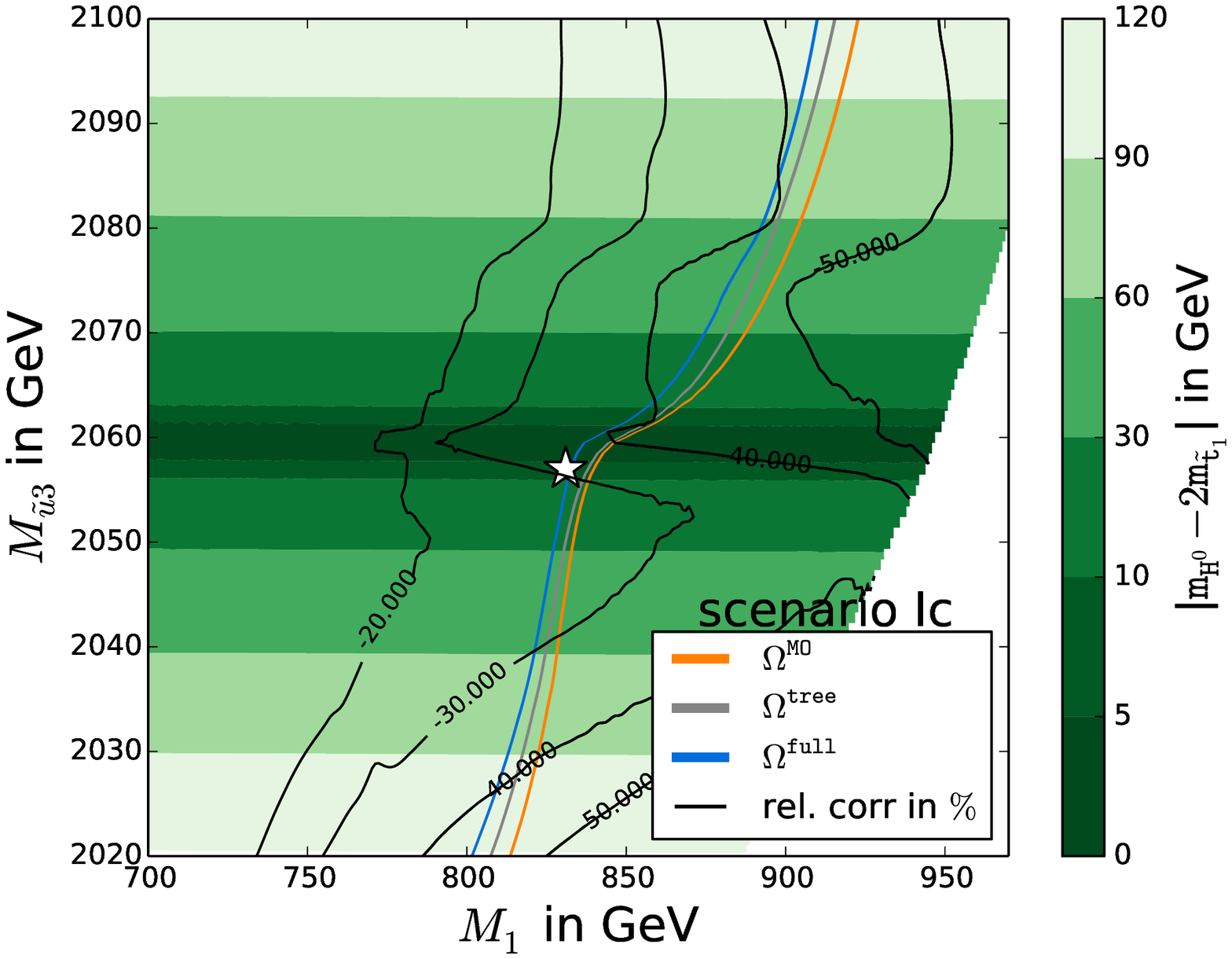}
        \includegraphics[width=0.49\textwidth]{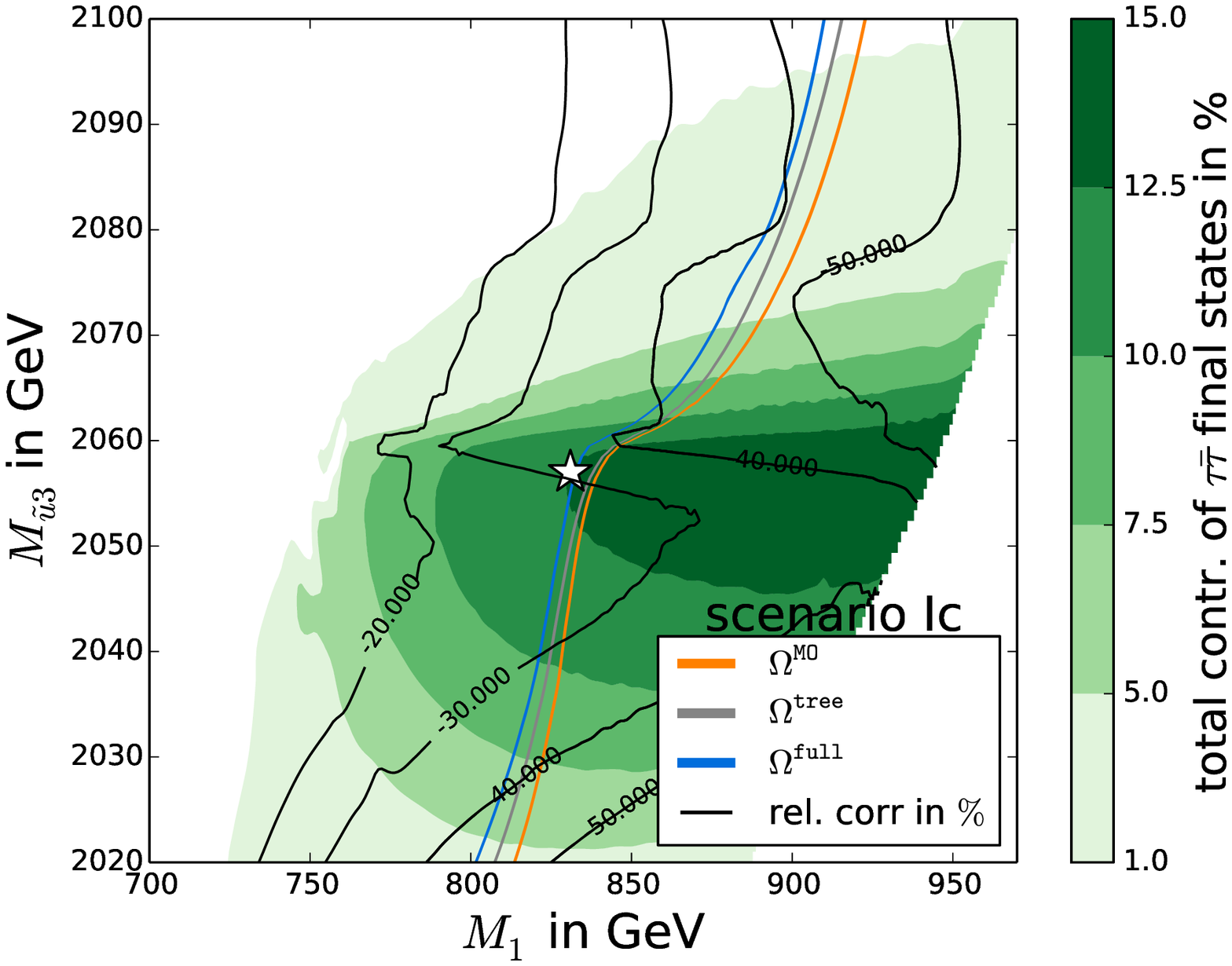}
	\caption{Scan over the scenarios \RM{1}a/b-plane. The white star marks the position of scenario \RM{1}c ($M_1=831$ GeV, $M_{\tilde{\mathrm{u}}_3}=2057$ GeV) further analyzed in Fig.\ \ref{fig13}.}
	\label{fig12}
\end{figure*}

\begin{figure*}
	\includegraphics[width=0.49\textwidth]{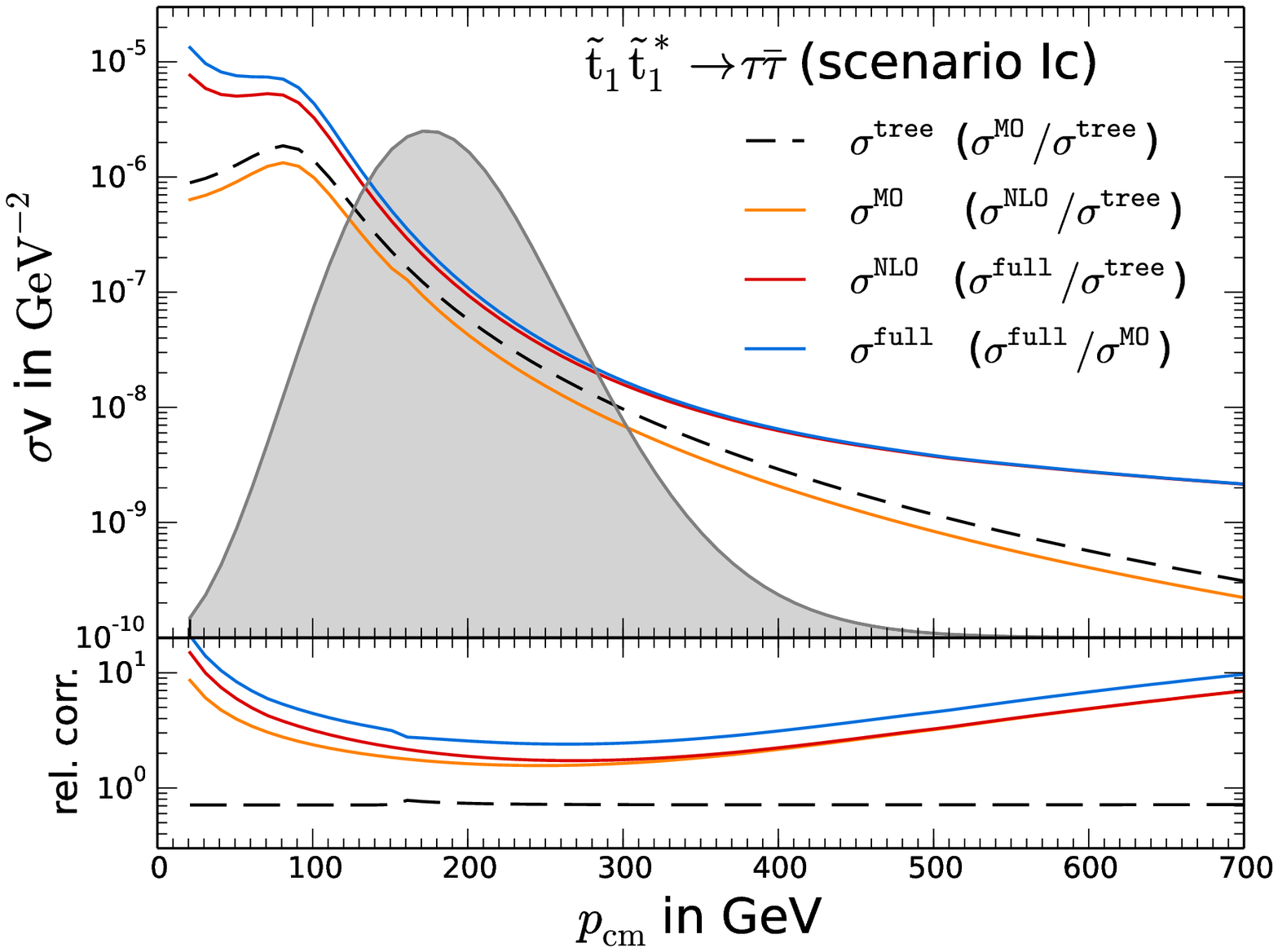}
	\includegraphics[width=0.49\textwidth]{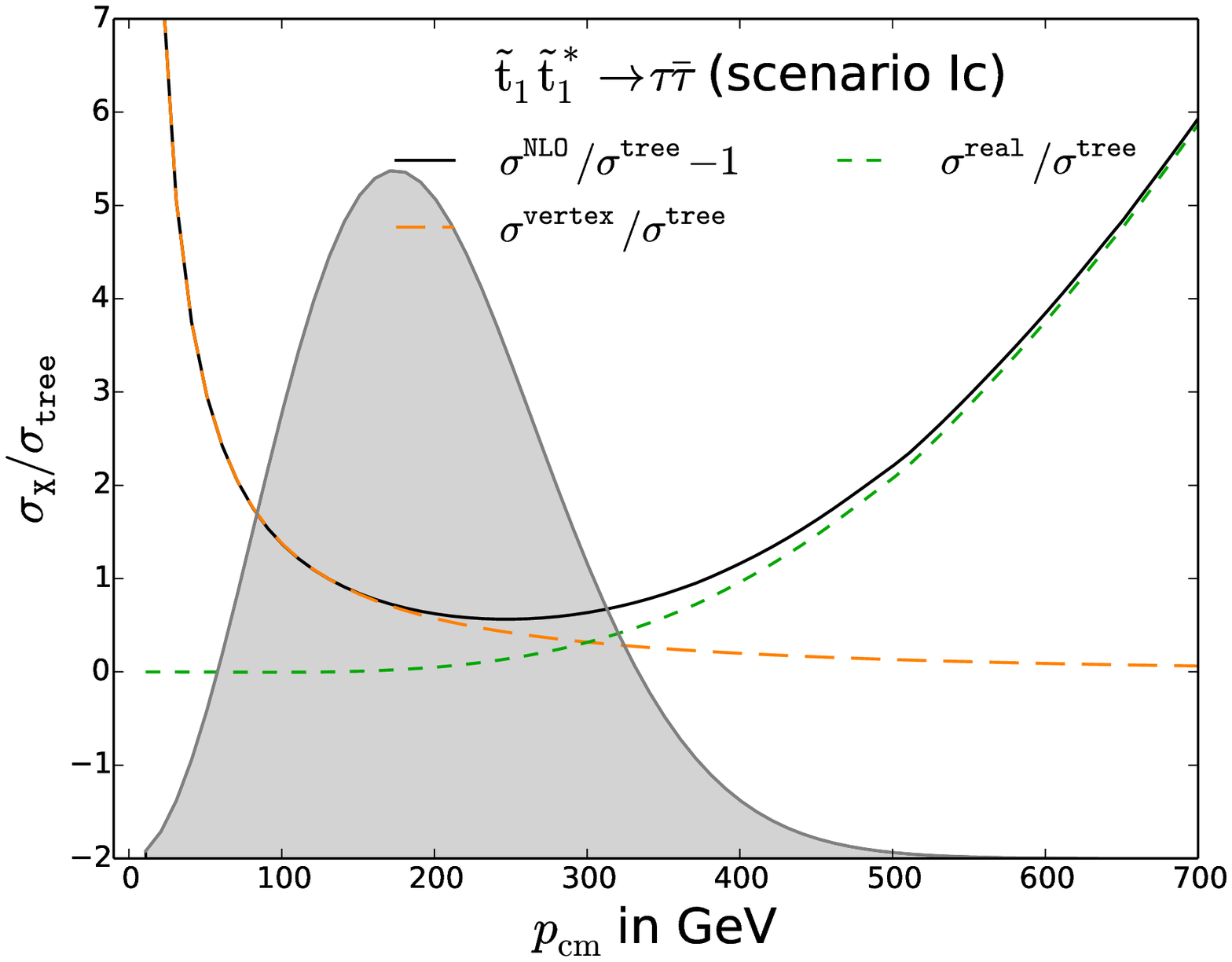}
	\caption{Cross sections and NLO contributions to scenario \RM{1}c of Fig.\ \ref{fig12}}
	\label{fig13}
\end{figure*}
Another interesting contribution with electroweak final states, which we have not mentioned yet, is the annihilation of scalar top quarks into 
lepton-antilepton pairs. Although this process is 
not the leading contribution to the total cross section in any of our scenarios, there is a region in the 
$M_{\tilde{u}_3}$-$M_{1}$ plane shown in the bottom-right plot of Fig.\ \ref{fig10}, where 
the process with $\tau \bar{\tau}$ final state contributes as much as 13\%. 
In Fig.\ \ref{fig12} we show a zoom into this area of enhanced $\tau \bar{\tau}$ contributions. 
It can be observed that the enhancement of the $\tau \bar{\tau}$ final state is due to an $s$-channel resonance caused 
by the heavier Higgs $H^0$  together with the Yukawa coupling, which for $\tan \beta=16.3$ favors the down-type fermions. 
Interestingly, the corrections to this process are significant enough to cause a shift of the relic density of
more than 20\% relative to our tree-level and even of more than 30\% relative to \MO\ despite the fact that its contribution 
is comparatively low. The reason is that the annihilation into 
$\tau \bar{\tau}$ proceeds only through an $s$-channel exchange of vector and Higgs bosons. As can be seen in Fig.\ \ref{fig9}, for all other final states the corrections from 
the vertex and the box diagrams cancel each other and lead to a reduction in the total correction. This is, however, not the case for $\tau$-leptons in the final state as no box 
diagrams exist, and thus this cancellation cannot take place. 
For further discussion we introduced a representative scenario \RM{1}c marked by the white star in Fig.\ \ref{fig12}. The relevant cross section contributions for this parameter point are shown in more detail in Fig.\ \ref{fig13}. 
We see that the corrections to the annihilation into $\tau \bar{\tau}$ are dominated by the vertex corrections and the 
real correction with the corresponding large Coulomb enhancement of the vertex corrections for small $p_{\mathrm{cm}}$.
One observes that starting at the $H^0$-resonance at around $p_{\mathrm{cm}}=80$ GeV (first plot of Fig.\ \ref{fig13}) the 
corrections comprise large Coulomb corrections stemming from the vertex diagrams. Later for larger $p_{\mathrm{cm}}$, 
the corrections are dominated by the relatively large contributions of the $2 \to 3$ processes (see the second plot of 
Fig.\ \ref{fig13}) due to the phase-space enhancement of the $2\to 3$ final states, which sets in already for much 
lower $p_{\mathrm{cm}}$ because of the small $\tau$-mass. Finally note that the s-wave contribution to the 
stop-annihilation cross section into $\ell \bar{\ell}$ final states is suppressed by a factor 
$(m_{\ell}/m_{\tilde{t}_1})^2$. Therefore a more elaborate treatment, which takes the full 
Coulomb corrections for the $p$-wave into account, may lead to relative corrections on the particular cross section, which 
are less suppressed than $\mathcal{O}(\alpha_s^2)$ compared to the leading-order (see Sec.\ \ref{Coulomb corrections}). 
However, as the leptons unfold their main impact on the relic density in the vicinity 
of the $H^0$-resonance, this in turn decreases the impact of the $p$-wave contributions (see the left plot in Fig.\ (\ref{fig13})). 
Hence, we leave this for further investigations.
\section{Conclusions}
\label{Conclusion}
An important mechanism for enhancing the annihilation cross section of the lightest neutralino 
in order to meet the experimentally determined value for the relic density $\Omega_{\tilde{\chi}_1^0} h^2$ is 
(co)annihilation processes of nearly mass degenerate particles. 
A theoretically well-motivated candidate for such (co)annihilation processes is the lightest stop $\tilde{t}_1$. 
Motivated by previous analyses \cite{ntnt2bb,ChiChi2qq,DMNLO_Stop1}, we investigated the impact of 
$\tilde{t}_1 \tilde{t}^*_1$ annihilation into electroweak final states on the neutralino relic density including the full $\mathcal{O}(\alpha_{\mathrm{s}})$ 
corrections as well as the Coulomb corrections due to the exchange of soft gluons between the incoming stop-antistop pair.

We further explored their impact on the neutralino relic density $\Omega_{\tilde{\chi}_1^0}h^2$ within
the phenomenological MSSM. For this purpose, we chose three reference scenarios, which are allowed by current 
experimental constraints and possess a rich variety of stop annihilation channels contributing to the relic 
density $\Omega_{\tilde{\chi}_1^0}h^2$. We performed large scans around these scenarios and compared the resulting 
$\Omega_{\tilde{\chi}_1^0}h^2$ by using the public code \MO $\hspace{1mm}$ with our results. We found that within these 
scenarios our results can change the neutralino relic density $\Omega_{\tilde{\chi}_1^0}h^2$ within the 
cosmologically favored region by more than 50$\%$, shifting the relic band by a few tens of GeV within some of the 
considered pMSSM parameters. They are therefore larger than the current experimental uncertainty coming from the latest 
Planck data. In these cases, both the full $\mathcal{O}(\alpha_{\mathrm{s}})$ corrections as well as the 
Coulomb corrections of $\mathcal{O}(\alpha_s^2)$ and beyond turned out to have a sizable impact on the cross sections within the kinematically relevant region. 
Further, we have split the annihilation cross section into contributions stemming separately from different types of final states and analyzed 
vector-vector, vector-Higgs, Higgs-Higgs and lepton-antilepton final states. Although the Higgs-Higgs final 
states turned out to be enhanced by large couplings due to a large  $A_{t}$ favored by scenarios containing a light 
stop, we also found regions within the parameter space where vector-vector and vector-Higgs final states contribute sizably to 
$\Omega_{\tilde{\chi}_1^0}h^2$. The lepton-antilepton final states do not contribute as much as the other final 
states, but nevertheless their corrections are sizable and can lead to a significant change in $\Omega_{\tilde{\chi}_1^0}h^2$ due to the absence of large cancellations between 
box and vertex corrections.

We conclude, that the identification of cosmologically favored regions at the currently available level of precision 
requires taking into account the next-to-leading-order as well as the Coulomb corrections including those investigated in this work.

\acknowledgments

The authors would like to thank A.~Pukhov for providing us with the necessary functions to implement our results into the {\MO} code 
and P.~Steppeler for useful discussions. This work is supported by the Helmholtz Alliance for Astroparticle 
Physics. The work of J.H.\ was supported by the London Centre for TeraUniverse Studies (LCTS), using funding from 
the European Research Council via Advanced Investigator Grant No.\ 26735.
\\
\appendix
\section{Hypergeometric function} \label{Appendix:Hypergeometric function}

The hypergeometric function is defined as 
\begin{align}
	\label{Coulomb_8}
	& _pF_q(a_1,a_2,...,a_p;b_1,b_2,...,b_q;z) = \\ 			
	& ~~~~~~~~~~~~~~~~~~~~~ \sum^{\infty}_{n=0} \frac{(a_1)_n(a_2)_n \dots (a_p)_n}{(b_1)_n(b_2)_n \dots (b_q)_n}\frac{z^n}{n!} \nonumber
\end{align}
with the restriction $b_{\mathrm{i}}\neq0,-1,...$ for $i = 1, 2, ..., q$, where $(x)_n = \Gamma(x+n)/\Gamma(x)$ 
are the Pochhammer symbols.
\begin{widetext}
The series defined by Eq.\ (\ref{Coulomb_8}) converges for 
$_4\mathrm{F}_3(1,1,1,1;2,2,1-\kappa;1)$, if 
\beq
	\Re \Biggr\{ \sum_{\mathrm{n}=1}^{\mathrm{q}} \mathrm{b}_{\mathrm{n}} - \sum_{\mathrm{n}=1}^{\mathrm{q+1}}\mathrm{a}_{\mathrm{n}} \Biggr\} > 0.
\eeq
To improve on the convergence of this series we have repeatedly employed
\begin{align}
	\label{Coulomb_9}
	_4\mathrm{F}_3(1,1,1,1;a,a,x;1) = & 
		\frac{1}{a^2x(x-2(2-a))(a-x)^2} \bigg[ a^2(x-1)^4 ~ _4\mathrm{F}_3(1,1,1,1;a,a,x+1;1) \\
	& +a(a-1)^3 x( 3a + 1 - 4x) ~ _4\mathrm{F}_3(1,1,1,1;a+1,a,x;1) \nonumber\\
	& +(a-1)^4x(x-a) ~ _4\mathrm{F}_3(1,1,1,1;a+1,a+1,x;1) \bigg] , \nonumber\\
	_4\mathrm{F}_3(1,1,1,1;a,b,x;1) = & 
		\frac{1}{a+b+x-4} \bigg[ \frac{(a-1)^4}{a(a-b)(a-x)} ~ _4\mathrm{F}_3(1,1,1,1;a+1,b,x;1) \\
	& + \frac{(b-1)^4}{b(b-a)(b-x)} ~ _4\mathrm{F}_3(1,1,1,1;a,b+1,x;1) 
	  + \frac{(x-1)^4}{x(x-a)(x-b)} \vphantom{X}_4\mathrm{F}_3(1,1,1,1;a,b,x+1;1) \bigg], \nonumber
\end{align}
which is valid for $x\neq -1,-2,\dots$ and 
$a, b \in \mathbb{N}/ \{0,1\}$, $a \neq b$ \cite{Sommerfeld1,Coulomb5}.
\end{widetext}

\newpage
\bibliographystyle{apsrev}

\end{document}